\documentclass[%
 reprint,
 superscriptaddress,
 amsmath,amssymb,
 aps,
prx,
floatfix,
noeprint,
longbibliography
]{revtex4-2}

\usepackage{siunitx}
\DeclareSIUnit\clight{\text{\ensuremath{c}}}
\DeclareSIUnit\echarge{\text{\ensuremath{e}}}
\DeclareSIUnit\mhz{\mega\hertz}
\DeclareSIUnit\khz{\kilo\hertz}
\DeclareSIUnit\mm{\milli\metre}
\DeclareSIUnit\cm{\centi\metre}
\DeclareSIUnit\ms{\milli\second}
\DeclareSIUnit\um{\micro\metre}
\DeclareSIUnit\ns{\nano\second}
\DeclareSIUnit\mev{\mega\electronvolt}
\DeclareSIUnit\nevsc{\nano\electronvolt\per\clight\squared}
\DeclareSIUnit\mevc{\mega\electronvolt\per\clight}
\DeclareSIUnit\mevsc{\mega\electronvolt\per\clight\squared}
\DeclareSIUnit\kev{\kilo\electronvolt}
\DeclareSIUnit\gevc{\giga\electronvolt\per\clight}
\DeclareSIUnit\kevsc{\kilo\electronvolt\per\clight\squared}
\DeclareSIUnit\evsc{\electronvolt\per\clight\squared}
\DeclareSIUnit\gesv{\per\giga\electronvolt\squared}
\DeclareSIUnit\gev{\per\giga\electronvolt}
\DeclareSIUnit\gevcc{\giga\electronvolt\per\cubic\cm}
\DeclareSIUnit[inter-unit-product =\ensuremath{\!\cdot\!}]{\ecm}{\echarge\cm}

\newcommand{\ka}{\ensuremath{\kappa_a}}
\newcommand{\gaNg}[1]{\ensuremath{g_{a#1\gamma}}}
\newcommand{\gadg}{\gaNg{d}}
\newcommand{\cgfa}{\ensuremath{\frac{C_G}{f_a}}}

\newcommand{\cdfa}{\ensuremath{\frac{C_d}{f_a}}}
\newcommand{\rLDM}{\ensuremath{\rho_{\mathrm{LDM}}}}
\newcommand{\dAC}{\ensuremath{d_{\mathrm{AC}}}}
\newcommand{\dACd}{\ensuremath{\dAC^d}}
\newcommand{\frev}{\ensuremath{f_\mathrm{rev}}}
\newcommand{\fsol}{\ensuremath{f_\mathrm{sol}}}
\newcommand{\fspin}{\ensuremath{f_\mathrm{spin}}}

\newcommand{\subit}[2]{\ensuremath{#1_\mathrm{#2}}}
\newcommand{\Alr}{\ensuremath{A_\mathrm{LR}}}
\newcommand{\Aip}{\ensuremath{A_\mathrm{IP}}}
\newcommand{\tstep}{\ensuremath{t_\mathrm{step}}}
\newcommand{\aest}{\ensuremath{\hat{A}}}
\newcommand{\pest}{\ensuremath{\hat{P}}}
\newcommand{\fc}{Feldman-Cousins}

\usepackage{comment}
\usepackage{graphicx}
\graphicspath{{figures/}}
\usepackage{dcolumn}
\usepackage{bm}
\usepackage[T1]{fontenc}
\usepackage{hyperref}
\usepackage{xcolor}

\usepackage[normalem]{ulem}

\begin{document}

\title{First Search for Axion-Like Particles in a Storage Ring   \\ Using  a Polarized Deuteron Beam} 

\author{S.\,Karanth}
\email{swathi.karanth@doctoral.uj.edu.pl}
\affiliation{Marian Smoluchowski Institute of Physics, Jagiellonian University, 30348 Cracow, Poland}

\author{E.J.\,Stephenson}
\email{stephene@iu.edu}
\affiliation{Indiana University, Department of Physics, Bloomington, IN 47405, USA}

\author{S.P.\,Chang}
\affiliation{Department of Physics, KAIST, Daejon 34141, Republic of Korea}
\affiliation{Center for Axion and Precision Physics Research, IBS, Daejon 34051, Republic of Korea}

\author{V.\,Hejny}
\affiliation{Institut f\"ur Kernphysik, Forschungszentrum J\"ulich, 52425 J\"ulich, Germany}

\author{J.\,Pretz}
\affiliation{Institut f\"ur Kernphysik, Forschungszentrum J\"ulich, 52425 J\"ulich, Germany}

\affiliation{III.\ Physikalisches Institut B, RWTH Aachen University, 52056 Aachen, Germany}

\affiliation{JARA--FAME (Forces and Matter Experiments), Forschungszentrum J\"ulich and RWTH Aachen University, Germany}

\author{Y.K.\,Semertzidis}
\affiliation{Department of Physics, KAIST, Daejon 34141, Republic of Korea}
\affiliation{Center for Axion and Precision Physics Research, IBS, Daejon 34051, Republic of Korea}

\author{A.\,Wirzba}
\affiliation{Institut f\"ur Kernphysik, Forschungszentrum J\"ulich, 52425 J\"ulich, Germany}
\affiliation{Institute for Advanced Simulation, Forschungszentrum J\"ulich, 52425 J\"ulich, Germany}

\author{A.\,Wro\'{n}ska}
\affiliation{Marian Smoluchowski Institute of Physics, Jagiellonian University, 30348 Cracow, Poland}

\author{F.\,Abusaif}
\affiliation{III.\ Physikalisches Institut B, RWTH Aachen University, 52056 Aachen, Germany}
\affiliation{Institut f\"ur Kernphysik, Forschungszentrum J\"ulich,  52425 J\"ulich, Germany}

\author{A.\,Aksentev}
\affiliation{Institute for Nuclear Research, Russian Academy of Sciences, 117312 Moscow, Russia}

\author{B.\,Alberdi}
\affiliation{III.\ Physikalisches Institut B, RWTH Aachen University, 52056 Aachen, Germany}
\affiliation{Institut f\"ur Kernphysik, Forschungszentrum J\"ulich,  52425 J\"ulich, Germany}

\author{A.\,Aggarwal}
\affiliation{Marian Smoluchowski Institute of Physics, Jagiellonian University, 30348 Cracow, Poland}

\author{A.\,Andres}
\affiliation{III.\ Physikalisches Institut B, RWTH Aachen University, 52056 Aachen, Germany}

\affiliation{Institut f\"ur Kernphysik, Forschungszentrum J\"ulich,  52425 J\"ulich, Germany}

\author{L.\,Barion}
\affiliation{University of Ferrara and Istituto Nazionale di Fisica Nucleare, 44100 Ferrara, Italy}

\author{I. Bekman}
\affiliation{Institut f\"ur Kernphysik, Forschungszentrum J\"ulich,  52425 J\"ulich, Germany}

\affiliation{{\rm now at:} Zentralinstitut f\"ur Engineering, Elektronik und Analytik, Forschungszentrum J\"ulich, 52425 J\"ulich, Germany}

\author{M.\,Bey{\ss}}
\affiliation{III.\ Physikalisches Institut B, RWTH Aachen University, 52056 Aachen, Germany}
\affiliation{Institut f\"ur Kernphysik, Forschungszentrum J\"ulich,  52425 J\"ulich, Germany}

\author{C. B\"ohme}
\affiliation{Institut f\"ur Kernphysik, Forschungszentrum J\"ulich, 52425 J\"ulich, Germany}

\author{B.\,Breitkreutz}
\affiliation{Institut f\"ur Kernphysik, Forschungszentrum J\"ulich,  52425 J\"ulich, Germany}

\affiliation{{\rm now at:}
GSI Helmholtzzentrum für Schwerionenforschung, 64291 Darmstadt, Germany}

\author{C.\,von Byern} 
\affiliation{III.\ Physikalisches Institut B, RWTH Aachen University, 52056 Aachen, Germany}

\affiliation{Institut f\"ur Kernphysik, Forschungszentrum J\"ulich, 52425 J\"ulich, Germany}

\author{N.\,Canale}
\affiliation{University of Ferrara and Istituto Nazionale di Fisica Nucleare, 44100 Ferrara, Italy}

\author{G.\,Ciullo}
\affiliation{University of Ferrara and Istituto Nazionale di Fisica Nucleare, 44100 Ferrara, Italy}

\author{S.\,Dymov}
\affiliation{University of Ferrara and Istituto Nazionale di Fisica Nucleare, 44100 Ferrara, Italy}

\author{N.-O. Fr\"ohlich}
\affiliation{Institut f\"ur Kernphysik, Forschungszentrum J\"ulich, 52425 J\"ulich, Germany}

\affiliation{{\rm now at}
DESY, Deutsches Elektronen-Synchrotron, 22607 Hamburg, Germany}

\author{R.\,Gebel}
\affiliation{Institut f\"ur Kernphysik, Forschungszentrum J\"ulich, 52425 J\"ulich, Germany}

\affiliation{GSI Helmholtzzentrum für Schwerionenforschung, 64291 Darmstadt, Germany}


\author{K.\,Grigoryev}
\affiliation{Institut f\"ur Kernphysik, Forschungszentrum J\"ulich, 52425 J\"ulich, Germany}

\affiliation{{\rm now at:}
GSI Helmholtzzentrum für Schwerionenforschung, 64291 Darmstadt, Germany}

\author{D.\,Grzonka}
\affiliation{Institut f\"ur Kernphysik, Forschungszentrum J\"ulich, 52425 J\"ulich, Germany}

\author{J.\,Hetzel}
\affiliation{Institut f\"ur Kernphysik, Forschungszentrum J\"ulich, 52425 J\"ulich, Germany}

\affiliation{{\rm now at:}
GSI Helmholtzzentrum für Schwerionenforschung, 64291 Darmstadt, Germany}

\author{O.\,Javakhishvili}
\affiliation{Department of Electrical and Computer Engineering, Agricultural University of Georgia, 0159 Tbilisi, Georgia}

\author{H.\,Jeong}
\affiliation{Department of Physics, Korea University, Seoul 02841, Republic of Korea}

\author{A.\,Kacharava}
\affiliation{Institut f\"ur Kernphysik, Forschungszentrum J\"ulich, 52425 J\"ulich, Germany}

\author{V.\,Kamerdzhiev}
\affiliation{Institut f\"ur Kernphysik, Forschungszentrum J\"ulich, 52425 J\"ulich, Germany}

\affiliation{{\rm now at:}
GSI Helmholtzzentrum für Schwerionenforschung, 64291 Darmstadt, Germany}

\author{I.\,Keshelashvili}
\affiliation{Institut f\"ur Kernphysik, Forschungszentrum J\"ulich, 52425 J\"ulich, Germany}

\affiliation{{\rm now at:}
GSI Helmholtzzentrum für Schwerionenforschung, 64291 Darmstadt, Germany}

\author{A.\,Kononov}
\affiliation{University of Ferrara and Istituto Nazionale di Fisica Nucleare, 44100 Ferrara, Italy}\

\author{K.\,Laihem}
\affiliation{III.\ Physikalisches Institut B, RWTH Aachen University, 52056 Aachen, Germany}

\affiliation{{\rm now at:}
GSI Helmholtzzentrum für Schwerionenforschung, 64291 Darmstadt, Germany}

\author{A.\,Lehrach}
\affiliation{Institut f\"ur Kernphysik, Forschungszentrum J\"ulich, 52425 J\"ulich, Germany}

\affiliation{JARA--FAME (Forces and Matter Experiments), Forschungszentrum J\"ulich and RWTH Aachen University, Germany}

\author{P.\,Lenisa}
\affiliation{University of Ferrara and Istituto Nazionale di Fisica Nucleare, 44100 Ferrara, Italy}

\author{N.\,Lomidze}
\affiliation{High Energy Physics Institute, Tbilisi State University, 0186 Tbilisi, Georgia}

\author{B.\,Lorentz}
\affiliation{GSI Helmholtzzentrum für Schwerionenforschung, 64291 Darmstadt, Germany}

\author{A.\,Magiera}
\affiliation{Marian Smoluchowski Institute of Physics, Jagiellonian University, 30348 Cracow, Poland}

\author{D.\,Mchedlishvili}
\affiliation{High Energy Physics Institute, Tbilisi State University, 0186 Tbilisi, Georgia}

\author{F.\,Müller}
\affiliation{III.\ Physikalisches Institut B, RWTH Aachen University, 52056 Aachen, Germany}

\affiliation{Institut f\"ur Kernphysik, Forschungszentrum J\"ulich, 52425 J\"ulich, Germany}

\author{A.\,Nass}
\affiliation{Institut f\"ur Kernphysik, Forschungszentrum J\"ulich, 52425 J\"ulich, Germany}

\author{N.N.\,Nikolaev}
\affiliation{L.D. Landau Institute for Theoretical Physics, 142432 Chernogolovka, Russia}
\affiliation{Moscow Institute of Physics and Technology,
School of Physics, Moscow region, 141707 Dolgoprudny,
Russia}

\author{S.\,Park}
\affiliation{Center for Axion and Precision Physics Research, IBS, Daejon 34051, Republic of Korea}

\author{A.\,Pesce}
\affiliation{Institut f\"ur Kernphysik, Forschungszentrum J\"ulich, 52425 J\"ulich, Germany}

\author{V.\,Poncza}
\affiliation{III.\ Physikalisches Institut B, RWTH Aachen University, 52056 Aachen, Germany}

\affiliation{Institut f\"ur Kernphysik, Forschungszentrum J\"ulich, 52425 J\"ulich, Germany}

\author{D.\,Prasuhn}
\affiliation{Institut f\"ur Kernphysik, Forschungszentrum J\"ulich, 52425 J\"ulich, Germany}

\affiliation{{\rm now at:}
GSI Helmholtzzentrum für Schwerionenforschung, 64291 Darmstadt, Germany}

\author{F.\,Rathmann}
\affiliation{Institut f\"ur Kernphysik, Forschungszentrum J\"ulich, 52425 J\"ulich, Germany}

\author{S.\,Siddique}
\affiliation{III.\ Physikalisches Institut B, RWTH Aachen University, 52056 Aachen, Germany}

\affiliation{Institut f\"ur Kernphysik, Forschungszentrum J\"ulich, 52425 J\"ulich, Germany}

\affiliation{{\rm now at:}
GSI Helmholtzzentrum für Schwerionenforschung, 64291 Darmstadt, Germany}

\author{A.\,Saleev}
\affiliation{University of Ferrara and Istituto Nazionale di Fisica Nucleare, 44100 Ferrara, Italy}

\author{N. Shurkhno}
\affiliation{Institut f\"ur Kernphysik, Forschungszentrum J\"ulich, 52425 J\"ulich, Germany}

\affiliation{{\rm now at:}
GSI Helmholtzzentrum für Schwerionenforschung, 64291 Darmstadt, Germany}

\author{D.\,Shergelashvili}
\affiliation{High Energy Physics Institute, Tbilisi State University, 0186 Tbilisi, Georgia}

\author{V.\,Shmakova}
\affiliation{University of Ferrara and Istituto Nazionale di Fisica Nucleare, 44100 Ferrara, Italy}

\author{J.\,Slim}
\affiliation{III.\ Physikalisches Institut B, RWTH Aachen University, 52056 Aachen, Germany}

\affiliation{{\rm now at:}
GSI Helmholtzzentrum für Schwerionenforschung, 64291 Darmstadt, Germany}

\author{H.\,Soltner}
\affiliation{Zentralinstitut f\"ur Engineering, Elektronik und Analytik, Forschungszentrum J\"ulich, 52425 J\"ulich, Germany}

\author{R. Stassen}
\affiliation{Institut f\"ur Kernphysik, Forschungszentrum J\"ulich, 52425 J\"ulich, Germany}

\author{H.\,Ströher}
\affiliation{Institut f\"ur Kernphysik, Forschungszentrum J\"ulich, 52425 J\"ulich, Germany}
\affiliation{JARA--FAME (Forces and Matter Experiments), Forschungszentrum J\"ulich and RWTH Aachen University, Germany}
\affiliation{{\rm now at:}
GSI Helmholtzzentrum für Schwerionenforschung, 64291 Darmstadt, Germany}

\author{M.\,Tabidze}
\affiliation{High Energy Physics Institute, Tbilisi State University, 0186 Tbilisi, Georgia}

\author{G.\,Tagliente}
\affiliation{Istituto Nazionale di Fisica Nucleare sez.\ Bari, 70125 Bari, Italy}

\author{Y.\,Valdau}
\affiliation{Institut f\"ur Kernphysik, Forschungszentrum J\"ulich, 52425 J\"ulich, Germany}
\affiliation{{\rm now at:}
GSI Helmholtzzentrum für Schwerionenforschung, 64291 Darmstadt, Germany}

\author{M.\,Vitz}
\affiliation{Institut f\"ur Kernphysik, Forschungszentrum J\"ulich, 52425 J\"ulich, Germany}
\affiliation{III.\ Physikalisches Institut B, RWTH Aachen University, 52056 Aachen, Germany}

\author{T.\,Wagner}
\affiliation{Institut f\"ur Kernphysik, Forschungszentrum J\"ulich, 52425 J\"ulich, Germany}
\affiliation{III.\ Physikalisches Institut B, RWTH Aachen University, 52056 Aachen, Germany}
\affiliation{{\rm now at:}
GSI Helmholtzzentrum für Schwerionenforschung, 64291 Darmstadt, Germany}

\author{P.\,W\"ustner}
\affiliation{Zentralinstitut f\"ur Engineering, Elektronik und Analytik, Forschungszentrum J\"ulich, 52425 J\"ulich, Germany}

\collaboration{JEDI Collaboration}
\noaffiliation

\date{\today}

\begin{abstract}

Based on the notion that the local dark-matter field of axions or axion-like particles (ALPs) in our Galaxy induces oscillating couplings to the spins of nucleons and nuclei
(via the electric dipole moment of the latter and/or the paramagnetic axion-wind effect),
{we establish the feasibility of} a new method to search for ALPs in storage rings.
{Based on previous work that allows us to maintain the in-plane polarization of a stored deuteron beam for a few hundred seconds, we performed a first proof-of-principle experiment at the Cooler Synchrotron COSY to scan momenta near 970\,MeV/c.}
This entailed a scan of the spin precession frequency. At resonance between the spin precession frequency of deuterons and the ALP-induced EDM oscillation frequency there will be an accumulation of the polarization component out of the ring plane. Since the axion frequency is unknown, the momentum of the beam and consequently the spin precession frequency were ramped to search for a vertical polarization change that would occur when the resonance is crossed. At COSY, four beam bunches with different polarization directions were used to make sure that no resonance was missed because of the  unknown relative phase between the polarization precession and the axion/ALP field. A frequency window of 1.5--kHz width around the spin precession frequency of 121 kHz was scanned. We describe the experimental procedure and a test of the methodology with the help of a radiofrequency Wien filter located on the COSY ring.
No ALP resonance was observed. As a consequence
an upper limit of the oscillating EDM component of the deuteron as well as its axion coupling constants are provided.
\end{abstract}

\maketitle

\section{\label{sec:intro}Introduction}
In 1977, Peccei and Quinn proposed an extension of the Standard Model (SM) of particle physics to include a global chiral symmetry in order to explain the small if not vanishing  magnitude of the CP violation in quantum chromodynamics (QCD)~\cite{Peccei:1977hh,Peccei:1977ur}. Since this so-called Peccei-Quinn (PQ) symmetry is necessarily spontaneously broken, the existence of the associated Nambu-Goldstone boson was conjectured by Weinberg~\cite{Weinberg:1977ma}
and Wilczek~\cite{Wilczek:1977pj} -- the latter coining  the name axion for this pseudoscalar particle
which acquires a small mass term via non-perturbative QCD effects.
{Since it was initially assumed that the order parameters of the spontaneous breaking of the PQ symmetry, $f_a$, would be the electroweak (Fermi) vacuum expectation value $v_{\text{F}}$~\cite{Peccei:2006as}, the original axion model by Peccei and Quinn
could be rather quickly ruled out by beam-dump experiments, see, {\it e.g.},
Ref.~\cite{Asano:1981nh}. The
focus then changed to the class of the so-called invisible axions with $f_a \gg v_{\text{F}}$,
which are limited by two types of models,
the KSVZ-model due to Kim~\cite{Kim:1979if} and Shifmann, Vainshtein and Zakharov~\cite{Shifman:1979if}, and the DFSZ-model
due to Dine, Fischler and Srednicki~\cite{Dine:1981rt} and Zhitnisky~\cite{Zhitnitsky:1980tq}}.
For the canonical QCD axion, which solves the strong CP problem, a relation between its mass $m_a$ and the order parameter $f_a$ can be determined~\cite{Gorghetto:2018ocs},
{which to leading order reads
\begin{equation}
    m_a   = \frac{\sqrt{z}}{1+z}m_\pi f_\pi \frac{1}{f_a}\,.
    \label{eq:QCD-axion}
\end{equation}
Here $m_\pi$ and $f_\pi$ are the mass and axial decay constant (chiral order parameter) of the pion, while $z=m_u/m_d \approx 0.474$, see Ref.\,\cite{Work22}, is the ratio of the $u$- and $d$-quark masses.
For the so-called axion-like particles (ALPs) which are not related to  the strong CP problem, there is not such a relation. Rather for a given value of the decay constant $f_a$ any value of the mass $m_a$, in particular, a smaller one than that determined by the relation (\ref{eq:QCD-axion}) is possible}, see {\it e.g.}, Chapter 90 ``Axions and Other Similar Particles'' of Ref.~\cite{Work22}.

{Recently, however, a lighter type of axion field than the canonical  one  was proposed~\cite{Hook:2018jle,DiLuzio:2021pxd} that is  based on the discrete $Z_\mathcal{N}$  shift symmetry suggested by Hook~\cite{Hook:2018jle} for a dark world extension
with $\mathcal{N}$  mirror and  degenerate worlds -- one of which is ours. These worlds are linked by the so-called  $Z_\mathcal{N}$  axion field. The latter  still  solves the strong CP problem if the integer $\mathcal{N}$ is an odd positive number, but has a ($\sim 2^{-\mathcal{N}/2}$) smaller  mass $m_a$ versus  $1/f_a$ relation than the canonical QCD axion, {\it i.e.},
\begin{equation}
   m_a  \simeq \left (  \frac{1-z}{\pi (1+z)} \right)^{{1}/{4}} m_\pi f_\pi\, \mathcal{N}^{\,{3}/{4}}\,
   z^{{\mathcal{N}}/{2}} \,\frac{1}{f_a}\,,
    \label{eq:ZN-axion}
\end{equation}
which in principle can be justified for $\mathcal{N}\gg 1$, but in practice already works
for $\mathcal{N}\geq 3$~\cite{DiLuzio:2021pxd,DiLuzio:2021gos}.
}

If sufficiently abundant, the {canonical  QCD axion, ALPs or $Z_\mathcal{N}$ axions} might be candidates for cold dark matter in the universe,
see {\it e.g.}, Refs.~\cite{Work22,Sikivie:2020zpn} for recent reviews. {In Refs.~\cite{Gr11,Gr13,Stadnik:2013raa,Budker:2013hfa}} it has been suggested that even axions and/or  ALPs of mass
from $10^{-7}\,\SI{}{eV/c^2}$ down to $10^{-22}\,\SI{}{eV/c^2}$ could be such candidates.
This  mass range is very challenging to reach with any established
technique.
For instance, the cavities of the microwave (haloscope) method, scanning for resonance frequencies due to the inverse Primakoff effect in  strong magnetic fields, {as suggested by Sikivie~\cite{Sikivie:1983ip,Sikivie:1985yu}}, would have to be unwieldy large~\cite{Gr15}.
Still, axions/ALPs of this mass range
could be associated with cosmic dark matter created in the Big Bang via the so-called pre-inflationary
PQ symmetry breaking scenario~\cite{Work22}. In this case these particles would be present now in sufficient concentrations to be regarded as an oscillating classical field that, established primordially, would still exist without losing most of its
coherence.
Locally within the Milky Way galaxy, the population of axions/ALPs would be dominated by those bound gravitationally to the galaxy. Their speed is limited by the virial velocity of the stellar ensemble, or roughly $v=10^{-3}\,c$  ({\it cf.} chapter~27 ``Dark Matter'' of Ref.~\cite{Work22}), which is similar to the orbital speed of the solar system containing the Earth with respect to the center of the galaxy. This would result in a non-relativistic distribution of the axion/ALP velocities, producing spatially  coherent, but
time-dependent  oscillations summarized by the classical axion/ALP field
\begin{equation}
  a(t) = a_0 \cos\big(\omega_a(t-t_0) +\phi_a(t_0) \big)   \,.
  \label{eq:a(t)}\,
\end{equation}
Here $a_0$ is the amplitude of the field,  while $\omega_a$
is the pertinent angular frequency  which, up to $\mathcal{O}\big(\{v/c\}^2\big) \sim 10^{-6}$ dispersive corrections, is
determined by the axion/ALP mass $m_a$,
\begin{equation}
    \hbar\omega_a = m_a c^2 \,. \label{eq:axm}
\end{equation}
Finally,
$\phi_a(t_0)$ is the local phase of the axion/ALP field, which is not only unknown but even changes  depending on the respective  starting point $t_0$ of any  new measurement. The lifetime of validity of this phase can be deduced by the simple quantum estimate
\begin{equation}
  \tau_a = \frac{h}{m_a v^2}\,, \label{eq:t_a}
\end{equation}
while the spatial extent of the phase coherence is given  by the length
\begin{equation}
    l_a = \frac{h}{m_a v}\,. \label{eq:l_a}
\end{equation}
{Therefore, for axions/ALPs with mass less than $10^{-7}\,\SI{}{eV/c^2}$ considered here, any additional spatial dependence   on the right-hand side of Eq.\,(\ref{eq:a(t)}) can be safely neglected in laboratory experiments, {\it e.g.}, also in axion searches in storage rings, since according to Eq.\,(\ref{eq:l_a}) and $v\approx 10^{-3}c$ the pertinent coherence length would be about $\SI{12}{km}$ and even proportionally larger for smaller masses.}

Detection in the laboratory of the oscillating dark-matter field of axions/ALPs given in Eq.~(\ref{eq:a(t)}) has to overcome
the extremely weak nature of the axion/ALP interactions {with each other
and other subatomic particles. Since their gravitational component can be safely neglected,  these interactions} scale with the inverse of the PQ order parameter $f_a$ that empirically has to
be much larger than the {electroweak vacuum expectation value, as mentioned above}.
Nevertheless, the pseudoscalar nature of axions/ALPs allows {-- in accordance with the Wigner-Eckart theorem --} for
potential couplings to the {\em total} angular momentum
(spin) of nucleons and nuclei and therefore opens up
further avenues for the detection of the oscillating dark matter field from Eq.~(\ref{eq:a(t)})
-- in addition to utilizing the inverse Primakoff effect,
astrophysical constraints etc.\ as specified, {\it e.g.}, in Refs.~\cite{Work22,Sikivie:2020zpn}.
This holds especially for the mass region specified above as the Primakoff-based methods do not
apply there.
In fact, these spin couplings can occur in two different ways, either by a coupling to the electric dipole moment (EDM) of a
non-selfconjugate matter particle carrying nonzero spin, {see, {\it e.g.}, Ref.~\cite{Bernreuther:1991mn}}, or via the pseudomagnetic direct coupling of the gradient of the axion field to
the spin of the matter particle, the so-called axion-wind effect.
{There can be further CP-allowed and even CP-violating couplings of axions/ALPs to nucleons (or nuclei), see, {\it e.g.}, the second term in Eq.\,(8) and the first and third terms in Eq.\,(9) of \cite{Sikivie:2020zpn}. These interactions will not be discussed  here, since they are either only sensitive to the (in the considered  mass region) suppressed spatial variation of the axion field (\ref{eq:a(t)})  or they only couple to  scalar nucleon or nuclear densities which are \emph{spin-independent} to leading order.}

The first  {of these couplings to the nucleon or nuclear spin}, see Refs.~\cite{Gr11,Gr13}, is based on the introduction of an oscillating component $d_\mathrm{AC}$ to
the total EDM of the pertinent matter particle,
\begin{equation}
    d(t)=\subit{d}{DC} + \dAC \cos\big(\omega_a( t-t_0) + \phi_a(t_0)\big)\,, \label{eq:oscd}
\end{equation}
pointing parallel to the spin direction, by the oscillating axion/ALP field $a(t)$. Here $\subit{d}{DC}$ is the permanent (static) component of the pertinent EDM, while the other parameters follow from Eq.~(\ref{eq:a(t)}). In addition to astrophysical constraints, see Ref.~\cite{Work22}, there are first limits reported in Ref.~\cite{Ab17}
based on the upper bounds on the neutron EDM measurements (see also Ref.~\cite{Roussy:2020ily}). These limits, however, only apply to oscillations of frequency $f_\mathrm{AC}$ below $10^{-2}\,\SI{}{\hertz}$,
{\it i.e.}, axion/ALP masses below $10^{-17}\,\SI{}{eV/c^2}$ (below $10^{-15}\,\SI{}{eV/c^2}$ in the case of Ref.~\cite{Roussy:2020ily}). To search for oscillating EDM components of higher frequency other methods have to be utilized.

It has been proposed to search for oscillating
EDMs with the help of electric, hybrid or  magnetic
{storage rings~\cite{Ch18,Ch19,Ab19,Kim:2021pld}}.
Especially in the latter case, the charged particles in the  comoving beam frame are subject to a large electric field  ($c \vec\beta\times \vec B$) due to their relativistic motion $c\vec\beta$ in the magnetic field  $\vec B$ in the laboratory system.  This (effective) electric field closes the orbit  such that the resulting force always points toward the center of the ring. The EDM of the
charged particle, which is aligned with the spin, feels a torque from this electric field. This causes the spin to rotate about the electric field direction. If the EDM is static, this rotation only wobbles the polarization about its starting point as the polarization precession in the ring plane continually reverses the torque. If, however, the EDM oscillates at the same rate as the torque reversal, then the rotations accumulate, eventually creating a measurable polarization component perpendicular to the ring plane.
Note that the action of an oscillating EDM on the spin can be  mimicked by a radiofrequency (RF) Wien filter  when its magnetic field, pointing horizontally, acts on the corresponding magnetic dipole moment, {\it cf.} Eq.~(\ref{eq:O_MDM}).

The second coupling of the axion/ALP field to matter particles is based  on the ``axion-wind'' or ``pseudomagnetic'' effect~\cite{Weinberg:1977ma,Krauss:1985ub,Georgi:1986df,Raffelt:1987yt,Choi:1988xt,Carena:1988kr,Barbieri:1985cp,Vorobev:1989hb,Kakhidze:1990in},
causing a rotation of the spin of a nucleon or nucleus
around the gradient of the axion field which acts analogously to a magnetic field~\cite{Krauss:1985ub,Vorobev:1989hb,Kakhidze:1990in,Vorobev:1995pb}.
The term ``axion wind'' was
coined in Ref.~\cite{Vorobev:1995pb} (see Ref.~\cite{Gr13}
for the extension to ``ALP dark matter wind'' and, {{\it e.g.}, {Refs.~\cite{Graham:2017ivz,Stadnik:2017mid,Smorra:2019qfx}})} for the same pseudomagnetic field -- this time
manifestly proportional to the velocity of the Earth-bound
spins with respect to the galactic axion/ALP field. In that case the actual velocity is a superposition of the motion of the solar system with respect to the axion
field plus the rotation of the Earth around the Sun plus the rotation of the Earth
around its
axis plus the movement of the particle inside the pertinent
sample or experiment in the laboratory.
All the above complexity with the time-dependent orientation of the pseudomagnetic field becomes entirely irrelevant to the
in-flight spins of the beam particles in a storage ring when the velocity is close to the speed of light. Most
remarkably, the corresponding pseudomagnetic field is then always tangential to the beam orbit~\cite{Graham:2020kai,Silenko:2021qgc}, {\it i.e.}, it plays the role of an RF solenoid uniformly distributed
along the ring circumference~\cite{Silenko:2021qgc,Kolya22}.

{The two spin-dependent axion/ALP couplings (EDM and axion wind) can be expressed in the Langrangian formalism.
The Lagrangian for the EDM coupling to nucleons is given by the generic expression of Ref.~\cite{Work22} in terms of the axion coupling
$g_{aN\gamma}$ to the EDM operator,
\begin{eqnarray}
    \mathcal{L}_{aN\gamma}
    &=& - \frac{i}{2} g_{a N \gamma} \,a\,  \bar \Psi_N \sigma_{\mu\nu} \gamma_5 \Psi_N F^{\mu\nu}
    \nonumber  \\
    &=&
    -\frac{i}{2} \frac{\dAC}{a_0} \,a\,  \bar \Psi_N \sigma_{\mu\nu} \gamma_5 \Psi_N F^{\mu\nu}\,,
    \label{eq:lag_aN}
\end{eqnarray}
where $N$ = $n$, $p$ denotes
neutron or proton, respectively. Note that the first line of
Eq.~(\ref{eq:lag_aN}) refers to $\hbar=c=1$ units, while
the  second line is given in SI units.
After the Dirac spinors $\Psi_N$ are reduced to standard spinors, $\chi_N$,
and $a(t)$ of Eq.~(\ref{eq:a(t)}) is inserted for the generic ALP field $a(t,\vec x)$,
the  pertinent Hamiltonian assumes the structure
\begin{eqnarray}
   H_{aN\gamma} &=& - \frac{\dAC}{a_0}\, a(t) \left(\chi_N^\dagger \frac{1}{S} \,\vec{S} \,\chi_N\right) \cdot \vec E   \nonumber \\
  &\equiv& \vec{\Omega}_{\mathrm{EDM}} \cdot \left (\chi_N^\dagger \hbar \vec S \chi_N \right)\,,
  \label{eq:ham_aN}
\end{eqnarray}
defining the angular velocity
$\vec{\Omega}_{\mathrm{EDM}} $ for the axion-EDM coupling. In fact, $\chi_N$ can be extended to apply even for the ($2S{+}1$)--dimensional representations of nuclei, especially  for the 3-dimensional one of the deuteron $d$, {\it cf.} Refs.~\cite{Silenko:2014kia,Silenko:2017iyv,Silenko:2021qgc}. In the following the electric field will be interpreted as
the effective field $\vec E = c \vec \beta \times \vec B$.

The axion wind case can be derived from  the generic interaction
Lagrangian of the pseudomagnetic coupling
of an axion/ALP field $a(t,\vec x)$ to an arbitrary fermion field $\Psi_f$ (where $f$  can stand for the nucleon $N$, proton $p$, neutron $n$, etc.).
In the notation of reference~\cite{Work22},  this Lagrangian reads
\begin{equation}
  \mathcal{L}_{aff} = \frac{C_f}{2 f_a} \partial_\mu a
  \bar\Psi_f \gamma^\mu \gamma_5 \Psi_f
  \label{eq:Lwind}\,
\end{equation}
in terms of the dimensionless coupling constant $C_f$ and
the generic axion/ALP decay constant $f_a$ which is independent
of the fermion (Dirac) field $\Psi_f$. If the latter is reduced to standard
spinors, {\it cf.} Ref.~\cite{Silenko:2021qgc}, and the ALP field
$a(t)$ of Eq.~(\ref{eq:a(t)}) is inserted for the generic field $a(t,\vec x)$, the corresponding
Hamiltonian in SI units has  the structure
\begin{eqnarray}
H_{a NN} &=& - \frac{C_N}{2 f_a} \, \hbar\partial_0 a(t) \left (\chi_N^\dagger \frac{1}{S} \vec S \,\chi_N\right) \cdot \vec \beta \nonumber\\
&\equiv& \vec{\Omega}_{\mathrm{wind}} \cdot \left(\chi_N^\dagger \hbar \vec S \chi_N \right)\, ,
\label{eq:ham_aNN}
\end{eqnarray}
which is the axion-wind analog  of the axion-EDM
Hamiltonian~(\ref{eq:ham_aN}).}

Quantitatively the spin motion relative to the momentum vector in purely magnetic fields is governed
by the subtracted, EDM- and axion-wind extended Thomas-BMT equation of Refs.~\cite{Bargmann:1959gz}, \cite{Fukuyama:2013ioa} and \cite{Silenko:2021qgc}, respectively:
\begin{equation}
\frac{d \vec{S}}{dt} = (\vec{\Omega}_{\mathrm{MDM}} - \vec{\Omega}_{\mathrm{rev}} + \vec{\Omega}_{\mathrm{EDM}} + \vec{\Omega}_{\mathrm{wind}}) \times \vec{S},  \label{eq:tbmt}
\end{equation}
defined in terms of the angular velocities for the magnetic dipole moment (MDM) including the Thomas precession, the revolution of the
beam (rev), the electric dipole moment (EDM)  and the axion wind effect (wind):
\begin{eqnarray}
\vec{\Omega}_{\mathrm{MDM}} &=& -\frac{q}{m} ~ \left(G +\frac{1}{\gamma}\right)\vec{B} ,  \label{eq:O_MDM}\\
\vec{\Omega}_{\mathrm{rev}} &=&  -\frac{q}{\gamma m} \vec{B}, \label{eq:O_rev}\\
\vec{\Omega}_{\mathrm{EDM}} &=& -\frac{1}{S\hbar}
d(t)  \,c \vec{\beta} \times \vec{B} , \label{eq:O_EDM} \\
 \vec{\Omega}_{\mathrm{wind}} &=& -\frac{1}{S\hbar} \,\frac{C_N}{2 f_a} \,
  \big(\hbar \partial_0 a(t) \big) \,\vec \beta \, ,
  \label{eq:O_wind}
\end{eqnarray}
where, to simplify the notation, terms including $\vec{\beta} \cdot \vec{B}$  were omitted.
$\vec{S}$ in  the above equations denotes the spin vector in the particle rest frame, $t$ the time
in the laboratory system, $\beta$ and $\gamma$ the relativistic Lorentz factors of a particle of rest mass $m$,  and $\vec{B}$  the magnetic field in the laboratory system pointing perpendicular to the ring plane.
The magnetic dipole moment $\vec \mu$ and electric dipole moment $\vec d$
are both pointing along the axis of the particle's spin $\vec{S}$.
The dimensionless
quantity $G$ (magnetic anomaly) is related to the magnetic moment as follows:
\begin{equation}
\vec{\mu} = g \frac{q \hbar}{2 m} \vec{S} = (1+G) \frac{q \hbar}{m} \vec{S}\, .
\label{eq1}
\end{equation}
The oscillating axion or ALP field $a(t)$, see Eq.~(\ref{eq:a(t)}), generates the oscillating {(AC)} term in the electric dipole moment $d(t)$, {{\it cf.} Eqs.~(\ref{eq:oscd}) and (\ref{eq:ham_aN})}.
Through the time derivative $\partial_0 a(t)$ a second oscillating contribution in the term
$\vec{\Omega}_{\mathrm{wind}}$~\cite{Graham:2020kai,Silenko:2021qgc,Kolya22} is present, {{\it cf.} Eq.\,(\ref{eq:ham_aNN})}.
It depends on the specific coupling strength $C_N$, while $f_a$ is
the generic axion or ALP decay constant,  namely the order parameter of PQ breaking
mentioned above.
By just measuring a vertical build-up of a polarization component perpendicular to the ring plane, the EDM-induced axion coupling and the axion-wind pseudomagnetic coupling cannot be distinguished, since they have the same effect on vertical polarization. Moreover, for axion/ALP masses below $10^{-7}\,\SI{}{\evsc}$
the sensitivity in any foreseeable search is not expected to extend to the scale where the pseudoscalar bosons
 determining the $a(t)$ field would appear as a result of known QCD processes; thus, {if the possibility of $Z_\mathcal{N}$ axions is ignored here},  these particles should be referred to as ALPs rather than axions.

At COSY, which belongs to the class of purely magnetic storage rings, we have stored in-plane   beams of deuterons
of approximately $\SI{970}{MeV/c}$ beam momenta
 with a spin precession frequency of $f_{\mathrm{spin}}=\frev |G\gamma|\approx 120\,\mathrm{kHz}$ (where $G=-0.1429875424$ is  the deuteron magnetic anomaly, and $\frev$ and $\gamma$ the revolution frequency and the relativistic factor of the circulating deuterons, {respectively)~\cite{Stephenson:2020jzx}}. This corresponds to an ALP mass of about $\SI{5e-10}{\evsc}$.
Then the simple quantum estimate  of the lifetime
in Eq.~(\ref{eq:t_a})  gives
$\tau_a \approx \SI{8}{\second}$.
The time for the frequency scan to cross the ALP resonance should not be much greater than this or else the size of the polarization jump will be attenuated. In fact, the crossing times are less in this experiment.
At the same time, the ALP field must be able to act on all the particles in the beam simultaneously. A similar estimate of the spatial extent of phase coherence  as in Eq.~(\ref{eq:l_a}) gives the length
$l_a \approx \SI{2500}{\kilo\metre}$.
This coherence length well exceeds the size of the storage ring (\SI{183.6}{\metre} circumference). All parts of the beam and their particles should therefore be exposed to the same ALP field as given in Eq.~(\ref{eq:a(t)}).

Since the ALP oscillation frequency $\omega_a$ in Eq.~(\ref{eq:a(t)}) is unknown, it was necessary to slowly ramp the beam energy and thus the spin precession rate while continually monitoring the vertical polarization with the hope of detecting its sudden change (to which we refer as a \textit{polarization jump} in the following) as the resonance was crossed. {For this, it was crucial to be able to maintain the in-plane polarization over the entire measurement cycle \cite{Gu16} and to continuously monitor the spin-precession frequency with high precision \cite{Ev15}.}The phase of the EDM oscillation relative to the polarization precession was also unknown, {\it cf.} $\phi_a(t_0)$ in Eqs.~(\ref{eq:a(t)}) and (\ref{eq:oscd}), so four beam bunches were stored simultaneously in the ring with different polarization directions in order that all possible phases were adequately sampled. A novel waveguide RF Wien filter designed for EDM searches at COSY was successfully used to generate a test signal in the beam polarization as a confirmation of the method.

This paper describes the details of the first search for ALPs using a storage ring.
Section~\ref{sec:descexp} provides a description of the experiment with polarized beam. This includes the properties of the beam and the requirements for the search. Subsections deal with the problem of using multiple beam bunches to ensure that all phase possibilities are covered and describe the management of the scanning process in detail. Section~\ref{sec:wftest} describes how the RF Wien filter installed in the COSY ring was used to create a resonance that confirmed our model for the process of generating a polarization jump. The analysis of the data is covered in Section~\ref{sec:dataana}, and Section~\ref{sec:FC} discusses how we handled a systematic problem with false positive signals. There is also a description of the model used for the calibration of any polarization jump in terms of an oscillating EDM. The various upper limits of the ALP-deuteron couplings are discussed in Section~\ref{sec:Result}.
{Conclusions and an outlook are pursued in Section~\ref{sec:conc}}. Details about the four-bunch procedure and the calibration of the sensitivity of the measurement are relegated to the Appendices~\ref{App:A} and \ref{App:B}.

\section{\label{sec:descexp}The Experiment}
The search for ALPs was performed at the Cooler Synchrotron (COSY) located at
Forschungszentrum J\"ulich, Germany~\cite{Ma97}  {in the spring of 2019~\cite{Stephenson:2020jzx}}. The polarized deuteron beam (${\vec D}^-$) was generated in an atomic beam polarized ion source~\cite{Ha67}. A single polarized state was made using a weak field transition unit. The beam was then pre-accelerated in the JULIC cyclotron. The beam polarization was measured in the transfer line between the cyclotron and the COSY ring using a dedicated low energy polarimeter (LEP, see Ref.~\cite{Ch06}). Deuterons were scattered from a carbon target at a kinetic energy of \SI{75.6}{\mev}~\cite{Ch06}. The quantization axis for the spins of the beam particles was vertical, a direction imposed by the cyclotron fields. Elastically scattered deuterons were detected at \ang{40} in the lab on either side of the beam using plastic scintillator detectors. A description of deuteron spin polarization is given by Tanifuji \cite{Ta18} and is consistent with the Madison Convention \cite{MC71}. The analyzing power in this configuration is $A_y = 0.61 \pm 0.04$~\cite{Ch06}. A left-right asymmetry $(L-R)/(L+R)=(3p_y A_y / 2)$ of $-0.508 \pm 0.007$ was recorded. Repeating this measurement with the polarized source RF transition turned off produced an asymmetry of $-0.159 \pm 0.008$. This is a measure of the geometrical errors in the detector and data acquisition setup. The difference, $-0.349 \pm 0.011$, results in a polarization of $p_y = -0.38 \pm 0.03$.

\begin{figure}[!hbt]
\includegraphics[width=\columnwidth]{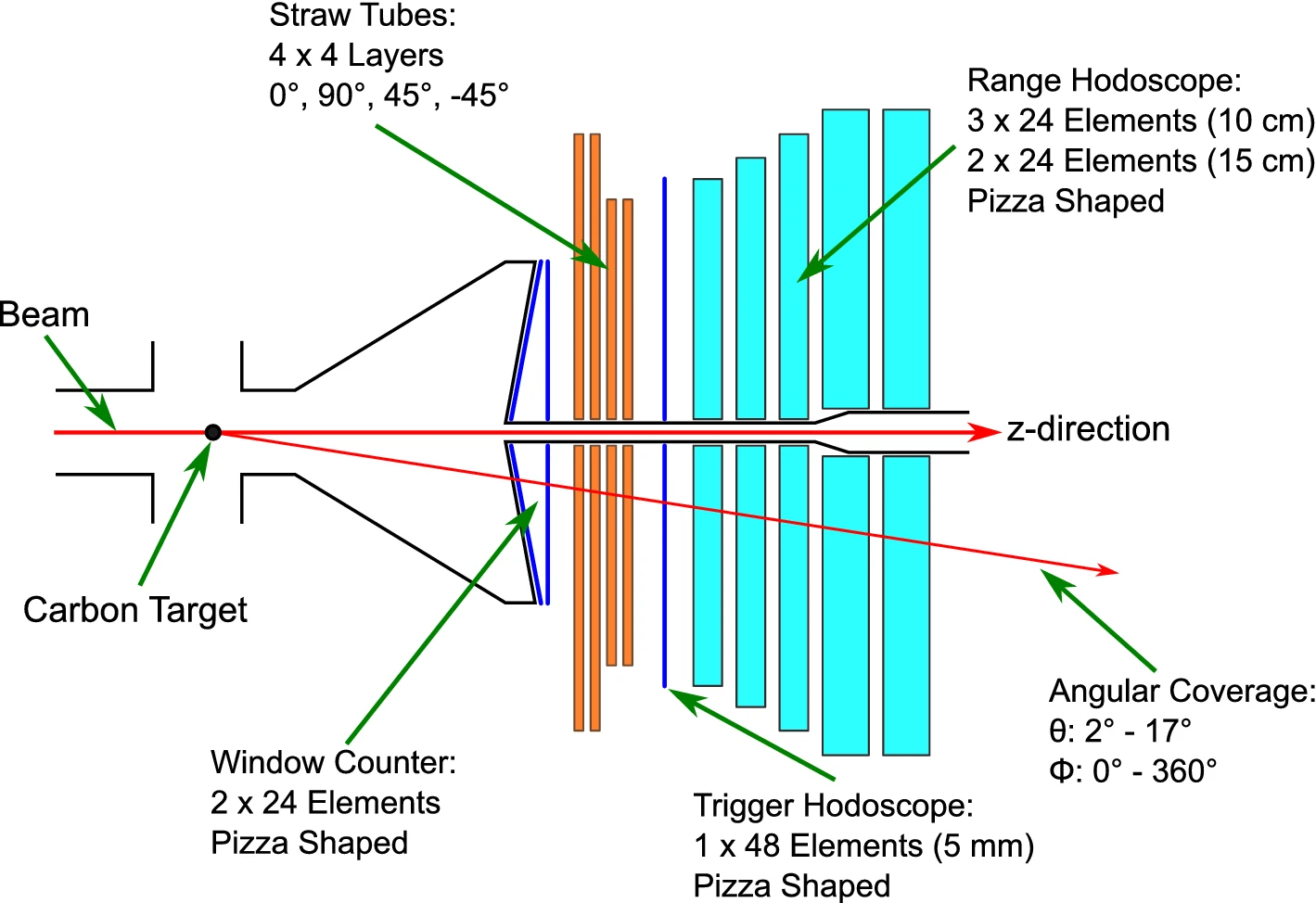}
\caption{\label{fig:wasa} Cross sectional diagram showing the layout of the WASA Forward Detector, reprinted from~\cite{Mu20}. The beam travels from left to right (horizontal red arrow), closely passing a 2-cm thick carbon block target along the way. The beam is heated vertically to bring beam particles to the front face of the target. Scattered particles moving forward exit the vacuum through a stainless steel window at angles between \ang{2} and \ang{17}. They then pass through two plastic scintillator window counters cut into pie-shaped segments, an array of straw tubes for position and angle tracking, a segmented trigger hodoscope, and five layers (light blue) of plastic scintillator calorimeter detectors. The trigger counter and calorimeter detectors are also divided into pie-shaped segments. All
the scintillator counters are read out using photomultiplier tubes mounted at the outer edge of each pie segment.}
\end{figure}

The deuteron beam was injected into the COSY synchrotron at \SI{75.6}{\mev} by stripping off the electrons in a thin foil, and ramped to an energy of \SI{236}{\mev} (\SI{0.97}{\gevc} momentum). At this energy, the polarization of the stored beam was measured using the Forward Detector from the WASA (Wide Angle Shower Apparatus) facility~\cite{Mu20}, as shown in Fig.~\ref{fig:wasa}.

A 2-cm thick carbon block was inserted from above the beam and brought into position with the bottom edge aligned along the center of the WASA detector. This required that the beam be locally lowered by about 3~mm as it passed the carbon target. At the start of data acquisition, the beam was heated vertically using RF white noise generated in a band about one of the harmonics of the vertical tune. This brought beam particles to the front face of the target. From there, deuterons passed through the target. Some were scattered into the WASA detector system. The observed event rate was dominated by elastic scattering which has a forward-peaked angular distribution. The relevant analyzing powers for elastic scattering are shown in Ref.~\cite{Mu20}. The detector acceptance was divided by software into four quadrants (left, right, down, and up). The sum of these four detector rates was fed back to control the power level of the white noise and maintain a constant event rate. Left-right and down-up asymmetries were computed in real time in four-second time intervals and made available for inspection as each beam store progressed. The down-up data stream was unfolded~\cite{Ba14} based on the spin tune frequency in order to generate a value of the magnitude of the rotating in-plane component of the beam polarization.

A typical left-right asymmetry with the vertically polarized beam running was $0.12 \pm 0.02$, where the error indicates the variation in this value during the experiment. This means that the cross section weighted average of the analyzing power over the WASA detector acceptance was $0.210 \pm 0.035$.

The beam was accelerated in less than a second to full energy. Then electron cooling was applied for \SI{71}{\second}. This reduced the phase space of the beam to a point where the polarization lifetime in the horizontal plane could become long~\cite{Gu16}. Once electron cooling was complete, the operating conditions for the beam were set. This required minimizing orbit deviations that would take the beam away from the centers of quadrupole lenses or having other unnecessary deviations. Finally, the polarization, which begins in the vertical direction, was rotated into the horizontal plane using an RF solenoid.

The timing list for a machine beam cycle is given in Table~\ref{tab:cyclestages}.

\begin{table}[!hbt]
\caption{\label{tab:cyclestages}
Times for various COSY operations during the beam cycle}
\begin{ruledtabular}
\begin{tabular}{ll}
 Event in the cycle & Time [s] \\
\hline
Acceleration off& 0.674 \\
E-cooling on& $4-75$ \\
Carbon target moved in& 75 \\
White noise extraction on& 77 \\
WASA flag (DAQ on)& 78 \\
RF Solenoid on (rotate $p_y$) & $83-86$ \\
Quick ramp to start of scan & $90.0-90.1$ \\
Constant frequency hold&$90.1-120.1$ \\
Ramp to search for ALP &$120.1-255.1$ \\
Constant frequency hold&$255.1-285.1$ \\
COSY RF stop& 287 \\
End of data taking& 288 \\
\end{tabular}
\end{ruledtabular}
\end{table}

In order to precess the deuteron polarization into the ring plane, the RF solenoid was operated for \SI{3}{s} on the $(1+G\gamma)\frev$ harmonic where $G=-0.1429873$ is the magnetic anomaly of the deuteron and $\frev=\SI{750602.6}{\hertz}$. At this frequency, the relativistic factor is $\gamma=1.126$. A search, made either as a scan or in fine steps, showed that the $(1+G\gamma)$ resonance for the RF solenoid occurred at $\fsol=\SI{629755.3}{\hertz}$. The difference of these two frequencies, $\frev-\fsol=\SI{120847.3}{\hertz}$, is the spin tune frequency, \fspin. The frequency generators were synchronized with the \SI{10}{\mega\hertz} signal
from GPS (Global Positioning System), thus the set values are precise and stable out to several~mHz. From these two frequencies and the assumption that the COSY ring is purely magnetic, it is possible to deduce the  kinematic parameters of the beam given in Table~\ref{tab:beam}. This parameter list is shown without errors since they lie beyond the range shown in the table. The values are typical of the initial conditions of the deuteron beam in the storage ring before ramping. The setup of the Wien filter in Section~\ref{sec:wftest} contains the results of the scan used to determine the $(1-G\gamma)\frev$ spin resonance frequency. In that case the resonance shape was measured as part of the matching process and found to have a fractional full width of about $2\times 10^{-9}$ which represents one estimate of its precision.

\begin{table}[!hbt]
\caption{\label{tab:beam}
Beam parameters}
\begin{ruledtabular}
\begin{tabular}{lll}
 Parameter& Symbol [Unit]&Value \\
\hline
Revolution frequency&\frev~[Hz]& $750602.6$ \\
Spin resonance frequency&\fsol~[Hz]& $629755.3$ \\
Spin tune frequency&\fspin~[Hz]& $120847.3$ \\
Lorentz factor&$\gamma$ [1]& $1.126$ \\
Beam velocity&$\beta$~[$c$]& $0.460$ \\
Orbit circumference&$l$ [m]& $183.57$ \\
{Number of deuterons per cycle} & {$N_d$ [1]} &
{$\approx 10^9$} \\
\end{tabular}
\end{ruledtabular}
\end{table}

As for the spin tune frequency~\fspin, the analysis of the polarization data allows us to measure it with a $10^{-10}$ precision~\cite{Ev15}, and the cycle-to-cycle variations are driven by the stability of the power supplies and the resulting orbit variations rather than by the precision of frequency setting.

It was important that the orbit does not deviate during the course of a ramp. This requirement was tracked with the use of beam position monitors (BPM).

\begin{figure}[!hbt]
\includegraphics[width=\columnwidth]{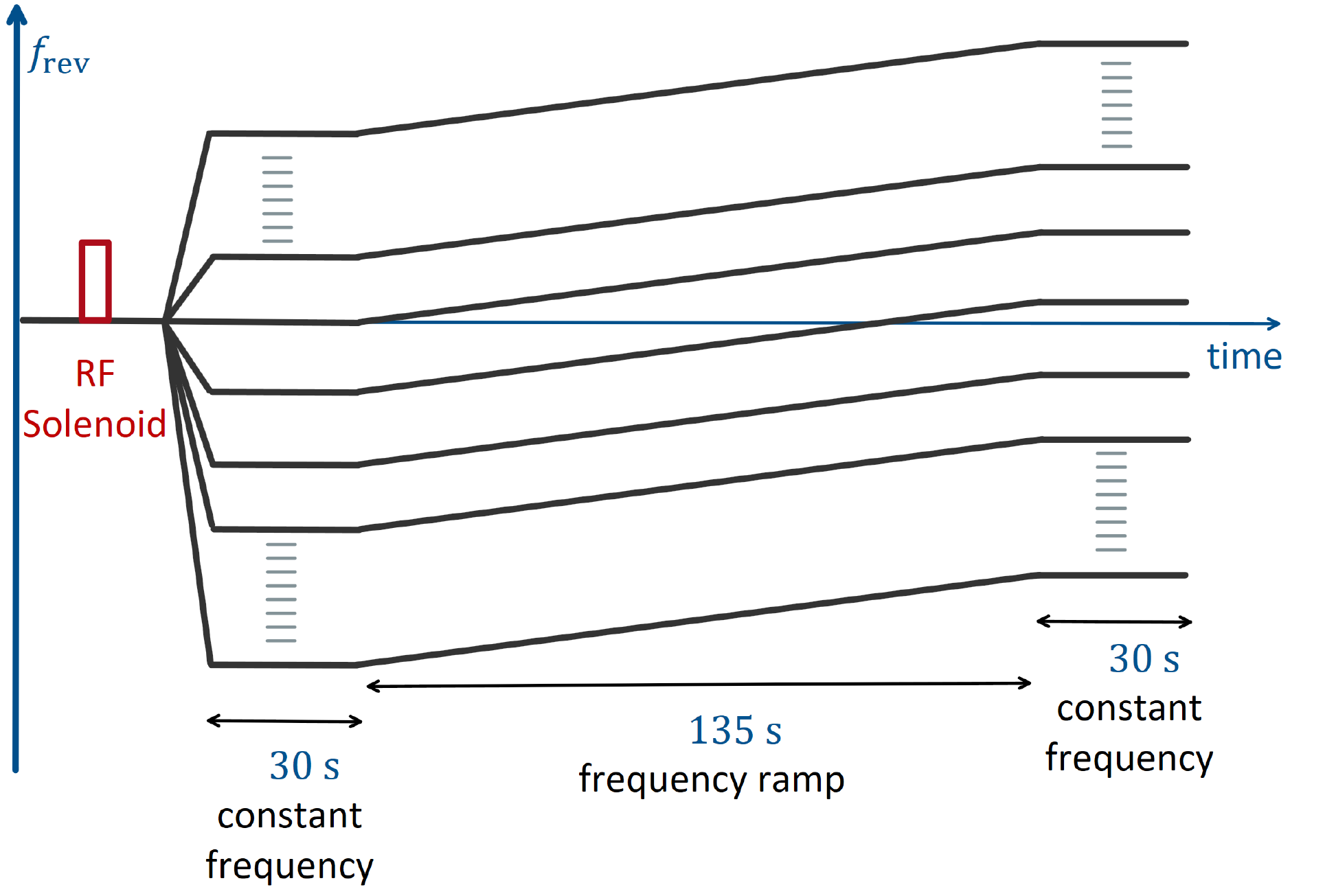}
\caption{\label{fig:rampscheme} Diagram of revolution frequency as a function of time (black lines) showing how scans for axions were organized. The diagram includes an early time (marked in red) when the RF solenoid rotated the polarization into the horizontal plane. Then quick ramps took the machine to the start point for each ramp. Flat parts were included at the beginning and end of each scanning ramp to allow checks of the polarization with enhanced statistical precision. The resonance jump was expected to appear at some time during the frequency ramp.}
\end{figure}

The strategy for making the ALP search contained the steps shown in Fig.~\ref{fig:rampscheme}. Because of the necessity to maintain reproducible conditions for the rotation of the polarization from the vertical into the horizontal plane, all machine cycles began at the same beam energy or revolution frequency, as indicated by the blue horizontal axis in the figure. Once the polarization rotation was complete, a quick ramp was made to take the machine to the starting point for the scan. This removed the necessity to search for a new resonance frequency before every new scan. The scans, indicated by the long sloping lines in Fig.~\ref{fig:rampscheme}, lasted for \SI{135}{\second}. Before and after, there was a \SI{30}{\second} period with no ramp. This was meant to provide extra data to characterize the starting and ending points under stable conditions. The ramps were planned to overlap at the ends, as shown in the figure. Since it was possible that an ALP-induced resonance might occur near the start or end, the overlap with the neighboring ramp was planned so that the resonance could be correctly characterized.

Altogether, there were 103 ramps covering a range from \SI{119.997}{\khz} to \SI{121.457}{\khz}, or an axion mass range of \SIrange{4.95}{5.02} {\nevsc}.

In detail, after completing the initial preparatory phase of the machine cycle, the beam was brought to an interaction with the polarimeter target by moving the target into the correct position, turning on the RF white noise, and initiating the feedback used to maintain the polarimeter count rate. At this stage the data acquisition was turned on. After a short period that was used to check the vertical polarization, the RF solenoid was activated to rotate the polarization into the horizontal plane. For each scan, the machine was first set to the starting conditions. Then the first 30-s holding time took place, followed by the ramp and the second 30-s holding time.

The RF solenoid, whose magnetic axis is along the beam axis, operates by giving a series of small kicks to the rotation of the polarization. The magnetic field, except for some mild focusing effects, does not steer the beam. It thus maintains the constant orbit length constraint. The RF solenoid was kept running for \SI{3}{\second}. If allowed to continue, it would drive the vertical polarization into an oscillating pattern~\cite{Be12, Be13}. The solenoid strength was adjusted until the vertical polarization component observed after the rotation was brought to zero. See the beginning of Appendix~\ref{App:A} for more details.

If the beam, once in the plane of the ring, remains polarized, then the down-up asymmetry in the WASA Forward Detector will oscillate with the precession frequency. However, this frequency is much too large to be able to observe even a single
polarimeter event per oscillation, thus a different technique has been developed~\cite{Ev15}. Namely, a value of the spin tune ($\nu=G\gamma$) was assumed and the data sorted among nine bins according to where the spin tune would predict it would lie along a single oscillation of the asymmetry. At the end of a preset time interval of \SI{4}{s}, the data accumulated in each of the time bins was used to calculate a down-up asymmetry for that bin. These asymmetries were reproduced with a sinusoidal curve from which the magnitude and phase of the oscillating horizontal asymmetry was obtained. The value of the spin tune was varied in small steps until a maximum in the amplitude of the sine wave was found \cite{Ba14}. The resulting magnitude and phase were recorded for that time bin. Data from the in-plane polarization measurement was recorded every \SI{4}{s} in order to provide the statistics to complete this search. The processing time was quick enough that values of the horizontal asymmetry were made available in real time during the experiment.

The horizontal polarization is subject to depolarization because betatron oscillations of the beam particles lead to variations in their spin tunes. It has been shown that the addition of sextupole fields to the ring allow for the compensation of this depolarizing effect~\cite{Gu16, Gu18}. Thus a part of the setup for this search involved optimizing these fields for the particular running conditions present in COSY at the time of this search. Online data was used to determine the horizontal polarization loss. Values for the strength of the three families of sextupole magnets were adjusted until the maximum polarization lifetime was obtained. During this experiment, the lifetime continued to vary because it was very sensitive to running conditions. At all times the slope of the horizontal polarization with time was maintained with a half-life greater than \SI{300}{\second}. Thus no more than a quarter of the polarization was lost during a typical scan.
\subsection{\label{sec:axionphase}Dealing with ALP Phase}

During the scan, the phase of the oscillating EDM with respect to the rotating in-plane polarization (with reference to the beam-frame electric field) is not known. With only a single beam bunch in the machine at one time, the amplitude of a jump is modulated by a sine function of this relative phase. This situation could easily allow an ALP to be missed during a single scan. To avoid this, our strategy was to use more than one beam bunch at the same time since the same ALP phase is shared among all of these bunches. With the equipment available at COSY, bunching the beam up through harmonic 4 was already available.

A model study was performed to see if this change would satisfy the requirement with no further additions to the COSY ring. The details of this calculation are described in Appendix~\ref{App:A}. Assuming a round rather than a race-track shaped ring, it was found that four equally spaced bunches offered four mutually perpendicular orientations of the polarization direction relative to the beam-frame electric field (see Fig.~\ref{fig:bunch2}). As the beam circulates in the ring, these four polarization directions rotate synchronously.

The COSY ring is race-track shaped, with arcs and straight sections of approximately the same length. This breaks the simple rotation pattern.
Two bunches on opposite sides of the ring
will have a rotating polarization while the two bunches in between will have a stationary phase without any electric field. This pattern swaps four times during each beam rotation.
The sensitivity to all possible ALP phases comes from the comparison between neighboring bunches. During a single turn, the angle between polarization vectors of two subsequent bunches oscillates from \ang{90} to \ang{61.2} and back again. In this case, the actual sensitivity must be averaged over the range of angles covered in this oscillation. This results in a reduction of the signal by 4.2\%, and the presented results have been corrected for this effect.

\subsection{\label{sec:scanman}Scan Management}

The approach to managing the scanning process was described in Section~\ref{sec:descexp} above and is shown in Fig.~\ref{fig:rampscheme}. In preparing this scheme, one of the most crucial parts was the precession of the polarization from the vertical to the horizontal. To do this, the resonant frequency for the RF solenoid on a harmonic of the revolution frequency must be located experimentally to within about \SI{0.1}{\hertz}. This level of precision requires several tests that demonstrate an efficient precession process and the vanishing of the vertical polarization component at the end. This procedure is usually time consuming, taking longer than the series of scans themselves at one frequency setting. It was decided to separate the RF-solenoid-driven spin  rotation from the rest of the scan procedure in order to save time and effort. After the RF-solenoid-driven spin precession was complete, the COSY operating frequency was ramped to the starting point for the scan and the scanning process began. In this way, the same resonance frequency is kept for all scans.

The ramping process itself must obey the constraint of preserving the orbit circumference while maintaining the optical properties of the beam. Even small variations can alter the way that sextupole corrections affect the beam. The resulting changes in the cancellation of depolarizing effects would make the lifetime of the horizontal polarization significantly smaller. For ramping, the two adjustable parameters on the machine are the magnetic field in the arcs and the frequency of the RF cavity that bunches (and accelerates) the beam. We chose to create a linear ramp in momentum. The field of the ring magnets is usually characterized by rigidity $B\rho$ which itself is proportional to the momentum. Thus, this requirement for the magnetic field is met straightforwardly using
\begin{equation}
    B\rho =\frac{pc}{q},
\end{equation}
where $\rho$ is the curvature radius of a particle track in field $B$, $c$ is the speed of light and $q$ is the electric charge of the nucleus. The bunching/accelerating cavity frequency must obey the same constraints and should follow \begin{equation}
    \frev = \frac{p}{\sqrt{m^2c^4+p^2c^2}} \frac{c}{\Lambda} \, ,
\end{equation}
where $m$ is the deuteron mass and $\Lambda$ is the orbit circumference. The value of $\frev$ used in the scan must be recalculated at each step of the ramp. The quality of the orbit control was checked by computing the RMS (Root Mean Square) deviation of the orbit summed over all of the beam position monitors in the ring. Control was adequate when this deviation was less than \SI{1}{\mm}.

The ramps were controlled by providing continuously changing momentum values to the COSY control system. The ramps were calculated from a common starting point of \SI{970}{\mevc}, the same momentum used for the RF solenoid on resonance. Then the machine settings were moved to the starting point for an ALP scan. Two speeds were employed during the experiment. For the faster ramps, the initial and final momenta were calculated according to
\begin{eqnarray}
p_{0}&=&970 (1+1.173\times 10^{-4} n)\,\SI{}{\mevc},\label{eq:pstart}\\
p_{f}&=&p_{0}+\Delta p,\label{eq:pstop}
\end{eqnarray}
where $n$ is the number of the scan (see Fig.~\ref{fig:rampscheme}), either positive or negative, away from \SI{970}{\mevc} and $\Delta p$ is the momentum change in \SI{135}{s}. Table~\ref{tab:ramps} gives the momentum change $\Delta p$ which was entered into the COSY control system and the corresponding change in \fspin\ and \frev\ calculated from the read back from the RF cavity frequency. For the faster of the two ramps, the overlap between ramps was about \SI{2.6}{\hertz}.

For the slower ramps (second row), the same formula was used for $p_{0}$, but $\Delta p$ was decreased. In this case, there was a small coverage gap between adjacent ramps. The original plan was to have the ramps overlap so that resonances near the beginning or end of a scan would not be missed because of reduced time near the center of the resonance. As it is, they barely touched for the slow scans. There were 85 of the faster scans made with $n$ ranging from $-42$ to 42. For the 18 slow scans, $n$ ranged from $-60$ to $-43$.

\begin{table}[!hbt]
\caption{\label{tab:ramps}
Change in beam parameters during the ramp}
\begin{ruledtabular}
\begin{tabular}{p{1.3cm}p{1.3cm}p{1.3cm}p{1.3cm}p{1.3cm}}
$\Delta p$ [\SI{}{\mevc}] & $\Delta \frev$ [\SI{}{\hertz}]& $\Delta \subit{\dot{f}}{rev}$ [\SI{}{\hertz \per \second}]& $\Delta \fspin $ [\SI{}{\hertz}] & $\Delta \subit{\dot{f}}{spin}$ [\SI{}{\hertz \per \second}]\\
\hline
0.138 & 81.0 & 0.600 & 16.8 & 0.124\\
0.112 & 66.15 & 0.490 & 13.5 & 0.100\\
\end{tabular}
\end{ruledtabular}
\end{table}

Each run usually consisted of ten separate beam fills. One out of five of the fills was deliberately unpolarized in order to provide a baseline for no polarization effect. Thus, each run usually consisted of eight polarized fills. The results from these fills were combined, as will be explained in more detail in the analysis section below, to yield sensitivity results for the range of the scan.

The goal for each fill was to begin with $10^9$ polarized deuterons. Due to changes in source and machine operation efficiency over time, this value could vary up or down by a factor of two. These changes are reflected in the final sensitivity as a function of ALP frequency.

\begin{figure}[!tb]
    \centering
    \includegraphics[width=\columnwidth]{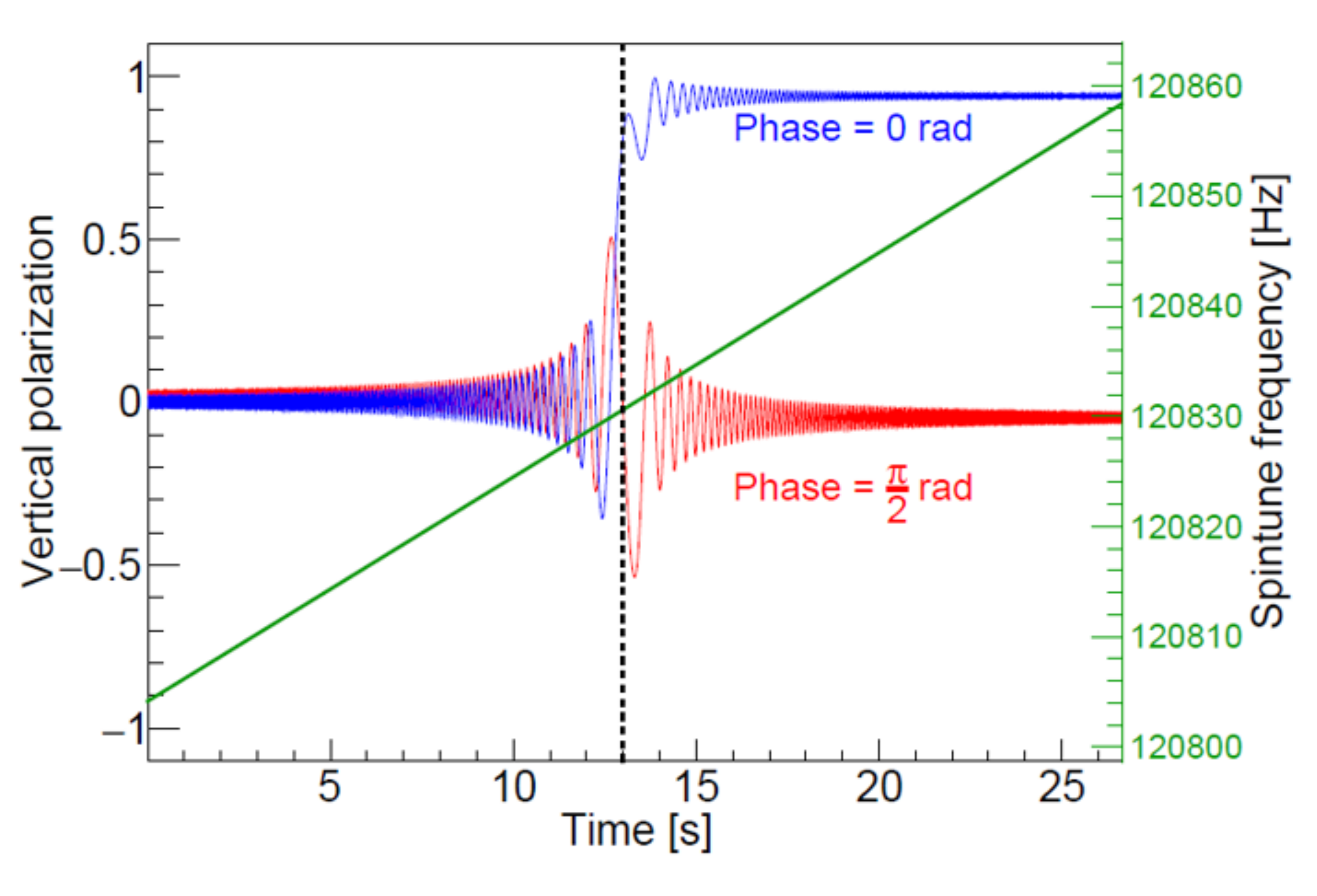}
    \caption{Graph of the vertical component of the polarization for a hypothetical resonance between the polarization rotation and the frequency of the ALP. Three quantities are shown as a function of time during a scan. The green line illustrates the changing spin precession frequency as time passes and the momentum of the beam is ramped up. At \SI{13}{\second} the spin tune frequency and the ALP frequency in this model are the same. The blue and red curves show the time dependence of the vertical polarization component for two different choices of the ALP phase. In this case, initial conditions are such that a large jump is seen for a phase of zero and a much smaller, negative jump is seen for a phase of $\pi /2$. These two cases would sample the ALP phase along perpendicular axes, thus the sum in quadrature of the jumps would represent the strength of the ALP coupling to the deuteron.}
    \label{fig:sample-jump}
\end{figure}

The calibration of the jump size in terms of the size of the oscillating EDM is presented in Appendix~\ref{App:B}. Here in the description of the experiment, it is useful to illustrate an example of the signal as it might appear for one bunch during the experiment. This is shown in Fig.~\ref{fig:sample-jump}.

In the analysis of the scans, the data were re-binned into 2-second bins, but the details of the resonance crossing as shown in Fig.~\ref{fig:sample-jump}, will not be apparent in the data. This same feature also applies to the Wien filter test described in the next Section. If the scanning speed is slower, then more time is available to make a polarization jump. Calibration calculations discussed in Appendix~\ref{App:B} show that the jump size scales as the square root of the reciprocal of the ramp rate.
\subsection{\label{sec:wftest}Wien Filter Test}
The COSY ring has been recently equipped with a spin manipulation tool that allows for spin rotations with minimal orbit disturbance, namely a waveguide RF Wien filter~\cite{Sl16, Slim_2020,Rat2020,Slim_2021}.
It was especially designed for precision experiments including the measurements of permanent EDMs in a magnetic storage ring,
which are performed at COSY in the framework of the JEDI collaboration.
The electric field is generated in sync with a perpendicular magnetic field so that the beam orbit is not perturbed. The Wien filter can be rotated \ang{90} around the beam axis without breaking the vacuum to adjust the field directions to the experimental needs. We chose to use this device to test the ability of our system to detect a vertical polarization jump when passing a resonance. For this it was operated at a fixed frequency on the $(1-G\gamma )\frev$ resonance with the magnetic field horizontal, such that the polarization is rotated about a sideways axis. By scanning the Wien filter frequency in small steps, the resonant frequency was found to be $871450.039\pm 0.002$~Hz. The scan of the RF frequency was done in the way established for the ALP scans. In this setup, the phase between the Wien filter oscillation and the rotation of the in-plane polarization was arbitrary for each fill of the machine. Thus jumps were expected to be of variable sign and magnitude in consecutive cycles. Nevertheless, the result of a random distribution of phases should be a series of jumps between a positive and negative limit of the same size with more cases located near the limits (projection of a sinusoidal function on the $y$ axis). As was the case for the ALP scans, the ramp operated between \SI{120}{\second} and \SI{255}{\second} in the machine cycle, producing a ramping time of \SI{135}{\second}. Ramps were made with the resonance in the middle of the ramp. The ramps went in both directions. There were two different ramp speeds, based on a total momentum change of either 0.056 or \SI{0.112}{\mevc} during the ramp.

\begin{figure}
    \centering
    \includegraphics[width=\columnwidth]{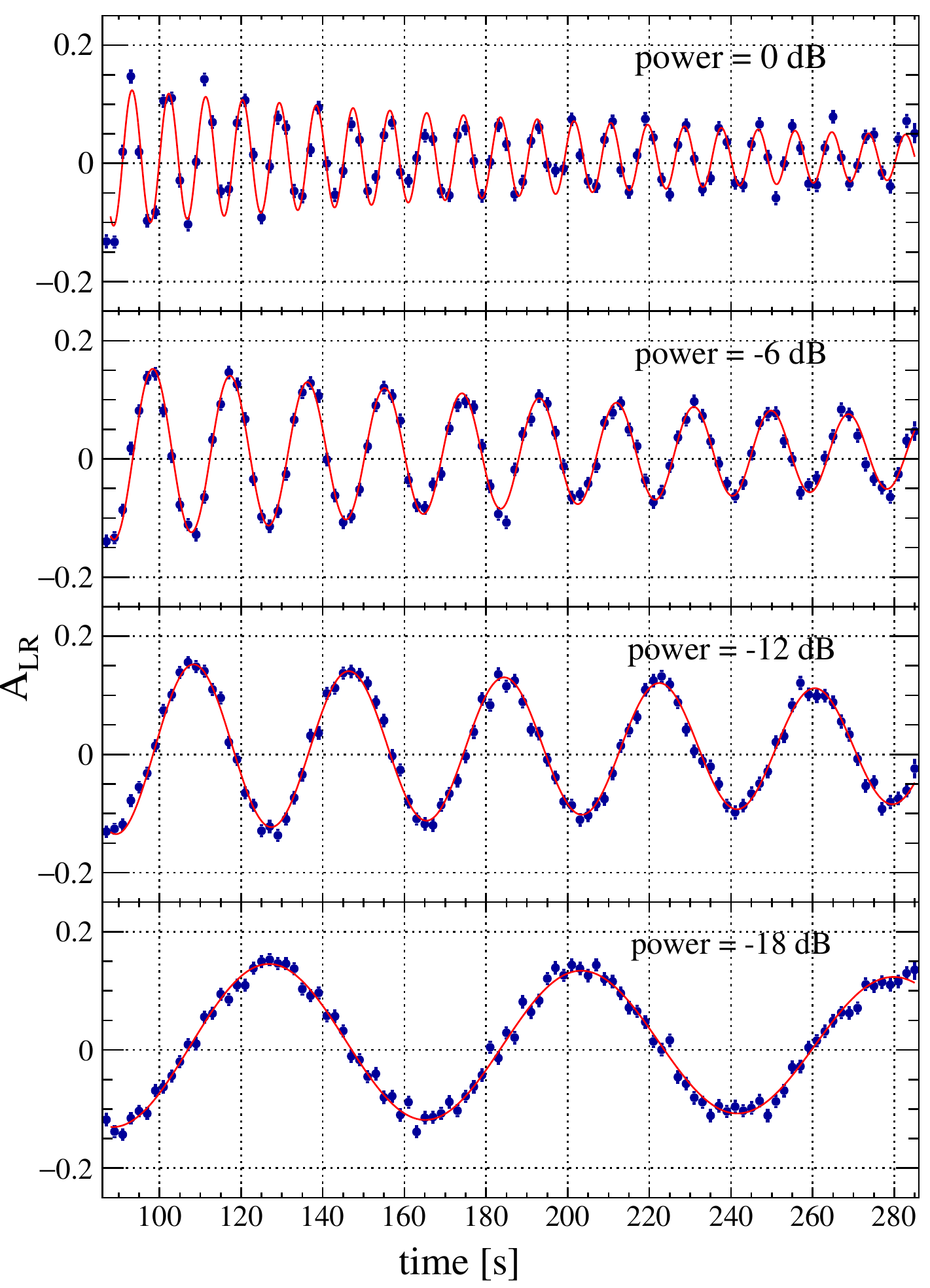}
    \caption{Measurements of the oscillating left-right asymmetry proportional to the vertical polarization produced by the continuous operation of the Wien filter at various power levels (noted in figure). The horizontal axis is time in seconds. The Wien filter was on continuously. Data from all four bunches were combined into a single asymmetry.}
    \label{fig:driven_osc}
\end{figure}

As an initial calibration of the strength of the Wien filter magnetic field, beam injected with a vertical polarization (no RF solenoid) was subjected to continuous operation of the Wien filter from \SI{88}{\second} to \SI{285}{\second} in the machine cycle. Thus time that was normally not a part of the scan in the machine cycle was added to the time to observe oscillations. This extra time came mostly from the two 30-second periods used previously for the non-ramp data. This setup should produce a continuous oscillation of the vertical polarization component. Four different power levels were used for the Wien filter, each differing from the previous by a factor of two in magnetic field. In Fig.~\ref{fig:driven_osc}, data is shown between \SI{81}{\second} and \SI{287}{\second}. The Wien filter is turned on at $t_0=\SI{88}{\second}$.

For the fits to the driven oscillations, the raw asymmetry data from the measurements were averaged across all four bunches and rebinned in 1-second intervals. Due to a slow depolarization arising from synchrotron oscillations \cite{Be12}, the oscillations are damped with time. These patterns were reproduced using the function
\begin{equation}
    \subit{A}{LR}(t)=A\left[e^{-\frac{t-t_0}{\tau}}
    \cos(2\pi \subit{f}{drv}(t-t_0)+\phi )\right]+k,
\end{equation}
where $\subit{A}{LR}(t)$ is the shape of the data, $A$ is the amplitude, $\tau$ is the decay constant, \subit{f}{drv} is the driven oscillation frequency, $\phi$ is the phase, and $k$ is the zero offset of the asymmetry data. The results for the frequency \subit{f}{drv}, which is a measure of the strength of the Wien filter magnetic field, are given in Table~\ref{tab:WFfreq}. The right-hand column shows the ratio between the frequency on that row and the previous row. Given the power settings, this ratio should be two. Within a few percent, this ratio is realized. Variations are due to the properties of the control system of the Wien filter. When scans were recorded for the size of their jumps, a power level of 0~db was used.

\begin{table}[!hbt]
\caption{\label{tab:WFfreq}
Driven oscillation frequencies.  The third column contains the ratio of the current row frequency to the one of the preceding row.}
\begin{ruledtabular}
\begin{tabular}{lll}
Power (db) & Frequency (Hz) & Ratio \\
\hline
-18 & 0.013084(19) & \\
-12 & 0.026326(21) & 2.0122(33) \\
-6 & 0.052816(25) & 2.0062(19) \\
0 & 0.110848(345) & 2.0988(66) \\
\end{tabular}
\end{ruledtabular}
\end{table}

\section{\label{sec:dataana}Data Analysis}

The data available from each run consisted of the left-right asymmetry and the unfolded down-up asymmetry as functions of time before, during, and after each scan. These measurements were available from each of the four beam bunches. Since the initial vertical polarization has already been precessed into the ring plane, the left-right asymmetry where the jump may appear is initially close to zero. Meanwhile, the down-up asymmetry, which is subject to depolarization due to spin tune spread, declines slowly with time. This behavior was usually linear. For each fill and bunch, a linear fit to these data provided values of the in-plane asymmetry (\Aip) as a function of time during the scan. Results from different bunches in the same fill were consistent. Jumps were observed only for the test case using the Wien filter to rotate the spins of the deuterons.

In line with the open science policy, the collected semi-raw experimental data have been made available in the J\"ulich~DATA repository~\cite{rawData}. Two independent analyses have been performed using somewhat different analysis algorithms to define confidence intervals. As they yielded consistent results, in the following we present only one of the approaches based on the well-known Feldman-Cousins~\cite{Fe98} procedure while the other can be found in~\cite{SPCthesis}.

Models, as described in the appendices, were used to relate the sizes of the jumps to the case where the beam polarization was unity and the effects were generated by the presence of an oscillating EDM. Subsequent examples of left-right asymmetries in this Section show the original measurements; any jumps recorded were then normalized by dividing by the linear fit to $\Aip(t)$ appropriate for the time of the observation.

This Section addresses in turn the calculation of \Aip\ in the presence of ramping, the general treatment of possible jumps, and results for the Wien filter scans and the ALP scans. The jumps for the four bunches were then combined into a single result by fitting them to a sinusoid as a function of the relative phase between the beam rotation and the ALP oscillation. This process produces non-vanishing amplitudes for the sine wave even in cases where no effect is expected. This leads to a more complicated interpretation, as will be explained in Section~\ref{sec:FC}.

\subsection{Calculation of the In-Plane Polarization}

The analysis of the down-up in-plane asymmetry \Aip~was described in Ref. \cite{Ba14}. Data consisting of down-up events were gathered into 2-second time bins. As a function of time the angle of the polarization $\alpha$ is given by
\begin{equation} \label{eq:st_phase}
    \alpha = \omega t = 2\pi \nu \frev (t-t_0)\,,
\end{equation}
where the spin tune $\nu=G\gamma$ and the revolution frequency \frev\ are assumed to be constant and $t$ is measured relative to $t_0$ at the beginning of the time bin. The events are divided into 12 angular bins according to the value of $\alpha$ modulo $2\pi$. The down-up asymmetry $A_{\rm DU}$ is calculated for the events in each bin. Finally, a sinusoidal curve is fit to $A_{\rm DU}(\alpha )$. The amplitude of the sine curve becomes the measure of \Aip, the in-plane asymmetry. The phase of the fit is tracked as a function of time bin. A constant value is interpreted as a validation that the initial choice of spin tune is correct.

In the case of a scan, the spin tune undergoes a linear ramp from time $t_1$ to $t_2$, as depicted in Fig.~\ref{fig:sample-jump}. At the same time the machine revolution frequency also ramps. This gives
\begin{equation} \label{eq:ang_vel}
    \centering
    \omega(t) =
        \begin{cases}
            2\pi\nu_0 \subit{f}{rev,0} \ \ \ \ \   \mathrm{for}\ t_0<t<t_1,\\
            2\pi[\nu_0+\dot{\nu}(t-t_1)] [\subit{f}{rev,0}+\subit{\dot{f}}{rev}(t-t_1)] \\\ \ \ \ \ \ \ \ \ \ \ \ \ \ \ \ \ \  \mathrm{for}\ t_1<t<t_2,\\
            2\pi\nu_f f_{\mathrm{rev,}f} \ \ \ \ \  \mathrm{for}\ t>t_2,
        \end{cases}
\end{equation}
\begin{equation}
\alpha(t) = \int_{t_0}^t \omega(t')\mathrm{dt'}
\end{equation}
where the subscripts 0 and $f$ correspond to the initial and final values. Dotted symbols denote time derivatives. Once the spin phase $\alpha (t)$ is known, the calculation of \Aip~proceeds as in the no-ramp case mentioned above.

\subsection{\label{sec:calc_jump} Calculation of Polarization Jump}

\begin{figure}[!hbt]
    \includegraphics[width=\columnwidth]{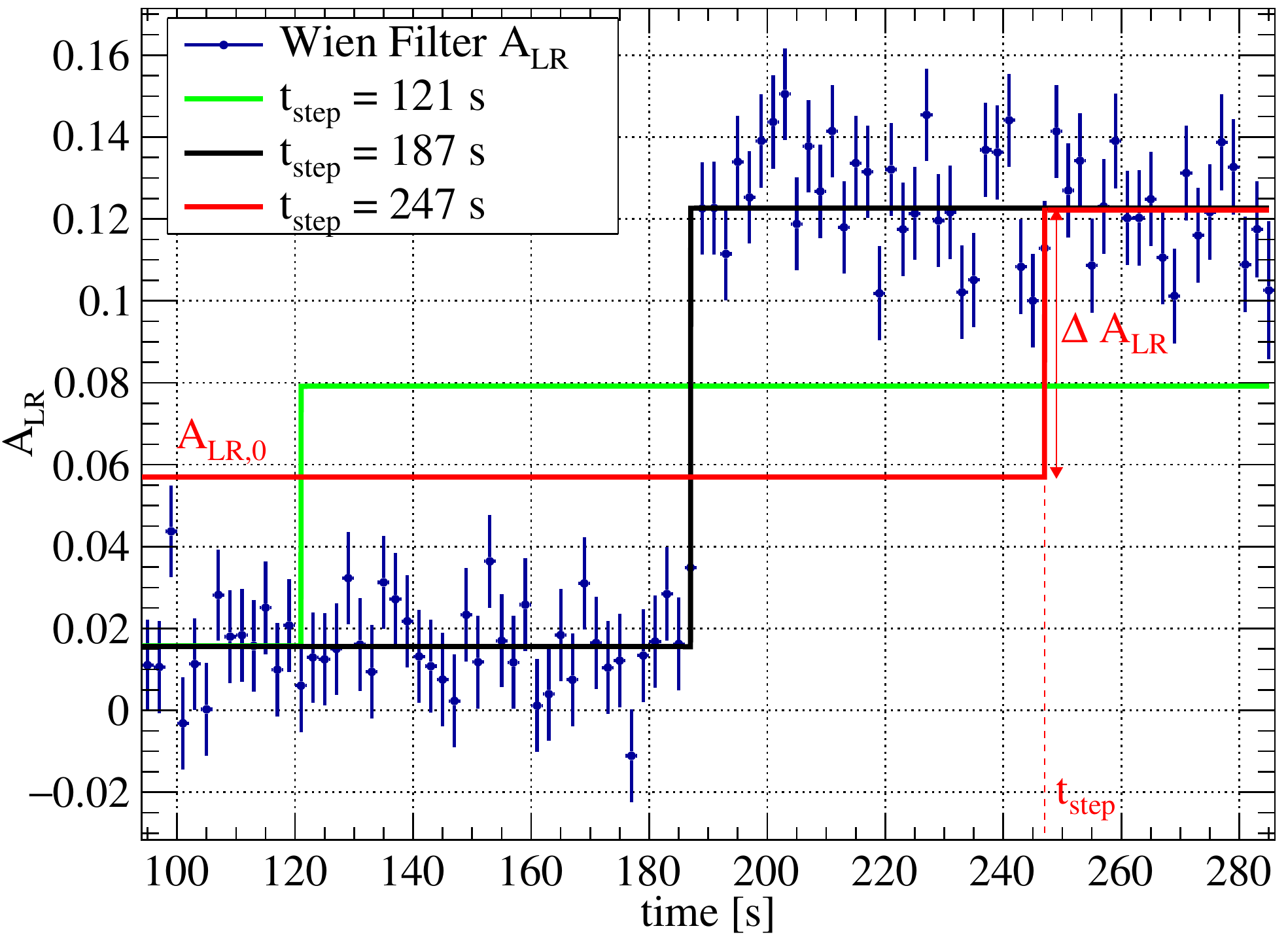}
    \caption{\label{fig:stepfn} Examples of step function fits for the Wien filter scan data for a single bunch from one cycle. The black line is fit with the jump ($\tstep$) at the resonance crossing. The red and green curves show the results for other choices of the jump time. In both cases, the jump size is smaller and the reduced chi-square of the fit is increased (see Fig.~\ref{fig:WF_chi2}).}
\end{figure}

The data to be used in the analysis come from the left-right asymmetries recorded during the scanning process as a function of time. An illustration based on data taken with the Wien filter is shown in
Fig.~\ref{fig:stepfn}. With the level of time binning used in this experiment, the jump appears to be instantaneous. We represent this process using the step function:
\begin{equation} \label{eq:stepfn}
    f(t) =
    \begin{cases}
        {\Alr}_{,0} & \mathrm{if}\ t < \tstep,\\
        {\Alr}_{,0} + \Delta \Alr & \mathrm{if}\ t \geq \tstep.
    \end{cases}
\end{equation}
In this equation, ${\Alr}_{,0}$ represents the left-right asymmetry before the scan as well as the asymmetry before the jump. The jump value, $\Delta \Alr$, is the size of the change in the asymmetry. In
Fig.~\ref{fig:stepfn}, the black curve uses $\tstep$ = 187~s, the time when the in-plane polarization rotation frequency and the Wien filter frequency were the same. The step function is a good representation of these data.

\begin{figure}[!hbt]
    \includegraphics[width=\columnwidth]{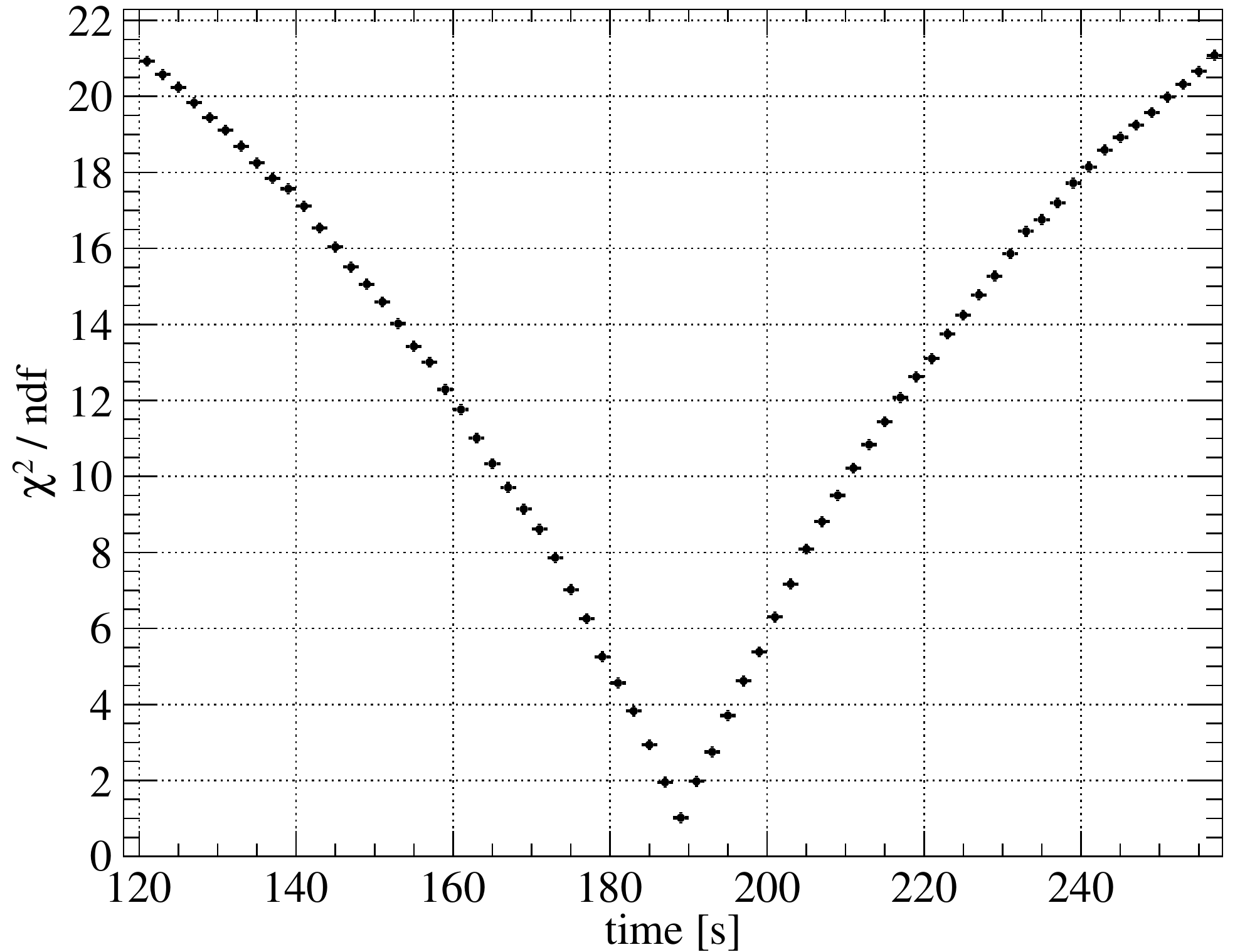}
    \caption{Reduced chi-squared plot of the step function fits for a Wien filter scan from one cycle as a function of the time assumed for the jump. The calculations are based on the data from
    Fig.~\ref{fig:stepfn}. The minimum corresponds to the time when the resonance takes place.}
    \label{fig:WF_chi2}
\end{figure}

In the case of normal scans for an ALP, we do not know a priori when the jump may have occurred, if at all. In this case, the fit is repeated as $\tstep$ is given the time of each bin from 121 to 257~s. When this is done for the Wien filter data of Fig.~\ref{fig:stepfn}, the fits away from the resonance, as shown by sample red and green curves, display a smaller value of $\Delta\Alr$. There is also a larger value of the fit $\chi^2$, as seen in Fig.~\ref{fig:WF_chi2} when \tstep\ is away from the resonant frequency. Each cycle and bunch of the ALP search data was scanned for such a feature. The results will be discussed in
Section~\ref{sec:dataaxion}.

\subsection{\label{sec:dataWF}Wien Filter Scan Analysis}

The Wien filter scan data consisted of 48 separate machine cycles. The four bunches within each cycle displayed oscillating jumps of the opposite sign.
In each different cycle, the phase between the Wien filter and the rotating in-plane polarization was random. This resulted in variations in the jump size from cycle to cycle that spanned the range of possibilities.
Statistical variations in the recorded asymmetries lead to errors in the jump of about 2\%. In addition, the phase uncertainty multiplies this size by the cosine of the unknown relative phase. This acts to reduce all jump sizes. But the peak of the distribution should be close to the maximum value of one for the cosine.

To get a better estimate of the maximum, the absolute values of the jumps for the two ramp speeds were placed into separate distributions. In each case, there was a clear maximum. The jump amplitude was found by interpolating half way between the bin with the maximum number of cases in the upper 20\% of the distribution and the maximum in the distribution. This in part allows the downward bias of the cosine effect to be corrected by the possible upward bias of the jump statistical distribution. An evaluation of this procedure using a Monte-Carlo model showed that the scatter of the answers was 2\% given the number of jumps recorded, roughly the same as the statistical error in the jump size. The 2\% error overlapped with the model value of the maximum jump.

The experimental values obtained using this procedure are presented in Table~\ref{tab:WF_jump} along with the values from a dedicated simulation.
This simulation was performed similarly to the one for the sensitivity calibration presented in Appendix~\ref{App:B}, only using the actual working parameters of the Wien filter and probing various relative phases. In addition, as the Wien filter is a localized device, the rotations have not been combined but executed subsequently as described for the RF solenoid simulation in Appendix~\ref{App:A}.
The resonance strength of the Wien filter was derived from the ratio of the driven oscillation frequency (see Sect.~\ref{sec:wftest}) to the revolution frequency in the storage ring at resonance
\begin{equation}
    \subit{\epsilon}{WF} = \frac{\SI{0.110848}{\hertz}}{\SI{750602.6}{\hertz}} = 1.4768\times 10^{-7}\ .
\end{equation}
This converts to an oscillation amplitude of the spin rotation per turn by the Wien filter of $\subit{\psi}{WF} = 4\pi\subit{\epsilon}{WF}$ (see, {\it e.g.}, Ref.~\cite{Rat2020}).
Note that at this large resonance strength the linear dependence between rotation amplitude and polarization jump (see Appendix~\ref{App:B}) no longer holds and, thus, the ratio of the polarization signals for the two ramp rates does not reflect the discussed scaling behavior.

While neither pair of values agrees within errors, taken together there is a confirmation that the simulation program correctly models the sensitivity of the experiment. We will therefore assume that the calibration described in Appendix~\ref{App:B} may be used to determine our sensitivity to ALPs.

\begin{table}[!hbt]
\caption{\label{tab:WF_jump} Comparison of the maximum polarization jump $\Delta p_y$ from simulation and experiment for the Wien filter test.}
\begin{ruledtabular}
\begin{tabular}{ccc}
$\Delta p$ [\SI{}{\mevc}]& Simulation & Experiment \\
\hline
0.112 & 0.75 & 0.796(15)\\
0.056 & 0.93 & 0.892(18)\\
\end{tabular}
\end{ruledtabular}
\end{table}

\subsection{\label{sec:dataaxion}ALP Scan Analysis}

\begin{figure}[!hbt]
    \includegraphics[width=\columnwidth]{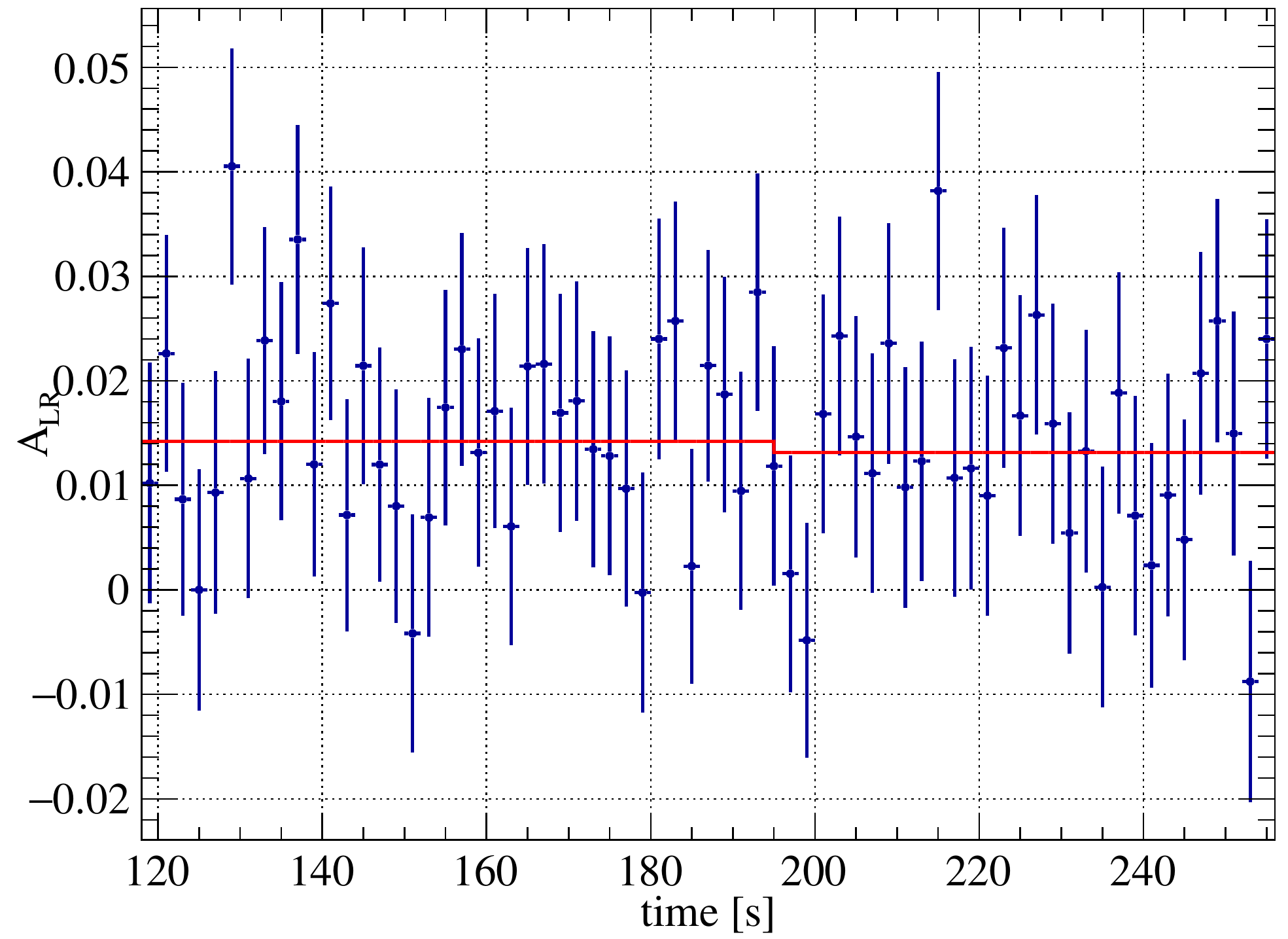}
    \caption{Example step function fit for an axion scan for a single bunch from one cycle. There is no jump in asymmetry since $\Delta {\Alr} =-0.00105(233)$ is consistent with zero.
    \label{fig:stepfn-axion}}
\end{figure}

\begin{figure}[!hbt]
    \includegraphics[width=\columnwidth]{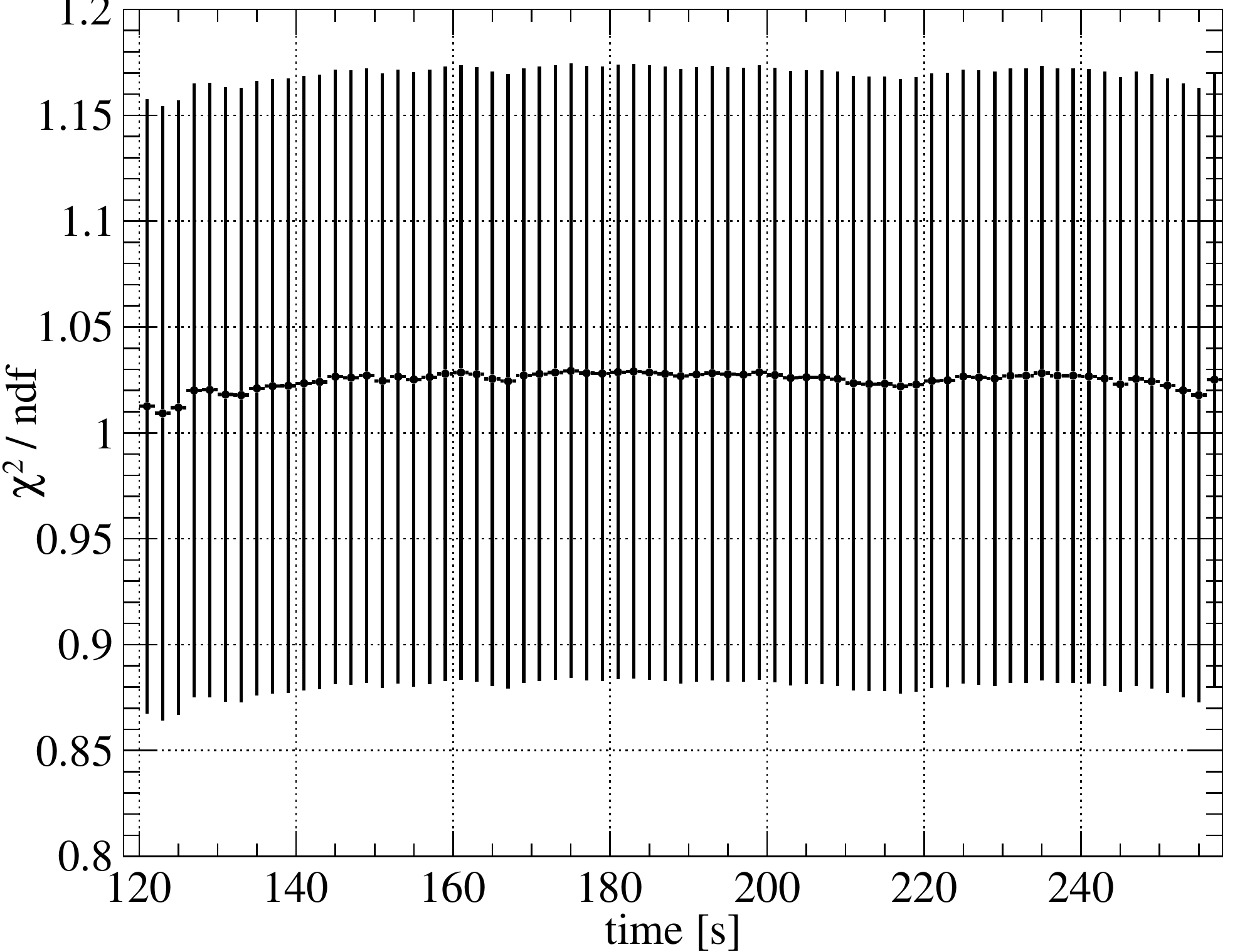}
    \caption{Reduced chi-squared plot of the step function fits for an axion scan from one cycle. The absence of a minimum indicates no resonance. The
    vertical bar shows the standard deviation of the chi square values  based on the number of degrees of freedom.
    \label{fig:axion_chi}}
\end{figure}

For the data generated during scanning for an ALP, the process just described to locate the most probable time and size for a jump was repeated for each machine cycle and bunch. A typical example is shown in Fig.~\ref{fig:stepfn-axion}. Unlike the Wien filter case, there is no apparent jump. The red curve shows the largest possible jump found. This result is consistent with the lack of a minimum in the associated chi-square versus time plot presented in Fig.~\ref{fig:axion_chi}. Together, these results point to the absence of a resonance between the spin tune frequency and any ALP frequency within the range of the scan. The vertical bars in Fig.~\ref{fig:axion_chi} indicate the standard deviation of the reduced $\chi^2$ distribution
given by  $\sqrt{2/\mbox{ndf}}$, where
\mbox{ndf} indicates the number of degrees of freedom in the fit with \mbox{ndf} = $2\times 15$ flat region points + 68 scan region points $-$ 3 fit variables = 95. (This corresponds to a standard deviation of 0.145). Figs.~\ref{fig:stepfn-axion} and \ref{fig:axion_chi} are for the scan data what Figs.~\ref{fig:stepfn} and \ref{fig:WF_chi2} were for the Wien filter data.

\begin{figure}[!hbt]
    \includegraphics[width=\columnwidth]{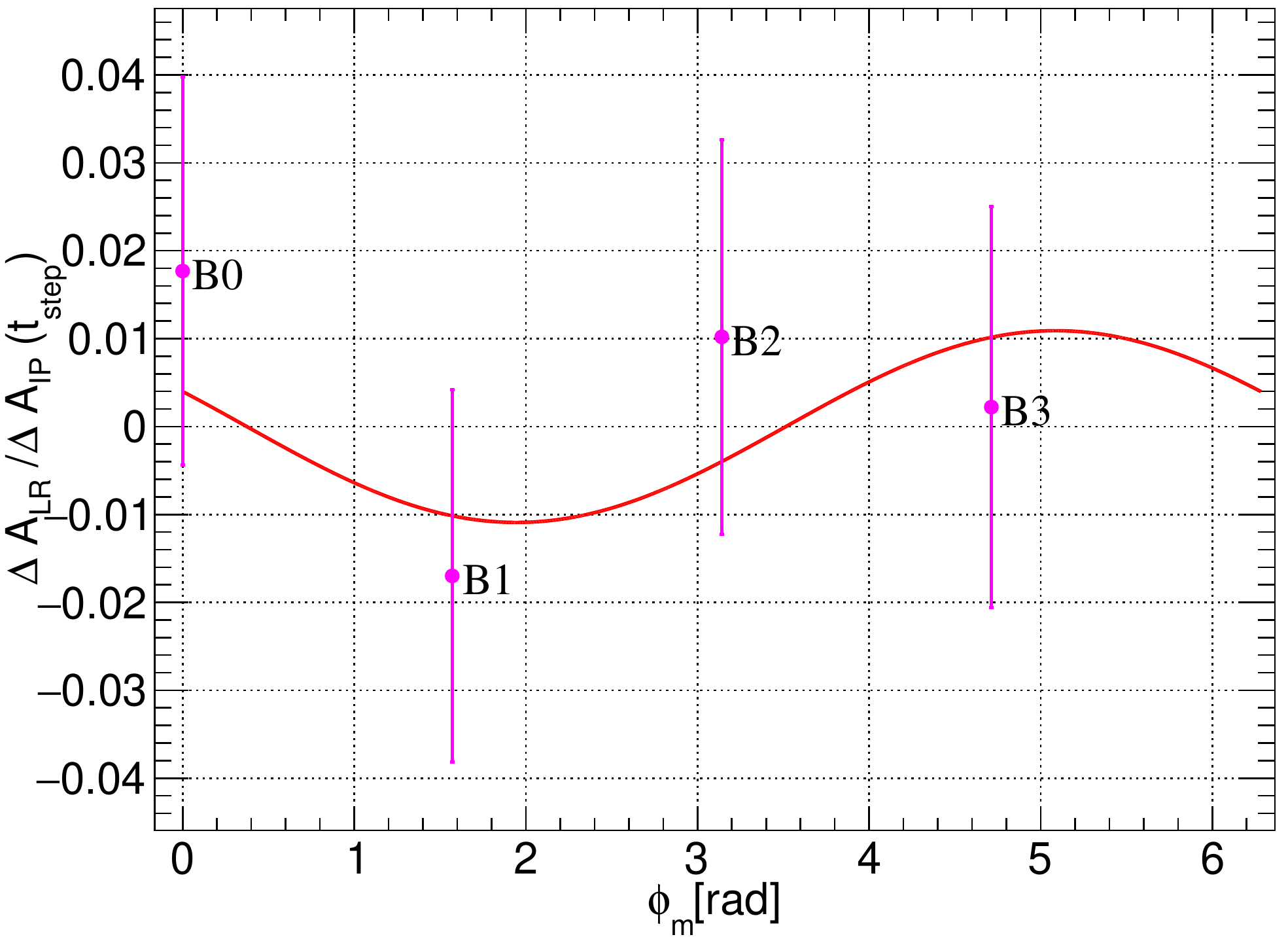}
    \caption{Left-right asymmetry jump for all four bunches, from one cycle, as a function of the angle between the bunch polarization and the axion phase~$\phi_m$ for a single time bin. The red curve is the sinusoidal fit from which the jump amplitude \aest\ is calculated. \label{fig:Bu_sin}}

\end{figure}

To avoid missing an ALP due to a phase mismatch, the search results from all four bunches in each time bin of a scan were combined to produce a single sinusoidal curve as a function of possible phase using the formula
\begin{eqnarray}
    f(\phi_m) &=& C_1 \sin{\phi_m}+ C_2 \cos{\phi_m} \label{eq:sin},\\
    \aest &=& \sqrt{C_1^2+C_2^2}, \label{eq:amp}
\end{eqnarray}
where $m$ denotes the bunch number. The $y$-axis has been renormalized and shows the amplitude of each jump analysis divided by the in-plane asymmetry \Aip~at the time of the tentative jump. The amplitude of the sinusoidal fit is given by
Eq.~(\ref{eq:amp}).
The spacing between the bunches on the $\phi_m$ axis is $\pi/2$ in Fig.~\ref{fig:Bu_sin}. As was discussed at the end of Section~\ref{sec:axionphase}, the spacing is not always equal but oscillates between two extreme values. A correction is made for this effect.

\begin{figure}[!hbt]
    \includegraphics[width=\columnwidth]{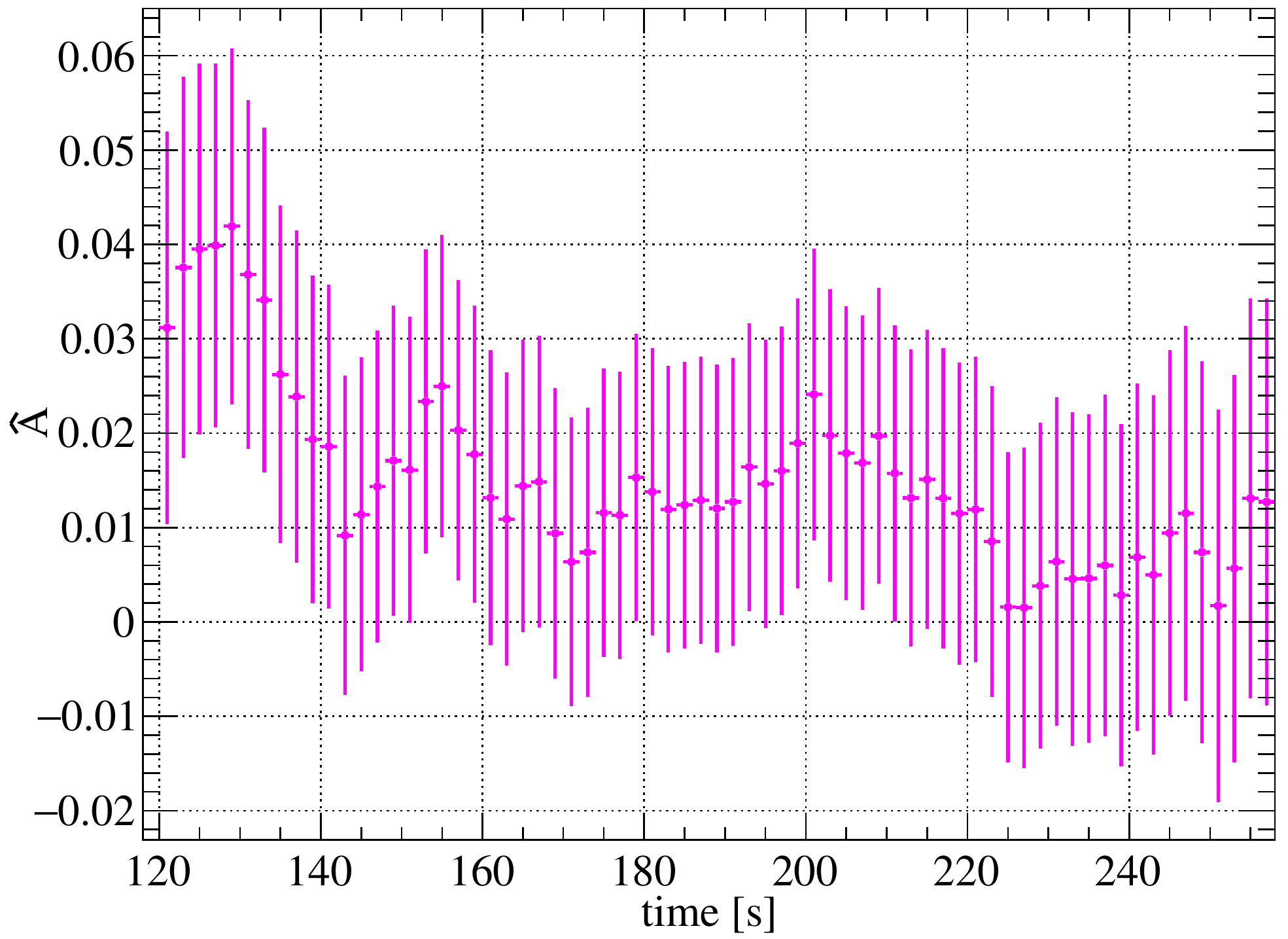}
    \caption{Amplitude \aest\ from sinusoidal fit for a single cycle.\label{fig:ampl}}

\end{figure}

The jump amplitude \aest~(\ref{eq:amp}) from the sinusoidal fit is calculated for each time bin and Fig.~\ref{fig:ampl} shows the time distribution of that amplitude for one cycle. For multiple cycles covering the same frequency region, the mean amplitude is calculated for each time bin as a weighted average of amplitudes from the individual cycles. That mean and its uncertainty enter the next stage of the calculation of the confidence limit.


\subsection{\label{sec:FC}Construction of Confidence Interval}

The use of Eqs.~(\ref{eq:sin}) and (\ref{eq:amp}) is designed to capture any possible ALP regardless of the ALP phase at the time of resonance crossing. The cost of using these equations is that near zero amplitude where most of the results will be, there is a systematic tendency to overestimate the size of the jump. Eq.~(\ref{eq:amp}) always generates a positive definite value. There is no distribution about zero that would allow zero as a mean. If $\hat A$ would happen to be large compared to its error, this effect would fade. To account for this positive bias and calculate a meaningful upper limit, the Feldman-Cousins procedure \cite{Fe98} will be used to construct the confidence interval. Refs. \cite{Pl14, Ev16} contain a detailed discussion on how the procedure is used for these cases.

In these references, the discussion describes how to interpret the estimated amplitude ($\hat A$) in terms of a true amplitude ($A$). In order to facilitate the application to a large number of time bins as a function of ALP frequency, we will switch to the amplitude normalized by the statistical error. This gives the normalized estimated value ($\hat P=\hat A/\subit{\sigma}{exp}$) and true value ($P=A/\subit{\sigma}{exp}$). The advantage is that we do not need to regenerate the interpretation for each time bin.

\begin{figure}[tbph]
    \includegraphics[width=\columnwidth]{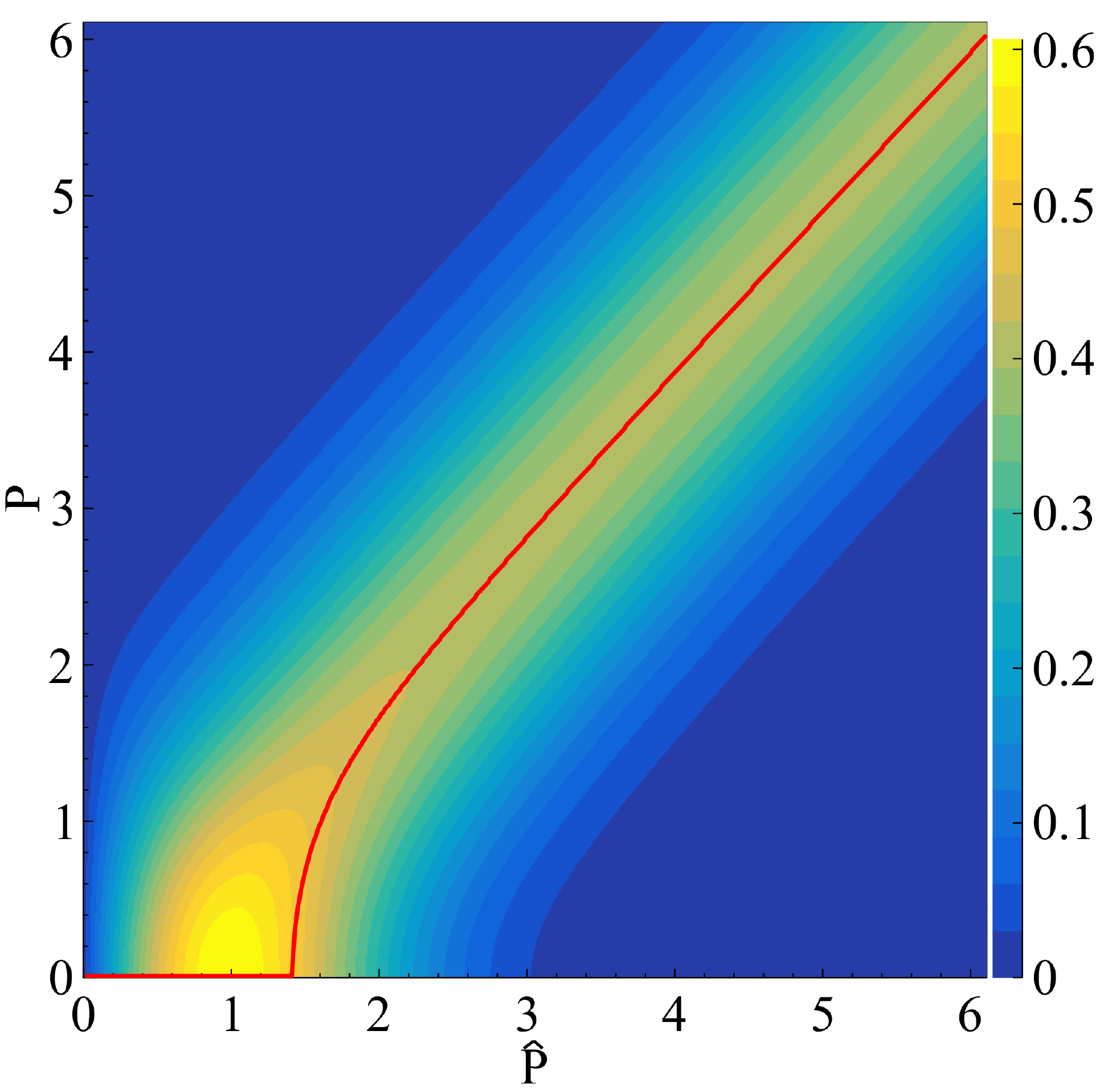}
    \caption{A 2-dimensional Rice plot for one cycle. The red line represents the value of $P$ for which the Eq.~(\ref{eq:pdf}) is maximum for a given value of \pest.
    \label{fig:Rice_1cyc}}
\end{figure}

The probability density function (PDF) for data distributed according to $\aest = \sqrt{{C}_1^2+{C}_2^2}$ is given in Eq.~(2.2) of \cite{Ev16}. Modifying this for the $P$ quantity we obtain:
\begin{equation}
    f(\pest|P)\,\mathrm{d}\pest = e^{-\frac{\pest^2+P^2}{2}} \pest I_0(\pest P)  \,\mathrm{d}\pest,
    \label{eq:pdf}
\end{equation}
where $I_0$ is the modified Bessel function of the first kind. Equation~(\ref{eq:pdf}) is called the Rice distribution. A 2-dimensional graphical representation of this distribution for \(0 \le P \le 6\) is shown in Fig.~\ref{fig:Rice_1cyc}. The red line denotes \subit{P}{best}, which is the value of $P$ for which $f(\pest|P)$ has the maximum probability in the physically allowed region for $P$.

The distortion of the distribution away from a typical Gaussian shape where $P\approx \hat P$ becomes clear below $\hat P=2.5$. For quantities that are near zero, the estimated or experimental value is about one. This means that the experiment seems to produce evidence of an effect even though it is only significant at the one standard deviation statistical level.

\begin{figure}[tbph]
    \includegraphics[width=\columnwidth]{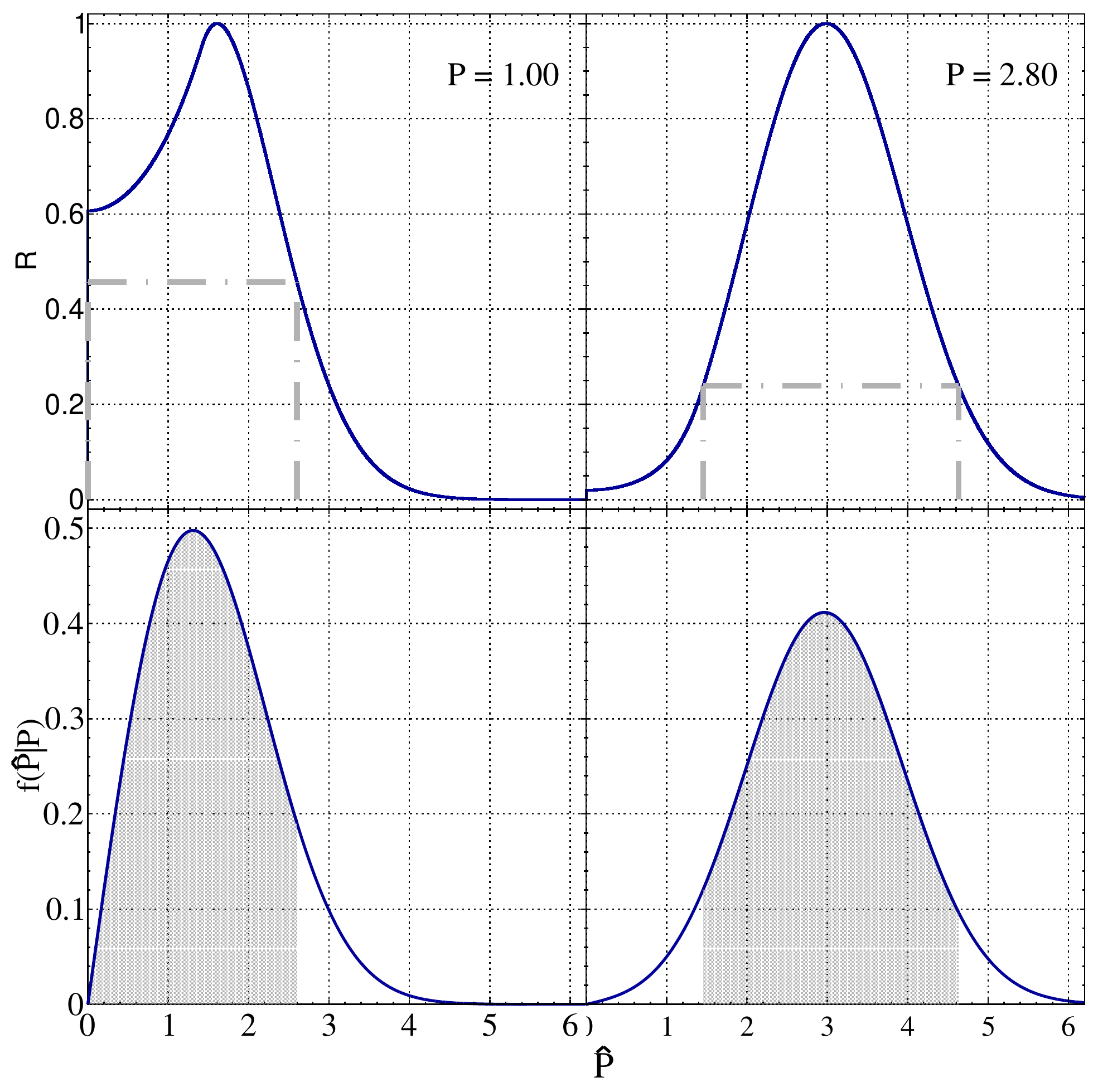}
    \caption{Two examples, left $P = 1.0$ and right $P = 2.8$, for the calculation of 90\% confidence limits using the likelihood ratio given by Eq.~(\ref{eq:LH_ratio}) (top row) and PDF given by Eq.~(\ref{eq:pdf}) (bottom row). The gray horizontal dashed line in the likelihood ratio curves denote the \(R\) value for which the corresponding \pest\ values (gray vertical dashed lines) forms the 90\% integral in the PDF curves. The gray shaded region marks the 90\% integral region.
    \label{fig:R_pdf_limit}}
\end{figure}

Next a likelihood ratio $R$ is calculated using the following definition,
\begin{equation}
    R = \frac{f(\hat{P}|P)}{f(\hat{P}|\subit{P}{best})}.
    \label{eq:LH_ratio}
\end{equation}
Two examples of the likelihood curve for \(P=1.0\) and 2.8 are shown, in blue, in the top row of Fig.~\ref{fig:R_pdf_limit}.  The \fc\ confidence interval is constructed by determining the bounds within which the integral of \(f(\pest|P)\) reaches the desired confidence interval, {\it e.g.}, 90\%. The bottom row of Fig.~\ref{fig:R_pdf_limit} shows the PDF for \(P=1.0\) and 2.8 along with the gray shaded region denoting the 90\% confidence limit for the two cases above.

The confidence limit bounds on $\hat P$ are determined by starting with the largest value of $R$ where it is one and following the two limit points given by the intersection of a straight horizontal line with the likelihood ratio curve as the line moves down the plot. In the upper left case where the curve ends at $\hat P=0$ as the horizontal line crosses $R=0.6$, the left axis where $\hat P=0$ replaces the lower limit of the $R$ curve. This process continues until the integral (gray shading) of the lower curve between the two limits reaches the desired confidence level. For the right-hand ``Gaussian’’ case, both limits are still on the $R$ curve and are roughly symmetric about the peak of $R$. The lower and upper limits of $\sim\!1.5$ and $\sim\!4.6$ represent the bounds of the 90\% confidence interval. For the ``left-hand’’ case there is only an upper limit at $\sim\!2.6$. Most of the data points in this experiment follow this example.

\begin{figure}[!hbt]
    \includegraphics[width=\columnwidth]{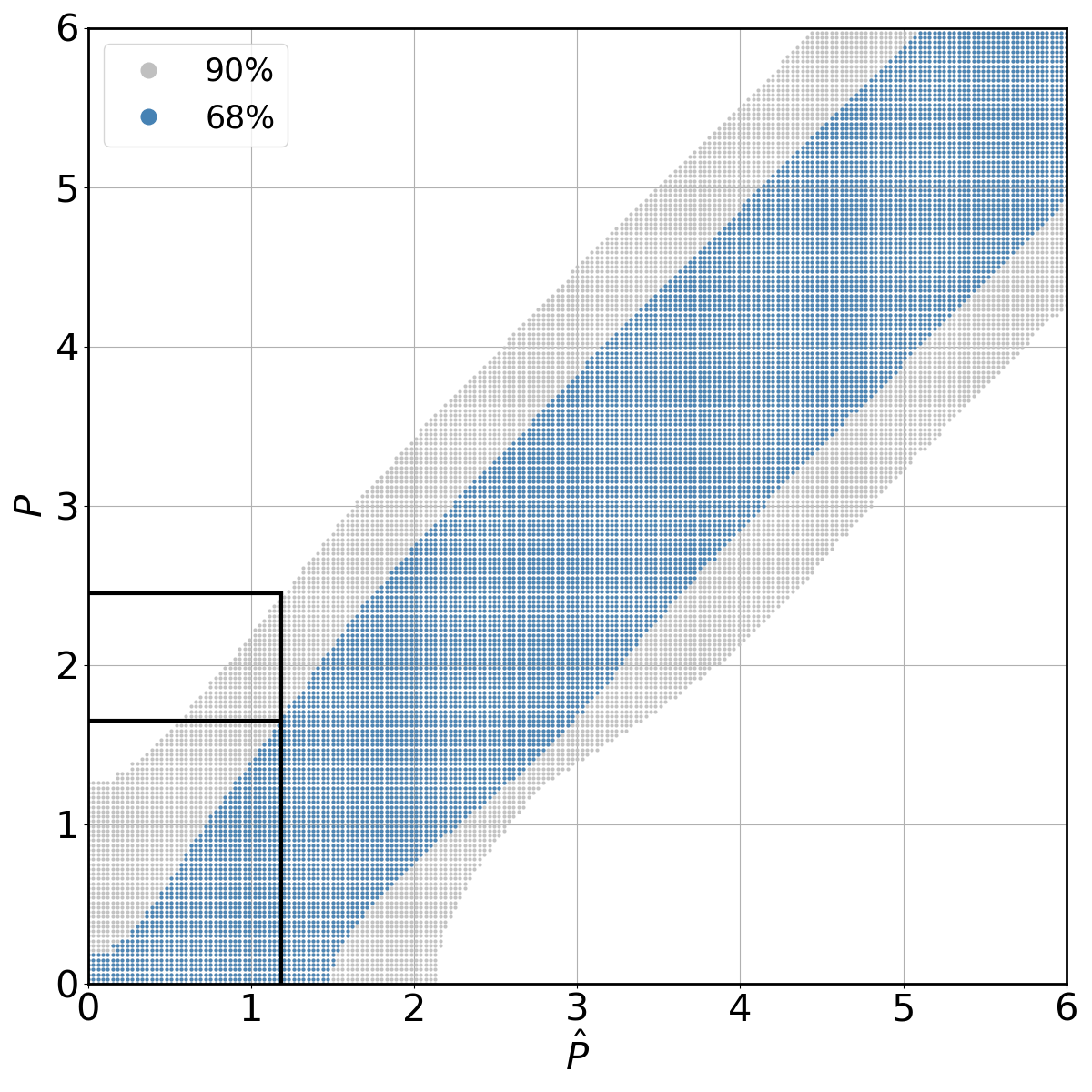}
    \caption{A 68\% (blue) and 90\% (gray) confidence interval for one cycle analysis. On the $x$-axis we have the estimated value $\hat{P}$ and on the $y$-axis is the true value $P$.
    \label{fig:conf_1}}
\end{figure}

A summary of all of the limits for $P$ may be found in Fig.~\ref{fig:conf_1}. Inside the blue band the confidence is 68\%. The gray band outside the blue band shows the edges for the 90\% limit. For a given value of $\hat P$ trace a line upward until it crosses the appropriate boundary. The case shown is for an upper limit only where there is no lower limit other than zero. These limits correspond to a single beam fill in the experiment with only one scan.

Since most scans comprise 8 cycles the confidence interval needs to be constructed taking this into account. A few scans comprised 7,  9, or 16 cycles. According to the central limit theorem, for $n$ cycles $\hat P$ follows a Gaussian distribution, the mean amplitude remains at the same $A$ value and the uncertainty is $\sigma_n = \subit{\sigma}{exp}/\sqrt{n}$. It is assumed that this is approximately true since, once set up, the beam current reproduced well from cycle to cycle for any particular scan. All \subit{\sigma}{exp} are the same for cycles being averaged this way. The PDF for $n$ cycles is a Gaussian with $P = A /\sigma_n$:
\begin{eqnarray}
    f(\pest|P) &=& \frac{1}{\sqrt{2\pi}\subit{\sigma}{Rice}} e^{\frac{(\pest-\subit{\mu}{Rice})^2}{2\subit{\sigma^2}{Rice}}},
    \label{eq:gauss_pdf}\\
    \subit{\mu}{Rice} &=& \sqrt{\frac{\pi}{2}} \sqrt{n} L_{1/2}\left(-\frac{1}{2} \frac{P^2}{n}\right) \label{eq:n_mu_Rice},\\
    \subit{\sigma^2}{Rice} &=&  2 + \frac{P^2}{n} - \frac{\pi}{2} L^2_{1/2}\left(-\frac{1}{2} \frac{P^2}{n}\right) \label{eq:n_var_Rice},
\end{eqnarray}
where $L_{1/2}$ is the Laguerre polynomial.

\begin{figure}[!hbt]
    \includegraphics[width=\columnwidth]{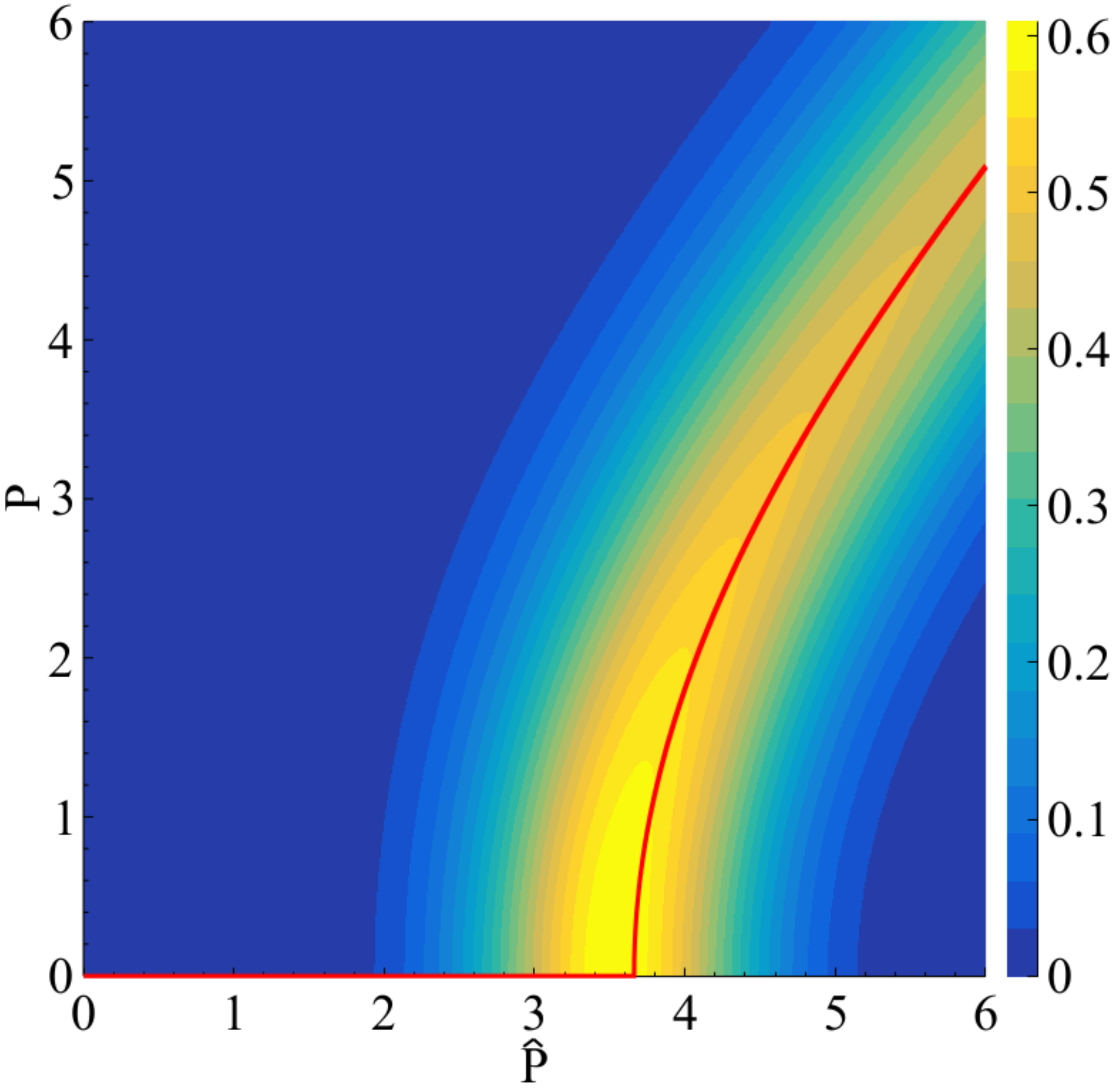}
    \caption{A 2-dimensional Rice plot for 8 cycles. The red line represents the value of $P$ for which the Eq.~(\ref{eq:gauss_pdf}) is maximum for a given value of \pest.
    \label{fig:Rice_8cyc}}
\end{figure}

\begin{figure}[!hbt]
    \includegraphics[width=\columnwidth]{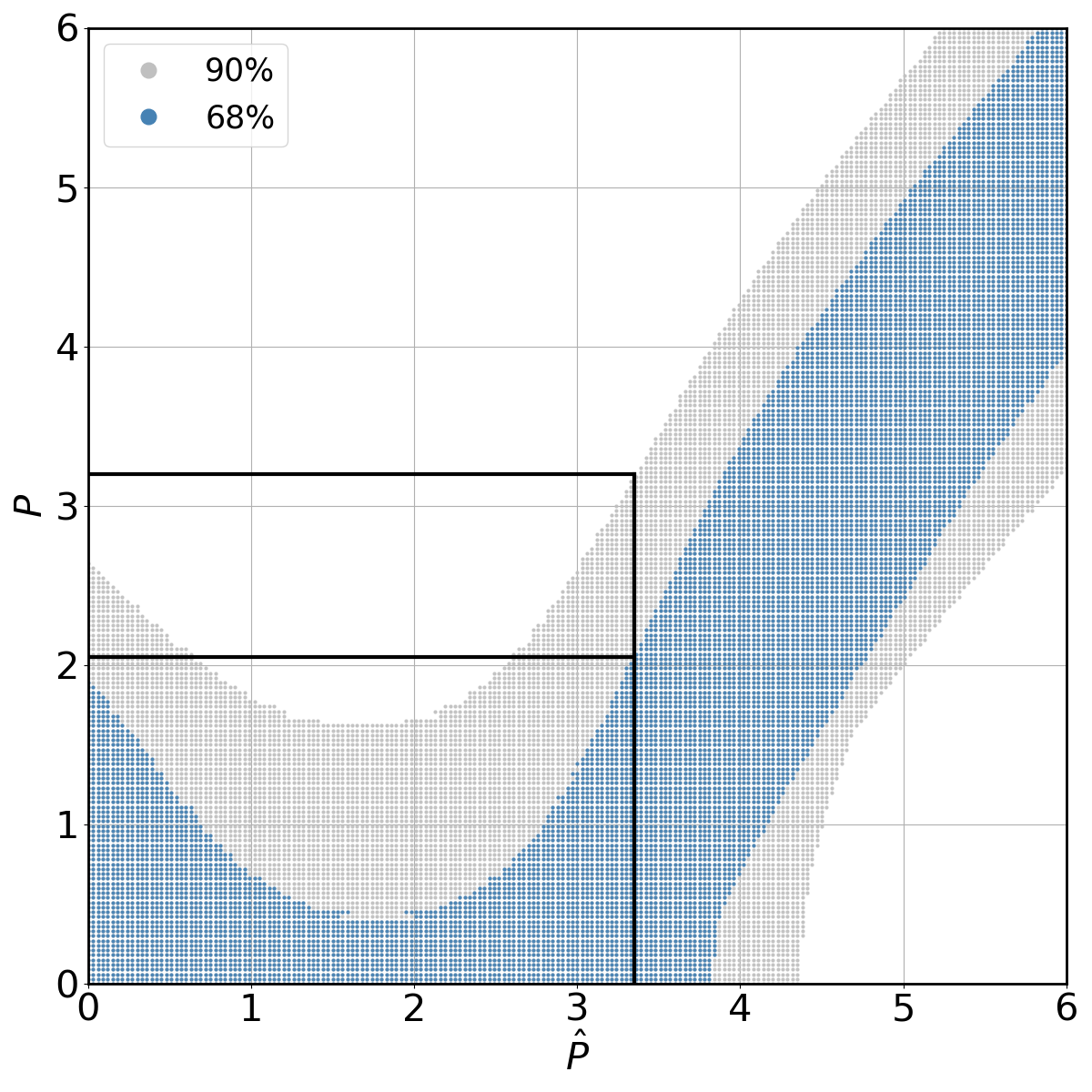}
    \caption{A 68\% (blue) and 90\% (gray) confidence interval for the multi-cycle analysis $(n=8)$. On the $x$-axis we have the estimated experimental value $\hat{P}$ and on $y$-axis is the true value $P$. For an experimental value of $\pest = 3.3$, the true value $P$ can be found between 0 and 3.15 with a confidence of 90\%.
    \label{fig:conf_8}}
\end{figure}

Figure~\ref{fig:Rice_8cyc} is the 2-dimensional plot for $n=8$ calculated using Eq.~(\ref{eq:gauss_pdf}). The construction of confidence interval follows the 1 cycle case. The confidence interval for $n=8$ is shown  in  Fig.~\ref{fig:conf_8}. The edges of the blue and gray bands  represent the 68\% and 90\% confidence levels, respectively.

\begin{figure}[!hbt]
    \includegraphics[width=\columnwidth]{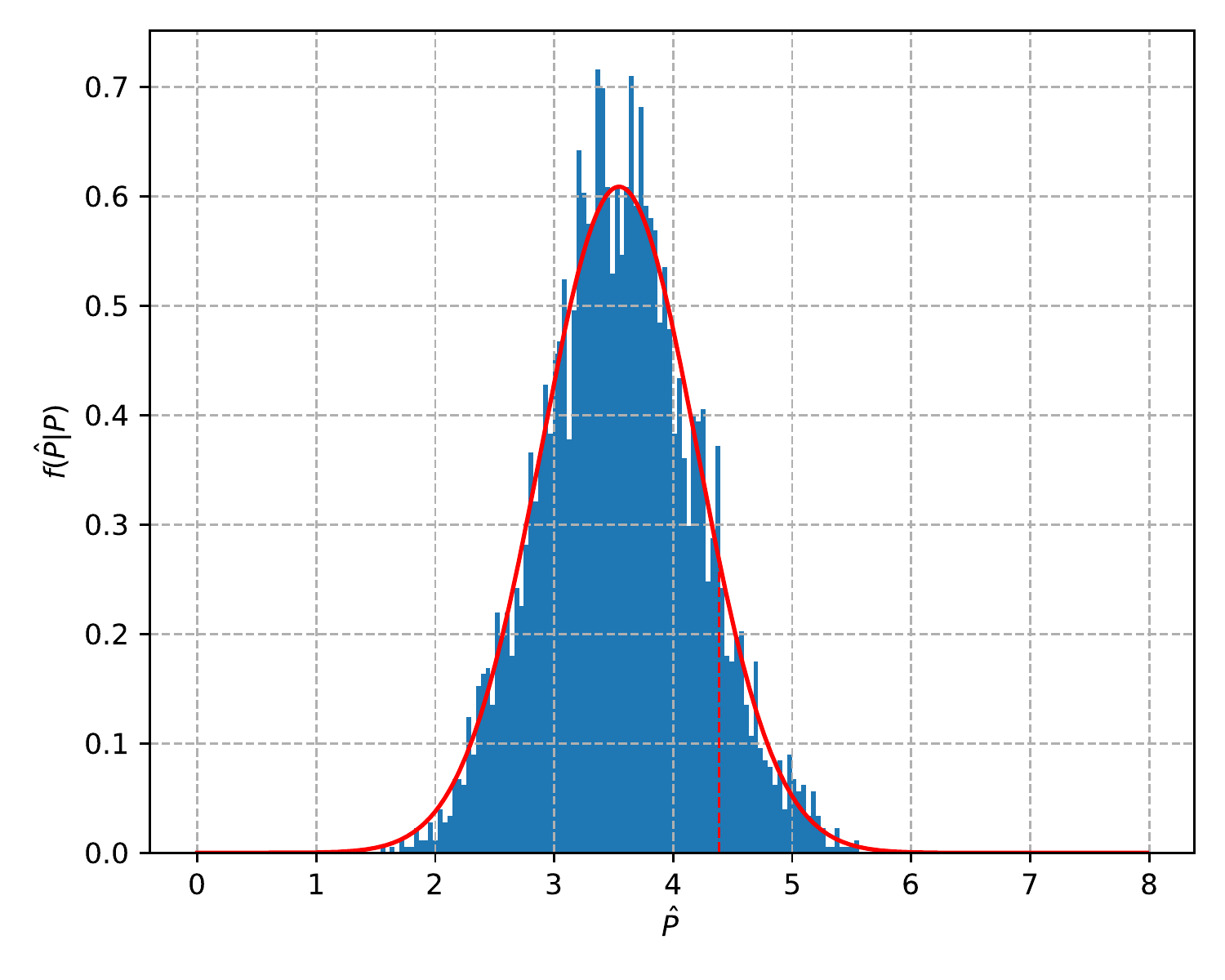}
    \caption{The blue histogram is the distribution of \pest, normalized such that the integral is one. The experimental data is in good agreement with the probability density function for $P=0$ (Eq.~(\ref{eq:gauss_pdf})) drawn in red. In both cases there are $n = 8$ cycles. The vertical red line at $\pest = 4.38$ corresponds to the lower limit of the 90\% confidence interval being greater than zero. This is the case for  9.63\% of the contributing data points.    \label{fig:dist_points}}
\end{figure}

Care must be taken if the observed \pest\ is less than the expected value \subit{\mu}{Rice}. These are considered to be from downward statistical fluctuations and $P$ is calculated at \subit{\mu}{Rice}\ as explained in \cite{Fe98}. For each experimentally obtained value of \pest\ the corresponding boundary values of $P$ are determined. This value is  multiplied by the experimental uncertainty $\sigma_n$ to give the true amplitude $A$.
In the frequency range or axion mass range
covered by the experiment, no signal was observed that could not be explained by a statistical fluctuation. Note that in setting a 90\% confidence interval, one expects that in 10\% of the cases a lower limit larger than zero even if no signal is present. This corresponds to our observation, as shown in Fig.~\ref{fig:dist_points}. From this we also conclude that at this level of precision there is no systematic effect resulting in a fake signal.

The conversion into   a limit of the oscillating EDM of the deuteron is done through the equation:
\begin{equation}
    |\dACd| = \lambda A \times 10^{-23}~\!\SI{}{\ecm},
    \label{eq:a_dac}
\end{equation}
where here and in the following we use the convention that the unit of electric charge $e$ is defined to be positive.
{The coefficient $\lambda = 316$ for the fast ramps and $286$ for the slow ramps, respectively}. $\lambda$ is based on a model of the polarization jump size for a particular ramp rate. The derivation of Eq.~(\ref{eq:a_dac}) is given in Appendix~\ref{App:B}, see Eq.~(\ref{eq:dAC-2}).

\section{\label{sec:Result}Result and Discussion}
\subsection{\label{sec:oEDM}Limits of ALPs signals}

According to references~\cite{Silenko:2021qgc,Kolya22} the
angular velocity
$\vec{\Omega}$ of the extended Thomas-BMT equation~(\ref{eq:tbmt})
of a beam  with particles of mass $m$, charge $q$, spin $S$, Lorentz factor $\gamma$ and velocity $\vec v = c \vec \beta$
acquires the following oscillating term
\begin{eqnarray}
 \vec{\Omega}_{a(t)}
 &=& -\frac {1}{S\hbar}\,\frac{\dAC}{a_0} \,a(t)\, c\vec \beta \times \vec B
 - \frac{1}{S\hbar}\,\frac{C_N}{2 f_a}\, \hbar \partial_0 a(t) \,\vec \beta
 \nonumber \\
 &=&\phantom{-} \dAC \frac{c \gamma m}{q \hbar S} \cos\big(\omega_a (t-t_0) + \phi_a(t_0) \big) \, \vec\beta \times \vec{\Omega}_{\mathrm{rev}} \nonumber\\
     & & \mbox{} + \frac{C_N}{2 f_a S}  \omega_a a_0
\sin\big(\omega_a (t-t_0) + \phi_a(t_0)\big) \, \vec \beta
\, , \label{eq:oma}
\end{eqnarray}
whenever
a classical ALP field, as in Eq.~(\ref{eq:a(t)}),
couples to the particles stored in the  beam, {{\it cf.} Eqs.~(\ref{eq:lag_aN}) to (\ref{eq:ham_aNN})}.
Note that the
magnetic field in the laboratory system can be
expressed as $\vec B=  (- m \gamma/q) \vec{\Omega}_{\mathrm{rev}}$
in terms of the angular revolution velocity of the particle beam,
$\vec{\Omega}_{\mathrm{rev}}$, as shown in Eq.~(\ref{eq:O_rev}).

According to Eq.~(\ref{eq:oma})
the spin rotation around the axis
$\vec\beta\times\vec {\Omega}_{\mathrm{rev}}$
(the latter always points radially outward, regardless of whether the beam is rotating clockwise or counterclockwise)
is generated
by  the AC part of the electric dipole moment
of the beam particle (see Eq.~(\ref{eq:oscd})) which in turn is induced by the ALP field, while the spin rotation with respect
to the longitudinal axis $\vec\beta$ of the beam, see  \cite{Silenko:2021qgc,Kolya22}, follows from the pseudomagnetic (axion-wind)  effect~\cite{Vorobev:1995pb,Gr13} of
strength $C_N/f_a$ in terms of the axion decay constant $f_a$~\cite{Work22}.
In the experiment we cannot distinguish these two rotation types around two orthogonal axes which both induce -- on resonance --  a polarization shift in the vertical direction but are $\pi/2$ out of phase with each other, so that the two rotation amplitudes add up
coherently.

Thus, to obtain an upper limit on $\dAC$
or $C_N/f_a$ one has to assume that the other term vanishes, such that the bound is
saturated by one term only.

\begin{figure}[!hbt]
    \includegraphics[width=\columnwidth]{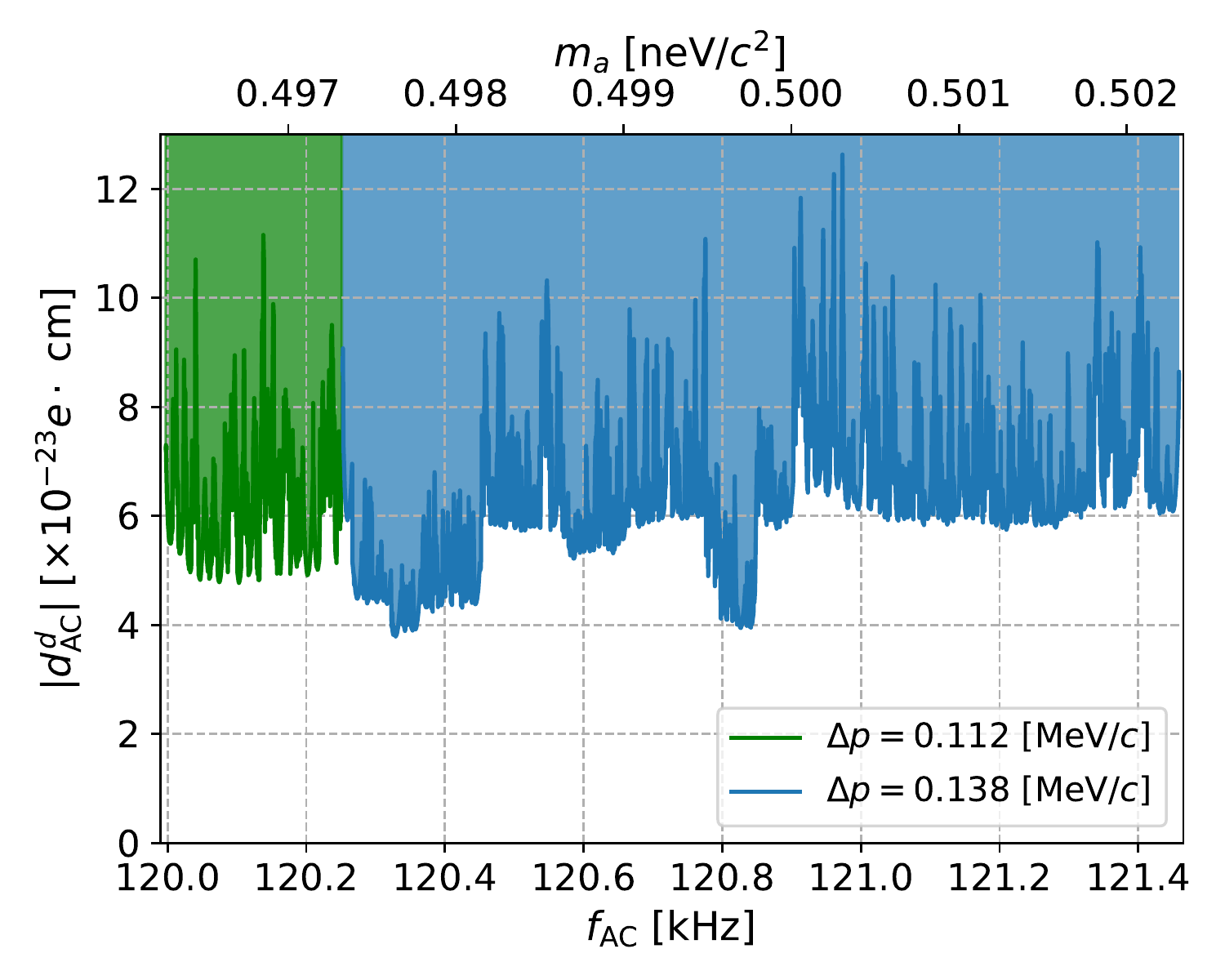}
    \caption {90\% confidence level sensitivity for excluding the ALPs induced oscillating EDM (\SI{}{\ecm}) in the frequency range \SIrange{120.0}{121.4}{\khz} ($m_a=$~\SIrange{0.495}{0.502}{\nevsc}). More explanation may be found in the text.
    \label{fig:Result}}
\end{figure}

First we assume that only the EDM-term is present, {\it i.e.}, $C_N/f_a =0$.
Figure~\ref{fig:Result} provides the 90\% confidence level sensitivity for excluding the ALPs induced oscillating EDM of the deuteron, $\dACd$, in the frequency range of \SIrange{120.0}{121.4}{\khz} and the corresponding axion mass range of \SIrange{0.495}{0.502}{\nevsc} represented on the upper axis.
The darker lines indicate the upper limit of the oscillating EDM and the lighter filled region above is the exclusion region. The green and blue colors differentiate the two different ramp rates mentioned in Section~\ref{sec:scanman}. The green indicates a momentum change $\Delta p = \SI{0.112}{\mevc}$ and the blue $\Delta p = \SI{0.138}{\mevc}$.
\begin{figure}[!hbt]
     \includegraphics[width=\columnwidth]{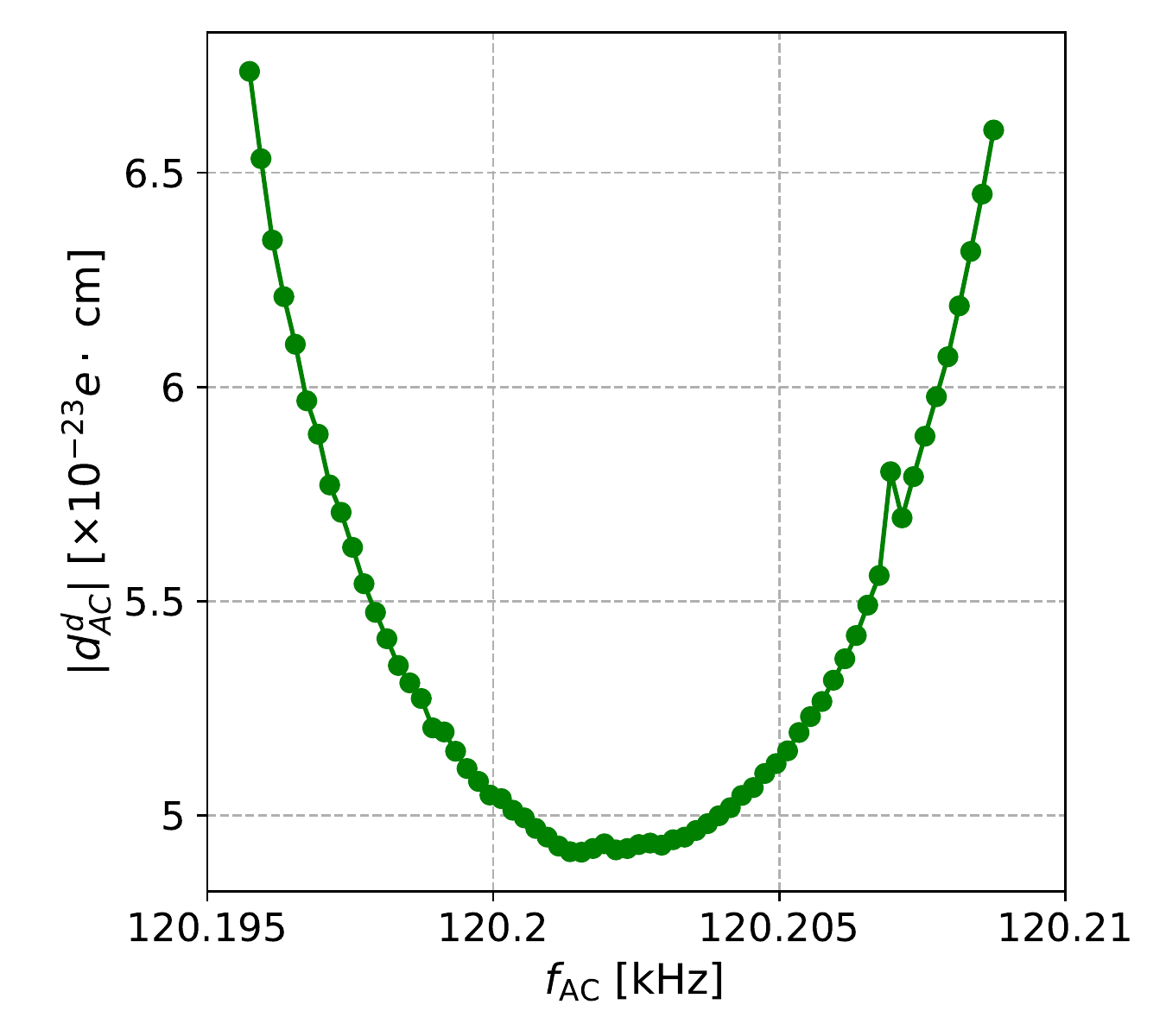}
    \caption {Sensitivity of a single scan including 8 cycles. The sensitivity decreases and the curve gets larger as one moves away from the center of the scan. There is no overlap between the scans in this example.
    \label{fig:Result_1cyc}}
\end{figure}

The fluctuations in the exclusion plot result mainly from two beam properties, intensity and polarization, as well as the clock time during the scan. Good beam properties mean better sensitivity.
The dependency of intensity is seen in a larger scale over multiple scans. If for a particular frequency range, a larger number of scans were performed, the obtained sensitivity is better. This can be observed in Fig.~\ref{fig:Result} around frequency \SI{120.8}{\khz}, for example.
The decline in sensitivity within a cycle is mainly due to beam depolarization.

A small contribution to these fluctuations arises from the way $\Delta {\Alr}$ is calculated in Eq.~(\ref{eq:stepfn}). As a consequence, the sensitivity becomes worse as one moves further from the middle of the scan because the imbalance in the number of points on both sides of the anticipated jump in the calculation of $\Delta {\Alr}$ leads to a larger uncertainty in the jump. An example of how this sensitivity appears for a single scan region comprising 8 cycles is shown in Fig.~\ref{fig:Result_1cyc}. However, the intensity of the beam and the polarization have the greatest influence and their combination is the reason for the higher values ($|\dACd| > \SI{8e-23}{\ecm}$).

This experiment to search for ALPs in the storage ring provides a 90\%~confidence level upper bound of
\begin{equation}
    |\dACd| <\SI{6.4e-23}{\ecm}. \label{our_bound}
\end{equation}
 This value is used to calculate ALP coupling constants in the next subsection and is based on the average of the individual limit points in Fig.~\ref{fig:Result}.

\subsection{\label{sec:couplings}Limits of Various ALP Couplings}

In this paper, we focus on the coupling of ALPs to the deuteron spin  via the oscillating
part of
the deuteron EDM $\dACd$
and/or via the axion wind effect proportional to $C_d/f_a$. For all these calculations, it is assumed that the local dark-matter density  $\rLDM = (0.55 \pm 0.17)\, \SI{}{\gevcc}$ (see, {\it e.g.}, Chapter 27 ``Dark Matter'' of Ref.~\cite{Work22}) contains only ALPs.

The bound on the amplitude of the  oscillating deuteron EDM \dACd,
{\it cf.} Eq.\,(\ref{our_bound}), can be interpreted as a bound on the  axion  coupling  to the deuteron EDM operator (in analogy to the axion coupling to the nucleon EDM operator, $g_{aN\gamma}$, of~\cite[Eq.\,90.38]{Work22}, {see also Eqs.\,(\ref{eq:lag_aN}) and (\ref{eq:ham_aN})}) in terms of the electromagnetic fine-structure constant $\alpha$:
\begin{equation}
  |\gadg| = \frac{|\dACd|}{a_0} \, \frac{\sqrt{4\pi\alpha}}{e\hbar c} <  \SI[per-mode=reciprocal]{1.7e-7}{\gesv}.
\label{gadg_val}
\end{equation}
Here we assume that $a_0 = \sqrt{2 \rLDM (\hbar c)^3}/(m_a c^2) = \SI{5.8}{MeV} $. {The occurrence of the axion/ALP amplitude $a_0$ in the denominator of
Eq.\,(\ref{gadg_val}) is typical when the calculation is based on axion/ALPs as
candidates for (local) dark matter particles.
The inverse proportionality of $a_0$ and $m_a$ implies that the exclusion limits derived from oscillating EDM measurements at similar experimental sensitivity have to be linearly increasing functions of the ALP mass $m_a$. }

\begin{figure}[tbph]
    \includegraphics[width=\columnwidth]{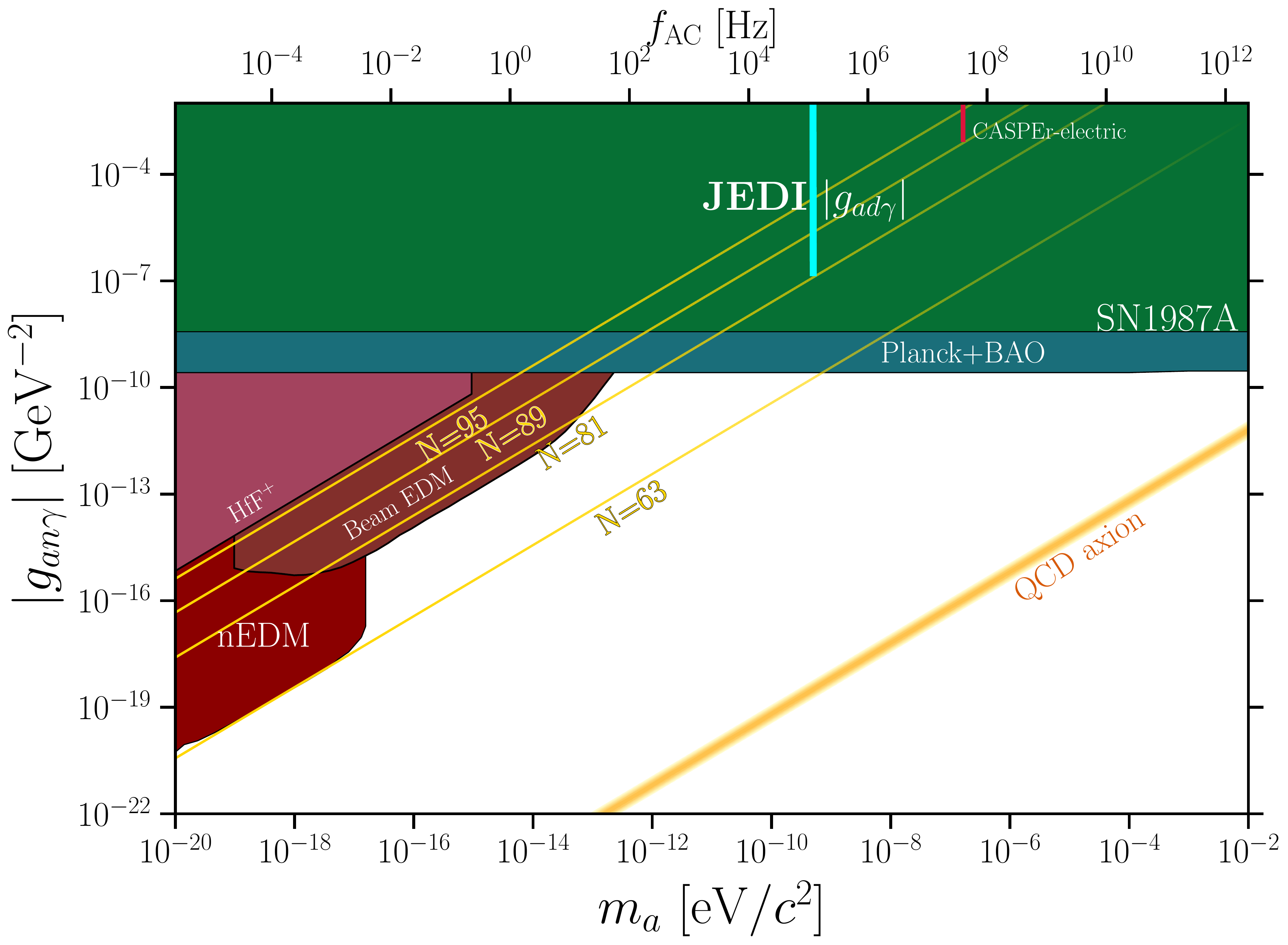}
    \caption {The 90\% upper bound on |\gadg| from this experiment (in cyan) is shown along with the bound on |\gaNg{n}| from experiments such as nEDM~\cite{Ab17}, CASPEr-electric~\cite{Ay21}, { HfF$^+$~\cite{Roussy:2020ily}, and Beam EDM~\cite{Schulthess:2022pbp}  (as presented in Ref.~\cite{AxLim}),} in different shades of red. Also, seen in green is the constraint, {calculated in \cite{Gr13}} from the SN1987A supernova energy loss, {which might be model-dependent~\cite{Bar:2019ifz}. In blue a  further constraint calculated in \cite{Caloni_2022} from the combined Planck 2018 and BAO (baryon acoustic oscillation) cosmological data is displayed {(as presented in Ref.~\cite{AxLim})}.}
    {Finally, the yellow lines {$\mathcal{N}=n$}, parallel to the  QCD axion band and plotted according to Eq.\,(\ref{eq:ZN-axion}) and Eq.\,(90.5) of \cite{Work22}, indicate that $Z_\mathcal{N}$ axions
    with {$\mathcal{N} > n$} are  excluded by the above-mentioned {\em measured} bounds {in their respective mass ranges}.}
    Figure courtesy \cite{AxLim, Work22}.
    \label{fig:AxionEDM}}
\end{figure}

    Figure~\ref{fig:AxionEDM} shows the limit on |\gadg| from this experiment in cyan along with bounds for $|g_{aN\gamma}|$ from the nEDM\,\cite{Ab17}, CASPEr-electric\,\cite{Ay21}, and Beam EDM\,\cite{Schulthess:2022pbp} experiments.
    In addition, the figure presents the $|g_{aN\gamma}|$ exclusion region as tabulated in~\cite{AxLim} via reformulating  the limits  of the electron-EDM HfF$^+$ experiment\,\cite[Fig.\,3]{Roussy:2020ily}.
    Furthermore,
    a constraint on $|g_{aN\gamma}|$  is shown that is derived in
    Ref.\,\cite{Gr13} from assuming
    $N +\gamma \to N +a$ cooling in SN1987A.
    Thus the latter bound is
    based on the strength of the coupling constant in
    the axion/ALP interaction with the nucleon EDM and is therefore independent of   $m_a$.
    Note, however, that Ref.~\cite{Bar:2019ifz} suggests
    an alternative collapse mechanism for supernovae SN1987A that would
    not place limits  on the emission of ALPs or axions.
    Moreover, {following  Ref.\,\cite{AxLim}, an exclusion region is shown that is based on}  a new constraint on the coupling of
    thermally-produced ALPs as calculated in Ref.\,\cite{Caloni_2022} from combined  data of
    cosmic microwave background spectra and baryon acoustic oscillations.
    {However, according to \cite{Caloni_2022}   these bounds were only derived for
    the mass range
    $10^{-4}\SI{}{eV/c^2} \lesssim m_a \lesssim \SI{100}{eV/c^2}$.}
   Our {{\em directly measured} upper bound (\ref{gadg_val})} at \(m_a = \SI{0.5}{\nevsc}\) falls within  {the model-dependent} constraint obtained from SN1987A, but is stronger than the CASPEr-electric result at \(m_a \approx \SI{100}{\nevsc}\).

    {Finally, in Figure~\ref{fig:AxionEDM},  $|g_{an\gamma}(m_a)|_\mathcal{N}$ lines of the $Z_\mathcal{N}$-axion model  are plotted which follow from the exclusion limits of the presented
    EDM-based experiments and which run parallel to the displayed QCD axion band.
    Each $\mathcal{N}$ must be an odd-valued positive integer as otherwise the pertinent model would not solve the strong CP problem, see \cite{Hook:2018jle,DiLuzio:2021pxd}, and one would be back at the ALP case.
    In detail, according to Eq.\,(\ref{eq:ZN-axion}) combined with
    Eq.\,(90.5) of \cite{Work22}, {\it i.e.},
    \begin{equation}
      g_{a n \gamma} = (3.7\pm 1.5) \times 10^{-3}\, {\rm GeV} \, \frac{1}{f_a(m_a)|_{\mathcal{N}} } \,,
        \label{eq:g_ang_vs_ma}
    \end{equation}
    $Z_{\mathcal{N}}$ axions with  {$\mathcal{N}>81$} are excluded by our bound~(\ref{gadg_val}) {at $m_a \simeq \SI{5e-10}{eV/c^2}$}.
    Even if the canonical QCD axion/ALPs scenario prevails, the straight lines (proportional to $m_a$) that can be derived from Eq.\,(\ref{eq:g_ang_vs_ma}) can still serve as  excellent guides for ``extrapolating''  exclusion limits to different ALP masses and therefore comparing  the experimental sensitivities of the underlying  measurements of hadron (neutron, proton, deuteron, etc.) electric dipole moments. For instance, the {line $\mathcal{N}=63$} shows that the  pertinent experimental
    sensitivity of the nEDM limits is approximately a factor $10^{3}$ better than in our experiment. This is
    compatible with a  factor $\sim 10^{-3}$ between
    the oscillating neutron EDM bounds
    between  $\SI{5.e-26}{\ecm}$ to $\SI{7.e-26}{\ecm}$  as shown
    in \cite[Fig.\,2]{Ab17}
    and our upper bound $|\dACd|< \SI{6.4e-23}{\ecm}$, see Eq.\,(\ref{our_bound}). The {line $\mathcal{N}=89$} indicates that the CASPEr-electric limit~\cite{Ay21} is roughly a factor 20 less sensitive than that of our experiment, while the sensitivity of the Beam EDM experiment~\cite{Schulthess:2022pbp}
    is  slightly better than ours.
    Finally the
    HfF$^+$ limits~\cite{Roussy:2020ily} exclude $Z_\mathcal{N}$ axions with {$\mathcal{N} >95$ in the specified mass range. The corresponding $\mathcal{N}=95$ line  implies
    that the sensitivity of  this {\em electron\/}-EDM based experiment
    to constrain oscillating {\em hadronic} EDMs is about $10^2$ times worse than
    in our case.}
}

The second coupling we considered is the ALP-gluon coupling $C_G/f_a$, generated from
the $\bar\Theta$ term  for the permanent EDM case, where  the use of $C_G/f_a$ instead of just $1/f_a$ takes into account that the ALP coupling strength might differ
from the axion one.
The coupling is given by \cite{Gr11,Pospelov:1999ha,Ab17}
\begin{eqnarray}
    \dAC^N(t)= S \cdot \ka\frac{e \hbar c}{2 m c^2}\cdot  \cgfa \cdot a_0  \cos \left(\omega_a (t-t_0) +\phi_a(t_0)\right) \nonumber \\
    \approx \SI{2.4e-16}{\ecm}\cdot  \frac{C_G}{f_a} \cdot a_0  \cos \left(\omega_a (t-t_0) +\phi_a(t_0)\right), \nonumber\\
\end{eqnarray}
where \(S\) and \(m\) are the spin and mass of the nucleon, respectively, and $\ka$ is the chiral suppression factor of the  $\bar\Theta$-term. Here the loop-enhanced value $\kappa_a \approx 0.046$ of \cite{Pospelov:1999ha,Work22,Ab17} was used.
Note that the numerical factor $\SI{2.4e-16}{\ecm}$ is the same for proton (or neutron)
and deuteron because the
ratio $S/m = (1/2)/m_p \approx 1/m_d$
is approximately the same for these  particle
species. Compared to the direct determination of $C_G/f_a$ in the case of the nucleon, however, corrections are expected in the deuteron scenario. From the isoscalar nature of the deuteron nucleus and the isovector nature of the
leading low-energy pion-loop contribution to {the nucleon EDM~\cite{Crewther:1979pi,Baluni:1978rf,Ottnad:2009jw}}, a severe cancellation between the contributions of its proton and neutron components is anticipated, see {\it e.g.}, Ref.~\cite{Work22}. Moreover, the small $D$-wave admixture of the deuteron wave function affects the weights of these individual nucleon components~\cite{Yamanaka:2015qfa,Bsaisou:2014zwa}.
Finally  $P$- and $T$-breaking meson-exchange terms contribute already
at leading tree-level order~\cite{Khriplovich:1999qr,Lebedev:2004va}. {The latter contributions to  the permanent EDM  of the deuteron, induced by the QCD-theta term or more generalized chromo-electric EDMs of quarks,} are shown to be of similar magnitude as the single nucleon ones, {see, {\it e.g.}, \cite{Yamanaka:2015qfa,Bsaisou:2014zwa,Khriplovich:1999qr,Lebedev:2004va,Liu:2004tq,Afnan:2010xd,deVries:2011re,Bsaisou:2012rg,Bsaisou:2014oka,Wirzba:2016saz}}.
The ALP-gluon coupling in the deuteron case is therefore denoted in the following by an upper index $d$, {\it i.e.}, $C_G^{d}$, to signal that this coefficient is likely to contain corrections of order one
relative to the coupling $C_G$ in the nucleon scenario.

So substituting \(S=1\) and $m_d$ for the deuteron, we get the bound on the coupling constant,
\begin{eqnarray}
    \left|\frac{C_G^d}{f_a} \right| &=& \left| \frac{\dACd}{\SI{2.4e-16}{\ecm} \times a_0} \right|  \nonumber \\
    &<& \SI[per-mode=reciprocal]{0.46e-4}{\gev}. \label{eq:CGfa}
\end{eqnarray}
{Note again the $a_0$ dependence in the denominator which
 implies a linear dependence of
the bound on ALP mass $m_a$ and is a signal that the calculation is based on axions/ALPs as dark-matter candidates.}

\begin{figure}[tbph]
    \includegraphics[width=\columnwidth]{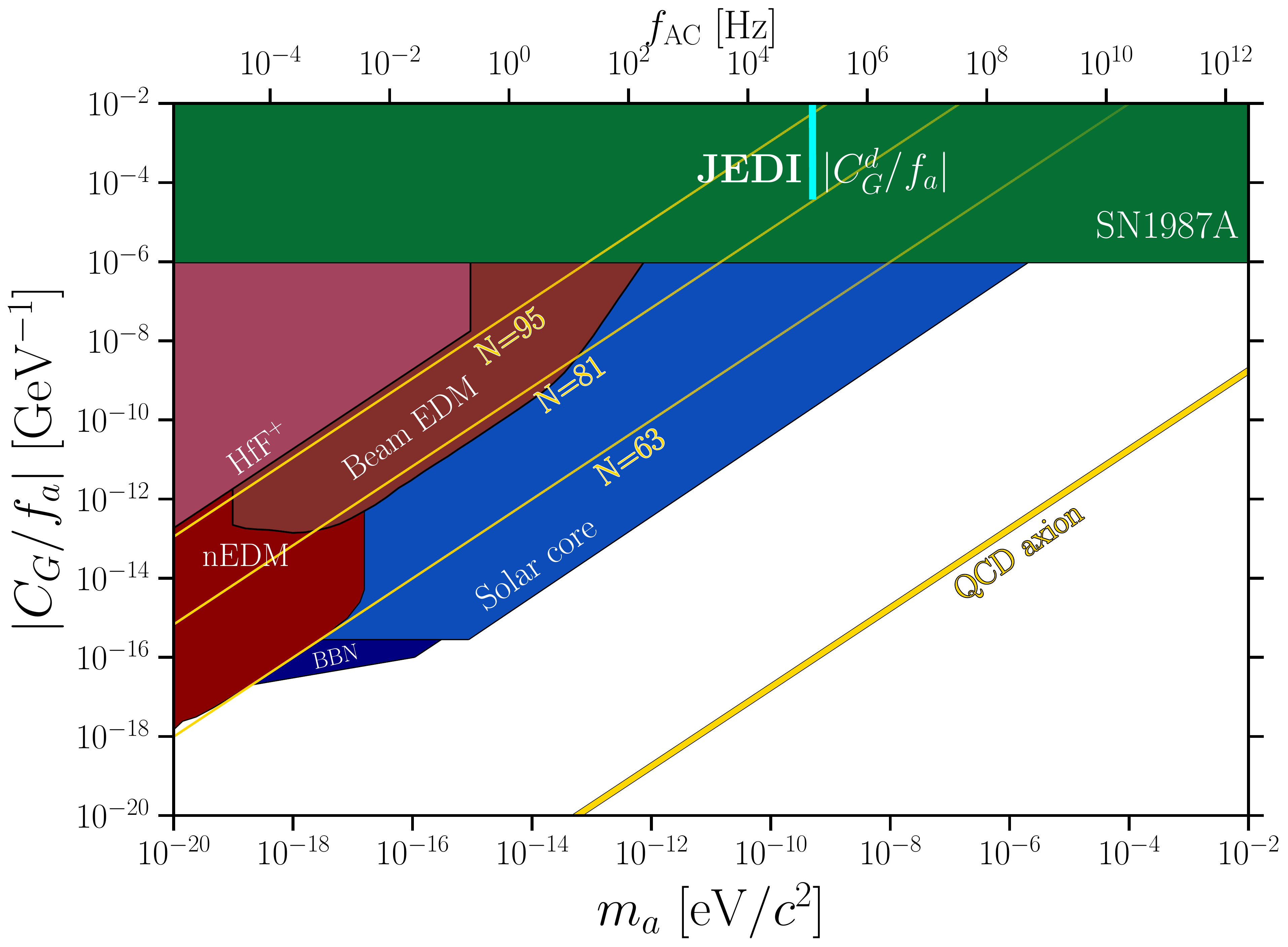}
    \caption {Figure showing the 90\% upper bound on $|{C_G^d}/{f_a} |$, in cyan, in comparison with the  $|C_G/f_a|$ nEDM~\cite{Ab17},  { HfF$^+$~\cite{Roussy:2020ily}, Beam EDM~\cite{Schulthess:2022pbp} results in various shades of red}. Also shown are the limits from supernova SN1987A,
    {as calculated in \cite{Gr13}} in green, solar core~\cite{Hook:2017psm,DiLuzio:2021pxd} in lighter blue and big bang nucleosynthesis~\cite{Blum:2014vsa} in darker blue. 
    {The three yellow lines {$\mathcal{Z}=n$}, which are parallel to the QCD axion band, have been calculated directly from Eq.\,(\ref{eq:ZN-axion}) and indicate that $Z_\mathcal{N}$ axions with {$\mathcal{N} > n$}   are excluded by the above-mentioned {\em measured} bounds {in their respective mass regions}.}
    Figure courtesy \cite{AxLim, Work22}.
    \label{fig:Axionfa}}
\end{figure}

Figure~\ref{fig:Axionfa} shows the upper bound on \(|\frac{C_G^d}{f_a} |\) in comparison with the results on $|\cgfa|$ from the nEDM experiment~\cite{Ab17}, the
{HfF$^+$ electron EDM\,\cite{Roussy:2020ily} and the Beam EDM\,\cite{Schulthess:2022pbp}} experiments
as well as  the limits obtained from astrophysical calculations such as Big Bang nucleosynthesis, solar core, and {supernova SN1987A~\cite{Gr13} -- the latter  based on the $N+\gamma\to N+a$ cooling mechanism}. Details can be found in \cite{Work22,AxLim}. Our result is within the limits obtained from the supernova emission.
{In addition, {three
    $Z_{\mathcal{N}}$ lines} are plotted, given directly by $1/f_a(m_a,\mathcal{N})$, as calculated in  Eq.\,(\ref{eq:ZN-axion}). They show that $Z_\mathcal{N}$ axions
    with {$\mathcal{N}> 81$}  and {$\mathcal{N}> 63$} are again excluded by the JEDI and
    nEDM~\cite{Ab17} experiments {in their respective mass regions}, while  the eEDM-based HfF$^+$ exclusion region, see Ref.\,\cite{Roussy:2020ily},
    vetoes $Z_\mathcal{N}$ axions
    with   {$N>95$ in the specified mass range.
    Note that the apparently better fitting line $\mathcal{N}=96$ can be excluded for another reason, since $Z_\mathcal{N}$ axions with even $\mathcal{N}$ do not solve the strong CP problem.
    }
    }

{Next we consider the axion wind case.}
By ignoring the EDM-term
in Eq.~(\ref{eq:oma})
we can provide a bound on the ALP (pseudo-magnetic) coupling
to the deuteron spin, $|C_d/f_a|$.
On resonance and using Eq.~(\ref{eq:oma}),  this limit
can simply be  expressed
in terms of the limit  on the oscillating EDM (\ref{our_bound}),
|$\dACd| < \SI{6.4e-23}{\ecm}$, as
\begin{eqnarray}
 \left |\cdfa\right| &=&
 \left| \frac{2  \gamma m_d c}{e \hbar \omega_a a_0} \right|
 \cdot \left| \vec{\Omega}_{\mathrm{rev}}\right| \cdot \left|{\dACd} \right|  \nonumber \\
&=&  \left| \frac{2   m_d c}{e \hbar G  a_0} \right|
 \cdot \left|\dACd\right|
 < 1.5 \cdot 10^{-5}\, \si{GeV}^{-1}  \ .
  \label{eq:Cdfa}
\end{eqnarray}
In the second line the ALP-resonance condition was applied, {\it i.e.}, $\omega_a =  \gamma | G \vec{\Omega}_{\mathrm{rev}}|$, where
$G$ is here the magnetic anomaly of the deuteron. {Thus, this limit also shows a linear dependence on the ALP mass $m_a$ indicating axions/ALPs as dark-matter candidates.}

\begin{figure}[tbph]
    \includegraphics[width=\columnwidth]{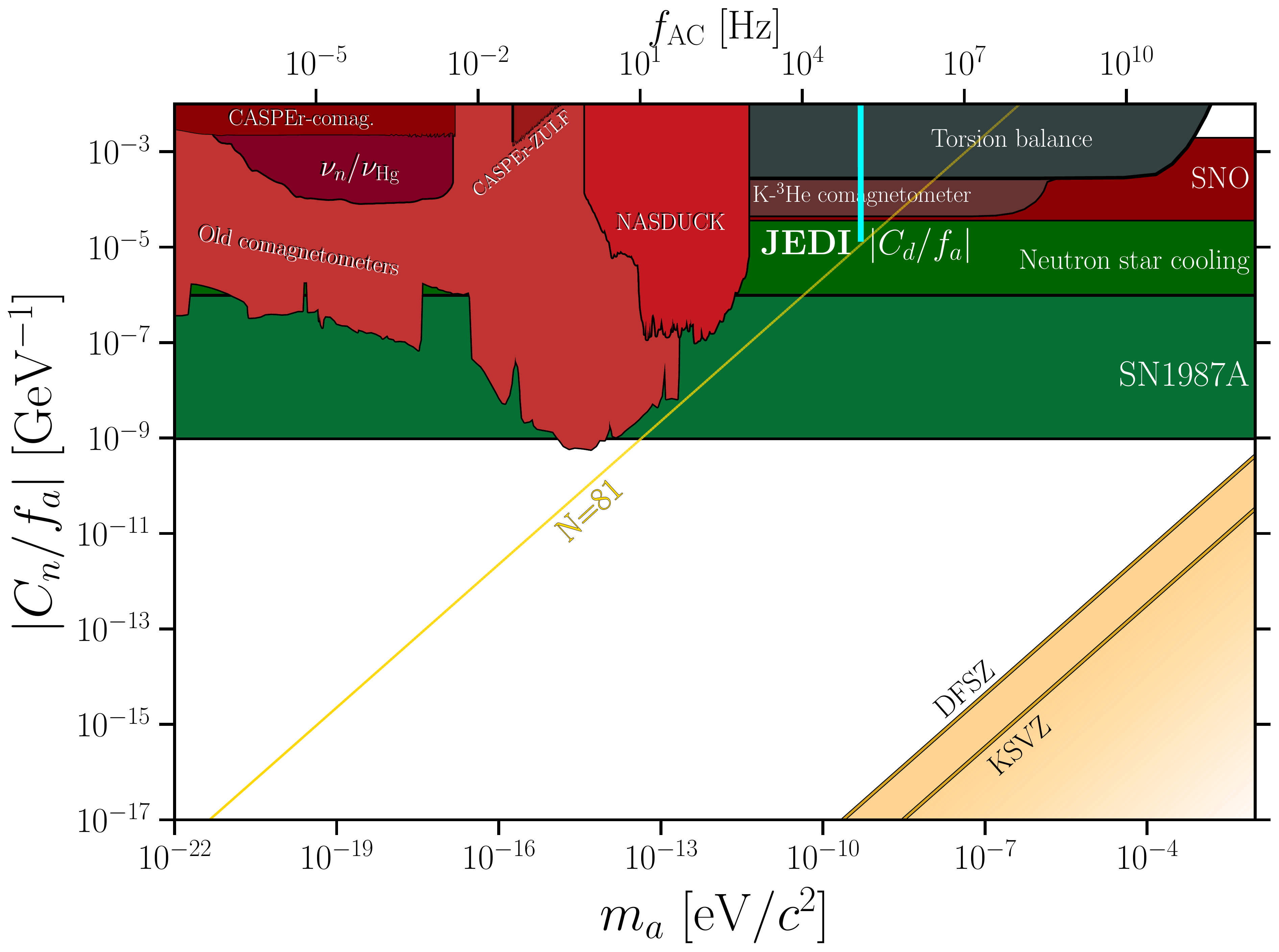}
    \caption {Figure displaying the ALP-neutron coupling  $|C_n/f_a|$ from various experimental results
    {
    (CASPEr-comag.~\cite{JacksonKimball:2017elr},
    $\nu_n/\nu_{Hg}$~\cite{Ab17},
    CASPEr-Zulf~\cite{Garcon:2019inh},
    Old comagnetometers~\cite{Bloch:2019lcy}, NASDUCK~\cite{Bloch:2021vnn},
    Torsion balance~\cite{Adelberger:2006dh},
    K$^3$He comagnetometer~\cite{Vasilakis2009},
    SNO~\cite{Bhusal:2020bvx}, Neutron star cooling~\cite{Buschmann:2021juv}, SN1987A~\cite{Carenza:2019pxu}).}
       The 90\% upper bound on the ALP-deuteron coupling,
    $|C_d/f_a|$ from the JEDI experiment, is shown in cyan,
    {and the corresponding $Z_\mathcal{N}$ axion line, labelled
    $\mathcal{N}=81$ and parallel to the DFSZ and KSVZ  axion lines, is displayed  in yellow}.
    {The limits from supernova SN1987A, as calculated in \cite{Carenza:2019pxu}
    (see also \cite{Raffelt:2006cw}), are presented in green.}
    Figure courtesy \cite{AxLim, Work22}.
        \label{fig:AxionNu}}
\end{figure}

The bound on the ALP-deuteron coupling  \(|\cdfa |\) is shown in Fig.~\ref{fig:AxionNu}. Other limits shown in this figure are bounds on ALP-neutron coupling.

{Moreover,
    constraints from supernova SN1987A on $|C_n/f_a|$  are shown in green that were calculated
    in Ref.\,\cite{Raffelt:2006cw} and recently updated  in Ref.\,\cite{Carenza:2019pxu}.
    Here the underlying cooling mechanism is assumed to be of
    bremsstrahlung type, {\it i.e.} $N N \to N N a$.
    The result is therefore
    based on the strength of the coupling constant in
    the axion/ALP  wind effect and  is  independent of   $m_a$.
    Remember, however, that Ref.~\cite{Bar:2019ifz} suggests
    an alternative collapse mechanism for supernova SN1987A that would
    not place limits  on the emission of ALPs or axions.}

{In addition, the {displayed $Z_\mathcal{N}=81$ line indicates that $Z_\mathcal{N}$ axions with $\mathcal{N}>81$ at $m_a \simeq \SI{5e-10}{eV/c^2}$ are excluded.  Here}
$1/f_a(m_a,\mathcal{N})$ of Eq.\,(\ref{eq:ZN-axion}) is now rescaled by a
factor of $\sim 1/3$ in order to follow the KSVZ axion line. The $\mathcal{N}=81$ line shows that the
underlying experimental sensitivity of the JEDI measurement is compatible with, if not better than, that of  the old comagnetometers~\cite{Bloch:2019lcy} and  NASDUCK~\cite{Bloch:2021vnn} experiments.}

This underlines the statement made earlier that the axion wind
effect in storage ring experiments is greatly enhanced relative to other
laboratory measurements
because it depends on the relative velocity
of the axions with respect to the particle under study (see Eq.~(\ref{eq:oma})).
In storage rings one has $v \approx c$, whereas for particles
at rest in the laboratory system~\cite{Ab17,Ay21},  the relative velocity is  given by the velocity of the Earth with respect to the center of our Galaxy, {\it i.e.}, $v \approx \SI{250}{km/s}\sim 10^{-3} c$. Since the latter contribution can be safely neglected in relativistic storage rings, the pertinent pseudomagnetic field of the axion wind always points tangentially to the beam trajectory.
Therefore, the direction of $\vec v$ is uniquely determined,
while in laboratory experiments it depends in a complicated way on a time-dependent superposition of a considerable number of non-negligible motions.

It should be noted that \cite{Ab17} has assumed \(\rLDM = \SI{0.4}{\gevcc}\) in contrast to  \(\rLDM = \SI{0.55}{\gevcc}\) assumed in this paper. Thus, quoted coupling constants in this paper are \(\approx 0.85\) times smaller compared to \cite{Ab17}.

\section{\label{sec:conc}Conclusions and Outlook}
This paper presents an experiment conducted to demonstrate a new method to search for ALPs using an in-plane polarized deuteron beam in a storage ring.
The polarization vector of the deuteron beam is influenced by ALPs due to two effects. First ALPs introduce an oscillating electric dipole moment (EDM) causing a spin rotation around a radial axis in the storage ring and second, the so-called axion wind  or pseudomagnetic effect resulting in a spin rotation around the longitudinal axis. Storage ring experiments are specifically sensitive to the second effect because it scales with the velocity of the particles with respect to the axion field which moreover always points tangentially to the beam, {\it i.e.}, in the same direction in the comoving (rest) frame of the beam particle.

The experiment did not see any ALP signal within the achieved sensitivity.
An upper limit on the deuteron oscillating EDM is quoted for the first time.
In the mass range $m_a=$~\SIrange{0.495}{0.502}{\nevsc}
oscillating EDM values above $\sim10^{-22}\,\SI{}{\ecm}$  are excluded by this experiment at least at a 90\% level, assuming a direct EDM coupling.  Constraints on other axion/ALP coupling strengths, like the ALP coupling to the EDM operator of the deuteron, $g_{ad\gamma}$, the ALP-gluon coupling of the deuteron, $C_G^d/f_a$ and the ALP (pseudo-magnetic) coupling to the deuteron spin $C_d/f_a$ were estimated as well.

{As a proof of principle for ALP searches in storage rings, this} experiment was just an exploratory study where the
actual data taking period for the axion search was only four days. In future experiments with extended beam times
and higher beam intensities, the sensitivity
can be increased by at least an order of magnitude. Systematic
effects are not expected to play an important role since
one is looking for an AC effect at a particular frequency.


In the future, new type of storage rings to measure the
permanent EDM of charged hadrons are planned \cite{Ab19}.
This would allow the search for axions essentially in the whole mass range
displayed in Figures~\ref{fig:AxionEDM}-\ref{fig:AxionNu}.
For this kind of storage rings it is proposed to use a combination
of {radial} electric and  {vertical} magnetic bending fields. In this case the spin
precession frequency $\vec{\Omega}_{\mathrm{MDM}}
-\vec{\Omega}_{\mathrm{rev}}$ is given by
\begin{eqnarray}
 \lefteqn{ \vec{\Omega}_{\mathrm{MDM}} - \vec{\Omega}_{\mathrm{rev}} =} \nonumber\\
 && -\frac{q}{m} ~ \left[G \vec{B}
 - \left(G-\frac{1}{\gamma^2-1} \right) \frac{\vec{\beta} \times \vec{E}}{c}\right] \, .
\end{eqnarray}
By using appropriate combinations of the electric and magnetic field, {the amplitude of}
$\vec{\Omega}_{\mathrm{MDM}}-\vec{\Omega}_{\mathrm{rev}}$ can be varied from 0 to the values corresponding
to {$\omega_a = m_a c^2/\hbar\simeq 10^{-9}\,\SI{}{eV/\hbar}$} as described in ~\cite{Pretz:2019ham}.
{In that case,
the relation of the EDM angular velocity ({\it cf.} Eq.\,(\ref{eq:O_EDM})) becomes
\begin{equation}
    \vec{\Omega}_{\mathrm{EDM}} = -\frac{1}{S\hbar} d(t) \left(\vec E+ c\vec \beta \times \vec B \right) \,,
    \end{equation}
but still agrees with the second line of Eq.\,(\ref{eq:oma}) when  expressed as a function of the
angular velocity of beam revolution.
Also the relation
of the axion-wind angular velocity remains unchanged, see Eq.\,(\ref{eq:O_wind})
and the third line of Eq.\,(\ref{eq:oma}).}

{Recently, various ideas have been discussed in the literature to extend  the here presented and established storage ring searches for axions/ALPs: namely, by
applying, {\it e.g.},  {\em static} Wien filters or
modulating radio-frequency {\em cavities}, see~\cite{Silenko:2021qgc}, or
by using radio-frequency Wien filters operating at the {\em sidebands} of the axion frequency $\omega_a$
and $\Omega_{\mathrm{MDM}}$ as discussed in~\cite{Kim:2021pld}.}

This kind of experiments can be further explored
at facilities like RHIC, NICA or GSI/FAIR where polarized hadrons beams are either available, planned or could be added to the physics program.
Using different particles (protons, deuterons, nuclei and even leptons) would allow to study the influence of spin
and isospin on various couplings. 
Moreover, it offers the possibility to perform measurements with different $G$-factors.

\begin{acknowledgements}
We would like to thank the COSY crew
for their support in setting up the COSY accelerator for the experiment. The work presented here has been performed in the framework of the JEDI collaboration and was supported by an ERC Advanced Grant of the European Union (proposal number 694340: \textit{Search for electric dipole moments using storage rings}), the Shota Rustaveli National Science Foundation of the Republic of Georgia (SRNSFG Grant No. JFZ\_18\_01), and partially { by IBS-R017-D1 of the Republic of Korea.}
This research is part of a project that has received funding from the European Union’s Horizon 2020 research and innovation programme under grant agreement STRONG-2020 - No 824093.
The work of N.~Nikolaev on the topic was supported by the Russian Science Foundation (Grant No. 22-42-04419).
{The presented results are based on the PhD projects of S.~P.~Chang and S.~Karanth.}
\end{acknowledgements}


\appendix

\section{Calculation of the Relative In-Plane Polarization Directions Using Four Bunches}
\label{App:A}
The signal of an ALP in a storage ring requires that the oscillation of the EDM be in phase with the rotation of the deuteron polarization in the ring plane. Specifically, the maximum value of the EDM must occur when the polarization is oriented perpendicular to the direction of the electric field in the particle frame of reference. During the search, the phase of the ALP is unknown. In order to make the effect visible for any phase, we chose to operate the COSY RF on the fourth harmonic $(h = 4)$ of the revolution frequency and store four beam bunches. This appendix will demonstrate that this choice provides beams with different phases between the oscillating EDM and the direction of the rotating beam polarization. Since the wavelength of the axion field is much larger than the physical size of the COSY ring, this allows the ALP signal to be observed regardless of the ALP phase.

The beam is loaded into COSY with the polarization oriented in the vertical direction. Rotation of the polarization into the ring plane is accomplished by operating an RF solenoid for a brief period of time. If the solenoid RF operates at the same frequency as the in-plane rotation of the polarization, then the small rotation induced by the solenoid will accumulate. Continuous running of the solenoid produces an oscillation of the vertical polarization component. If the solenoid is stopped when the polarization reaches the in-plane orientation, then the beam is prepared for the experiment.

This result may be calculated using a simple series of classical rotations, each associated with one turn of the beam around COSY. For this a comoving coordinate system is used with the $z$-axis pointing in momentum direction, the $y$-axis upwards parallel to the magnetic field, and, consequently, the $x$-axis from the center of the ring outwards as the beam is rotating clockwise. In the model, the polarization is described by a vector, $[p_x,\, p_y,\, p_z]$ with the initial polarization $[0,1,0]$. One turn around the ring is described by two rotations, one ($\theta$) for the precession in the ring magnets and another ($\alpha$) for the precession in the RF solenoid, as shown in Eq.~(\ref{eq:polprec}):

\begin{flalign}
&\!\label{eq:polprec}
  \begin{bmatrix}
    p'_x\\
    p'_y\\
    p'_z\\
  \end{bmatrix}
\!=\!
  \begin{bmatrix}
    \cos\alpha & -\sin\alpha& 0\\
    \sin\alpha & \cos\alpha & 0\\
    0 & 0 & 1\\
  \end{bmatrix}\!\!
  \begin{bmatrix}
    \cos\theta & 0 & \sin\theta\\
    0 & 1 & 0\\
   -\sin\theta & 0 & \cos\theta\\
  \end{bmatrix}\!\!
  \begin{bmatrix}
    p_x\\
    p_y\\
    p_z \\
  \end{bmatrix}.
&
\end{flalign}

The primed spin vector is the result of one revolution of the beam around the ring. The rotations may be treated separately since the length of the RF solenoid is very short compared with the circumference of the ring. For the purposes of a computer-based calculation, the precession of the spins about the $y$-axis in the ring magnets per turn is given by $\theta=-2\pi G \gamma$, where $G=-0.1429875424$ is the deuteron magnetic anomaly and $\gamma=1.1259762$ is the relativistic factor at the initial beam energy. This rotation is the same for every turn. The RF solenoid operates on a harmonic of the revolution frequency and with an adjustable strength $4\pi\subit{\epsilon}{sol}$, such that $\subit{\epsilon}{sol} = \subit{f}{sol}/\subit{f}{rev}$ with $\subit{f}{sol}$ being the frequency of the resulting driven spin oscillations. Thus $\alpha=4\pi\subit{\epsilon}{sol}\cos[2\pi n(1+G\gamma)+\phi]$ where $n$ is the turn count (or the number of times the two rotations have been applied) and $\phi$ is a phase that will be described later. For each turn of the beam, the operation shown in Eq.~(\ref{eq:polprec}) is repeated based on the result of the previous series of rotations. The solenoid rotation $\alpha$ is cumulative, adding another $4\pi\subit{\epsilon}{sol}\cos[2\pi n(1+G\gamma)]$ to the previous value on each turn.

 A program was written to complete the numerical sum of all rotations. In the model, $2\times10^6$ turns were used, and a value of $4\pi\subit{\epsilon}{sol}=1.5708 \times10^{-6}$ brought the vertical polarization very close to zero.

To simulate what happens for each of the four beam bunches, we need to repeat the calculation described above, but with an initial phase added to the RF solenoid angle $\alpha$ to describe the delay in the phase for each bunch. For the first bunch, denoted as B0, $\phi=0$. For the three subsequent bunches, the starting phase is $U(1+G\gamma)$ where $U=\pi/2,\ \pi ,\ \textrm{and}\ 3\pi/2$ for bunches B1, B2, and B3 respectively.

The orientation of the polarization at the end of this process can be described using the $x$ and $z$ coordinates as follows:

\begin{table}[!hbt]
\caption{Model calculation of bunch spin directions as measured at a fixed point in the ring, \it{e.g.}, at the
polarimeter.\label{tab:bunch1}}
\begin{ruledtabular}
\begin{tabular}{ccccc}
 Bunch&$x$&$z$&Angle [rad]& Angle B(n-1)-B(n)\\
\hline
B0 & -0.639562 & 0.768740 & -0.693928 & \\
B1 & -0.904313 & -0.426870 & -2.011825 & 1.317897\\
B2 & 0.187022 & -0.982356 & -3.328722 & 1.317897\\
B3 & 0.997903 & -0.064724 & -4.647619 & 1.317897\\
B0 (again) &  &  &  & 2.329493\\
\end{tabular}
\end{ruledtabular}
\end{table}

\begin{figure}[tbph]
\includegraphics[width=\columnwidth]{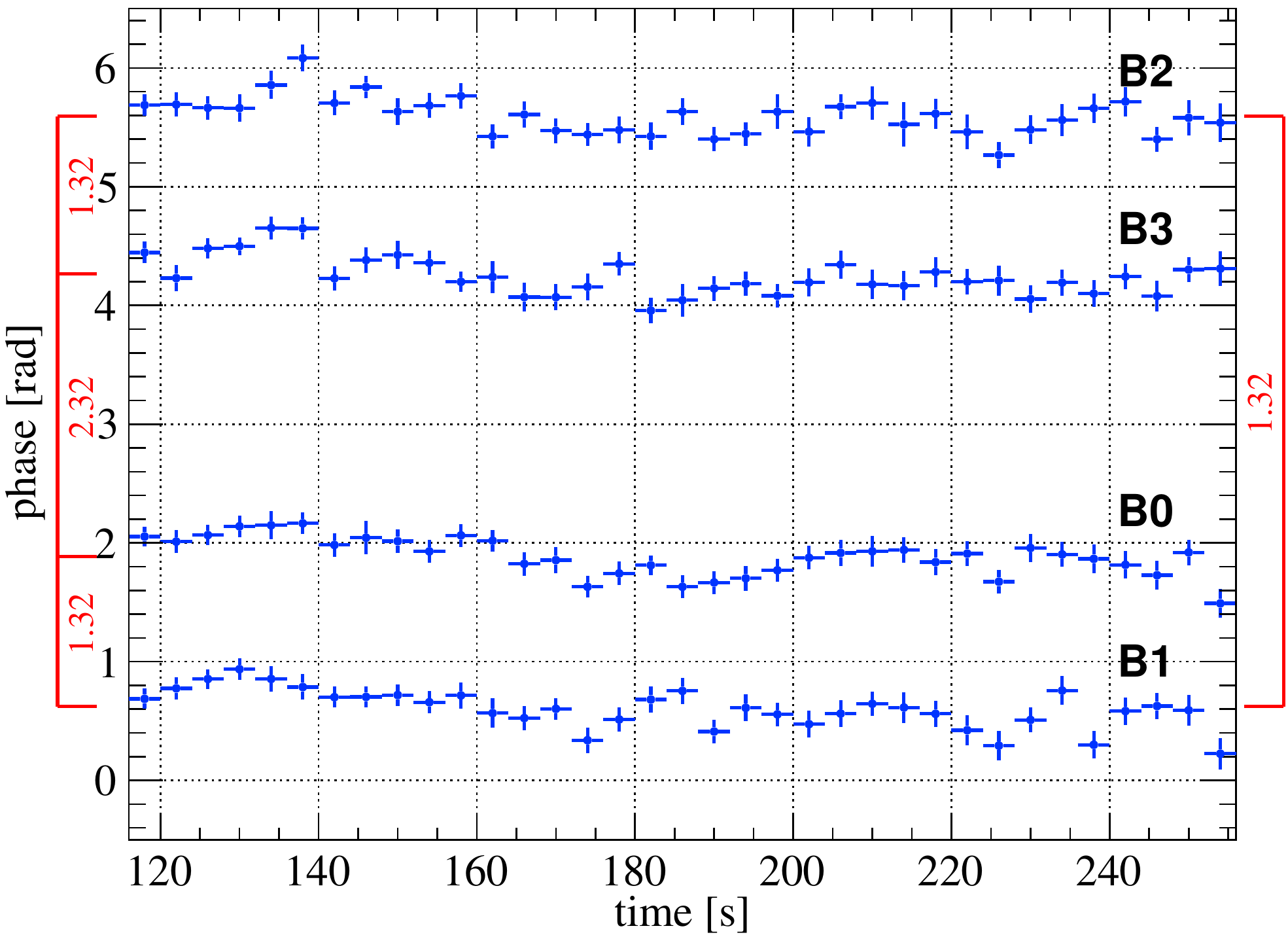}
\caption{\label{fig:bunch1} Measurements of the phase of the in-plane polarization of a four-bunch beam as a function of the time in the store after the RF solenoid is turned off. The four bunches are B0, B1, B2, and B3. The differences in phase angle are indicated by the red diagram that includes the relative bunch angles in radians. A fixed value of the spin tune $G\gamma$ is assumed during the analysis in order to freeze any phase drift with time.}
\end{figure}

The angle starts at the $z$ axis. The first four columns of Table~\ref{tab:bunch1} show the results at the end of $2\times 10^6$ turns. The last column shows the differences in the polarization directions between adjacent bunches, as predicted by the rotation model. The phase angles in the next to last column apply at the time that the bunch lands in the ring plane, which is different for each bunch. There is also a polarimeter in the COSY ring that is capable of measuring the phase at the beginning of each 4-second time interval. It is worth noting that the spacing between the bunches is not equal across the break from B3 to B0. Thus, we should be able to tell from the relative phases which bunch is the first. Sample results are given in Fig.~\ref{fig:bunch1}.

The match of the phase differences in Fig.~\ref{fig:bunch1} with the predictions in Table~\ref{tab:bunch1} (column 5) shows that the process of using an RF solenoid to rotate the spins into the horizontal plane matches the model.

Using the model results as a starting point, we can extrapolate forward in the calculation to the same point in time for each bunch. If we choose the moment when the rotation of B3 to the horizontal plane is complete, then the in-plane rotation of bunches B0, B1, and B2 move forward by $3\pi/2$, $\pi$, and $\pi/2$, respectively. This produces a final orientation of the polarization given by Table~\ref{tab:bunch2}.
\begin{table}[!hbtp]
\caption{Model calculation of bunch spin directions as measured at a fixed point in time, \textit{i.e}, when the bunch B3 is completely rotated into the horizontal plane.\label{tab:bunch2}}
\begin{ruledtabular}
\begin{tabular}{ccc}
 Bunch&$x$&$z$\\
\hline
B0 & 0.064724 & 0.997903 \\
B1 & -0.997903 & 0.064724 \\
B2 & -0.064724 & -0.997903 \\
B3 (no change)& 0.997903 & -0.064724\\
\end{tabular}
\end{ruledtabular}
\end{table}

An inspection of the $x$ and $z$ columns shows that these four polarization directions form right angles to each other in the beam coordinate system, thus mapping out a space in all directions. The rotation of the ALP EDM is generated by the presence of an electric field in the rest frame of the deuterons as they pass through the ring magnets and are subject to a vertical lab magnetic field. This induces a force on the deuterons, $\vec{F}=e\, c\vec{\beta}\times\vec{B}$, that bends them into the closed orbit around the ring. The resulting electric field in the comoving frame also creates a torque on the oscillating EDM to the extent that the latter is perpendicular to the field at the time that the EDM is at an extreme point in its oscillation. Note that the electric field points toward the center of the ring. With this assortment of polarization directions, all phases (represented by sine and cosine functions) will generate a measurable change in the vertical polarization and no ALP field will go undetected due to phase mismatch. In addition, the presence of polarizations of opposite sign ensures that any non-zero offset in the polarimeter that measures the size of the resonant jump will be offset by a jump on the opposite bunch that is of equal and opposite sign.

\begin{figure}[tbph]
\includegraphics[width=\columnwidth]{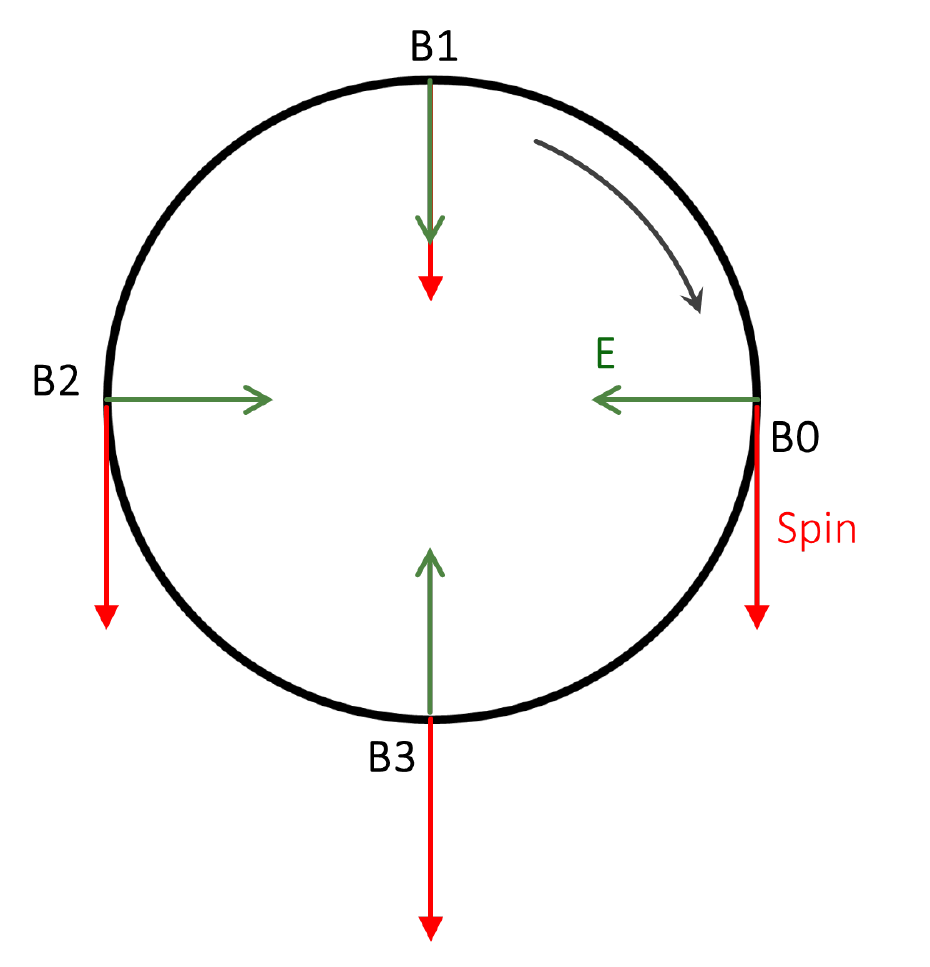}
\caption{\label{fig:bunch2} Diagram showing the orientations of the polarization relative to the electric field for the four bunches circulating in the storage ring, as given in Table~\ref{tab:bunch2}. The black arrow shows the direction of the clockwise rotating beam while the rotation of the spins of the deuterons in the comoving frame and the order of the bunches on the ring (B$i$, $i=0,1,2,3$) are counterclockwise (all viewed from above).}
\end{figure}

The orientations of the polarization relative to the electric field are illustrated by the diagram in Fig.~\ref{fig:bunch2}.
It then becomes clear that the polarization directions for each of the bunches relative to the local electric field is perpendicular to the bunch preceeding it. All the polarizations are either parallel to or perpendicular to their respective electric fields. This figure assumes a circular ring
without straight sections and  a clockwise rotating beam viewed from above.

These calculations may be repeated for the case of the $1 - G\gamma$ harmonic. Here, the resulting phase gaps have a different pattern, which was also confirmed experimentally using the phase measurements.
The modeling shows that a good polarization distribution among the four bunches is possible using either harmonic for the RF solenoid. There is a sort of symmetry between the two possibilities.
The sets of phases as measured by a fixed polarimeter looking at the bunches one at a time are distinctive and allow one to pick out the “first” bunch in each group from its location next to the single gap that is different from all the rest. This result is impervious to a number of potential issues, including whether or not the RF solenoid switching time is gradual (as is the case experimentally) or instantaneous (as it is in the model).

\section{Calculation of the Sensitivity Calibration}
\label{App:B}
\begin{figure}[htbp]
\includegraphics[width=\columnwidth]{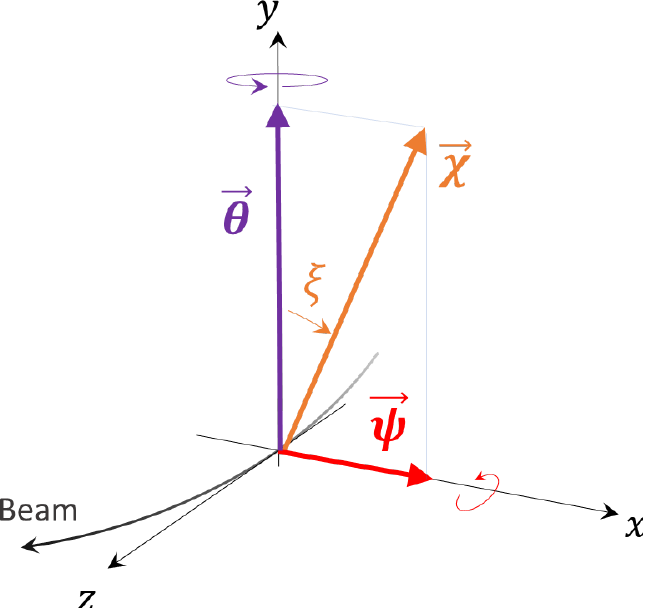}
\caption{Diagram showing the orientation of the rotation vectors associated with an EDM precession in the presence of bending in a storage ring. Coordinates and the bent particle path are shown. The total rotation $\vec\chi$ is the vector sum of $\vec\theta$ and $\vec\psi$. The angle between $\vec\theta$ and $\vec\chi$, denoted by $\xi$, is the angle of the coordinate system rotation (see text). The size of $\vec\psi$ in this figure is exaggerated to make it visible to the reader. \label{fig:combinedRotation}}
\end{figure}

Previous calculations of the response of the COSY storage ring have been made using a “no-lattice” model \cite{Be12,Be13} of successive rotations without a breakdown for each element of the ring. While the rotations in the bending magnets are continuous in this model, other devices such as the RF solenoid (see Appendix~\ref{App:A}) and the Wien filter are relatively short and may be treated as having zero length. The spin rotation per turn due to the bending magnet is about the vertical axis and given by the rotation vector $\vec\theta=-2\pi G\gamma\vec e_y$ $(=\theta\vec e_y)$. When we include the EDM, this introduces another continuous effect with a rotation about the radial (pointing outward from the center of the bend assuming $d>0$ in Eq.~(\ref{eq:O_EDM})) axis given by $\vec\psi=2\pi\vec\Omega_\mathrm{EDM}/|\vec\Omega_\mathrm{rev}|$ $(=\psi\vec e_x)$.
As a result, one gets a new, combined rotation about a new axis $\vec\chi$:
\begin{equation}
    \vec \chi = \vec\theta + \vec\psi .
\end{equation}
The situation is depicted in
Fig.~\ref{fig:combinedRotation}.

To calculate the result within the no-lattice model, we chose to tilt the reference frame about the $z$-axis so that the new $y$-axis lies along the total rotation vector $\vec\chi$. The angle of tilt becomes
\begin{equation}
    \xi = \arctan \frac{\psi}{\theta}
\end{equation}
One turn through the storage ring is represented by
\begin{equation}
\!\label{eq:combinedRotation}
  \begin{bmatrix}
    p'_x\\
    p'_y\\
    p'_z\\
  \end{bmatrix}
=
  \begin{bmatrix}
    \cos\xi & -\sin\xi& 0\\
    \sin\xi & \cos\xi & 0\\
    0 & 0 & 1\\
  \end{bmatrix}
  \begin{bmatrix}
    \cos\chi & 0 & \sin\chi\\
    0 & 1 & 0\\
    -\sin\chi & 0 & \cos\chi\\
  \end{bmatrix}\nonumber
  \end{equation}
  \begin{equation}
  \times
  \begin{bmatrix}
    \cos\xi & \sin\xi& 0\\
    -\sin\xi & \cos\xi & 0\\
    0 & 0 & 1\\
  \end{bmatrix}
  \begin{bmatrix}
    p_x\\
    p_y\\
    p_z \\
  \end{bmatrix} \\
\end{equation}
where the vector $[p_x, p_y, p_z]$ represents the initial projection of the polarization along the axes shown in Fig.~\ref{fig:combinedRotation}, and $[p'_x,p'_y,p'_z]$ represents the resulting polarization. The first and third square matrices handle the transformation of the coordinate system while the main rotation is described by the middle matrix.

In the simulation used to calibrate the response of the system to an axion, the revolution frequency of the deuteron was ramped. The ramp was centered at the nominal beam frequency of $\frev=$\SI{750602.6}{\hertz}, with a \SI{100}{Hz} scanning range. The ramp rate for \frev\ in the calculation was \SI{1}
{\hertz \per \second}. As the ramp was followed, the small changes to the relativistic factor $\gamma$ and the elapsed time of a single turn 1/\frev, were followed as discussed in the main text.

In Eq.~(\ref{eq:oscd}) the oscillating part of the EDM is described by
\begin{equation}
 \subit{d}{osc}(t)=\dAC\cos\big(\omega_a( t-t_0) + \phi_a(t_0)\big).\label{eq:oscd2}
\end{equation}

This can be expressed in terms of the EDM rotation angle $\subit{\psi}{osc}(t)$ as
\begin{equation}
 \psi_\mathrm{osc}(t)=\subit{\psi}{AC}\cos\big(\omega_a( t-t_0) + \phi_a(t_0)\big).\label{eq:oscedm2}
\end{equation}
In this equation, $t_0$ was assumed to be the time at the start of the scan. Thus, time accumulated with an ever-decreasing time step for each turn as the scan slowly ramped up the revolution frequency. This caused the EDM oscillation, initially out of step with the polarization rotation, to fall in step and then out of step as the scan proceeded. Depending on the exact conditions at the beginning, the individual accumulation of the vertical polarization $y'$ as the resonance is crossed may be any value between its positive and negative limits. Thus, the size of the calculated jump will vary similarly. In order to know the maximum jump possible, the calculation must be run with two orthogonal phases such as $\phi_a=0$ and $\pi/2$. Then the sizes of the jumps are added in quadrature to obtain the final value of the jump size.

The numerical simulation used in the calibration of the sensitivity began with a particular size of the oscillating EDM and scanning rate to calculate the expected polarization jump. For jumps that are much less than one (assuming complete polarization), the relationship between the EDM size and the jump is nearly linear. This allows us to use just the slope given by the calibration. One example of such a calculation begins with an EDM rotation of
$\subit{\psi}{AC}=\num{8e-9}$~rad/turn
and a scanning rate of 1~Hz/s. After calculating the jump for two orthogonal choices of the axion phase, the results were combined and gave a jump of $\Delta p_y=0.0066$, which is normalized to a beam polarization of one.  Tests with the calibration program demonstrate that the jump scales with the reciprocal of the square root of the ramping rate in the linear region. The ratio of EDM rotation to total polarization jump must be scaled by $w=\sqrt{\text{ramp(actual)}/\text{ramp(calib.)}}=0.775$ for the faster scans and 0.700 for the slower scans.

 The value of ramp(calib.) is 1.00 and the values of ramp(actual) are found in column 3 of Table~\ref{tab:ramps} where the rates are 0.600~Hz/s for the fast scans and 0.490~Hz/s for the slow scans.  The ratio or slope between $\psi_\mathrm{AC}$ and $\Delta p_y$ then becomes $\num{9.35e-7}$~rad/turn for the fast scans and $\num{8.48e-7}$~rad/turn for the slow scans.

Using the first (radially pointing) term of Eq.~(\ref{eq:oma})
we can describe the amplitude of the contribution to the determination of the oscillating EDM
in terms
of $\psi_\mathrm{AC} = 2\pi \Omega_a/|\vec \Omega_\mathrm{rev}|$ where $\Omega_a$ is the amplitude of
 the oscillating angular velocity $\vec \Omega_{a(t)}$, by
\begin{equation}
\label{eq:dAC-1}d_\mathrm{AC} =\frac{1}{2\pi}\frac{S\hbar q} {\beta\gamma m_d c}\,
\frac{w}{0.958}\,
\psi_\mathrm{AC}\,,
\end{equation}
where the spin $S$ is equal to one. The factor $w$ corrects for the ramp rate and 0.958 corrects for the alternating straight and curves sections in the COSY ring (see end of Section~\ref{sec:axionphase}). For ease of connecting with the parameters of the COSY ring, the charge $q$ and the denominator of the first fraction may be swapped for the  beam momentum expressed as $(B\rho )$. Then, in the usual EDM units of \SI{}{\ecm} and expressed in terms of the above
quoted slope between $\psi_\mathrm{AC}$ and the jump $\Delta p_y$
we have
\begin{equation}
\label{eq:dAC-2}
 |d_\mathrm{AC} |
     =
 \frac{1}{2\pi}
 \frac{\hbar}{B\rho }\,
 \frac{w}{0.958}\,
 \left| \frac{\psi_\mathrm{AC}}{\Delta p_y} \right|_\mathrm{calib.} A
\,,
 \end{equation}
where $A$ is the true value of the upper limit on the magnitude of the jump. The  second fraction in this expression has the value of $\hbar /B\rho =\num{3.26e-35}$~J$\cdot$s$\cdot($T$\cdot$m$)^{-1}  = \SI{2.03e-14}{\ecm}$
while the rest of the expression is dimensionless.
In this way, Eq.~(\ref{eq:a_dac}) was derived -- including the values $316$ and $286$, respectively, of the coefficient $\lambda$.
For typical values of the true $A$, values for $d_\mathrm{AC}$ usually lie below $10^{-22}\,\SI{}{\ecm}$.

The oscillating EDM has a period that is comparable in size to the revolution frequency. We made the approximation that the size of the EDM could be represented at any moment by its average value during a time interval that was chosen to be a fraction of a turn as the beam circulated in the storage ring.
The 3-matrix formula shown above in Eq.~(\ref{eq:combinedRotation})  was repeated $N$ times during each turn. For the calculations reported here, we chose $N = 15$ for which the calculations had converged to a precision of $0.1\%$.

The calculations were repeated for spin rotation with respect to the longitudinal axis of the beam arising from axion-wind effect and the calibration matched the rotation along radial axis as explained in this appendix.

\bibliographystyle{paper}

\bibliography{mainAxion}

\providecommand{\noopsort}[1]{}\providecommand{\singleletter}[1]{#1}%
\begin{thebibliography}{100}
\newcommand{\enquote}[1]{``#1''}
\providecommand{\url}[1]{\texttt{#1}}
\providecommand{\urlprefix}{URL }
\expandafter\ifx\csname urlstyle\endcsname\relax
  \providecommand{\doi}[1]{doi:\discretionary{}{}{}#1}\else
  \providecommand{\doi}{doi:\discretionary{}{}{}\begingroup
  \urlstyle{rm}\Url}\fi
\providecommand{\bibAnnoteFile}[1]{%
  \IfFileExists{#1}{\begin{quotation}\noindent\textsc{Key:} #1\\
  \textsc{Annotation:}\ \input{#1}\end{quotation}}{}}
\providecommand{\bibAnnote}[2]{%
  \begin{quotation}\noindent\textsc{Key:} #1\\
  \textsc{Annotation:}\ #2\end{quotation}}
\providecommand{\eprint}[2][]{{#2}}

\bibitem{Peccei:1977hh}
R.~D. Peccei and H.~R. Quinn.
\newblock \enquote{{CP Conservation in the Presence of Instantons}}.
\newblock \emph{Phys. Rev. Lett.}, \textbf{38}, 1440 (1977).
\newblock \doi{10.1103/PhysRevLett.38.1440}.
\bibAnnoteFile{Peccei:1977hh}

\bibitem{Peccei:1977ur}
R.~D. Peccei and H.~R. Quinn.
\newblock \enquote{{Constraints Imposed by CP Conservation in the Presence of
  Instantons}}.
\newblock \emph{Phys. Rev. D}, \textbf{16}, 1791 (1977).
\newblock \doi{10.1103/PhysRevD.16.1791}.
\bibAnnoteFile{Peccei:1977ur}

\bibitem{Weinberg:1977ma}
S.~Weinberg.
\newblock \enquote{{A New Light Boson?}}
\newblock \emph{Phys. Rev. Lett.}, \textbf{40}, 223 (1978).
\newblock \doi{10.1103/PhysRevLett.40.223}.
\bibAnnoteFile{Weinberg:1977ma}

\bibitem{Wilczek:1977pj}
F.~Wilczek.
\newblock \enquote{{Problem of Strong $P$ and $T$ Invariance in the Presence of
  Instantons}}.
\newblock \emph{Phys. Rev. Lett.}, \textbf{40}, 279 (1978).
\newblock \doi{10.1103/PhysRevLett.40.279}.
\bibAnnoteFile{Wilczek:1977pj}

\bibitem{Peccei:2006as}
R.~D. Peccei.
\newblock \enquote{{The Strong CP problem and axions}}.
\newblock \emph{Lect. Notes Phys.}, \textbf{741}, 3 (2008).
\newblock \doi{10.1007/978-3-540-73518-2_1}.
\newblock \eprint{hep-ph/0607268}.
\bibAnnoteFile{Peccei:2006as}

\bibitem{Asano:1981nh}
Y.~Asano \emph{et~al.}
\newblock \enquote{{Search for a Rare Decay Mode K+ ---\ensuremath{>} pi+
  Neutrino anti-neutrino and Axion}}.
\newblock \emph{Phys. Lett. B}, \textbf{107}, 159 (1981).
\newblock \doi{10.1016/0370-2693(81)91172-2}.
\bibAnnoteFile{Asano:1981nh}

\bibitem{Kim:1979if}
J.~E. Kim.
\newblock \enquote{{Weak Interaction Singlet and Strong CP Invariance}}.
\newblock \emph{Phys. Rev. Lett.}, \textbf{43}, 103 (1979).
\newblock \doi{10.1103/PhysRevLett.43.103}.
\bibAnnoteFile{Kim:1979if}

\bibitem{Shifman:1979if}
M.~A. Shifman, A.~I. Vainshtein, and V.~I. Zakharov.
\newblock \enquote{{Can Confinement Ensure Natural CP Invariance of Strong
  Interactions?}}
\newblock \emph{Nucl. Phys. B}, \textbf{166}, 493 (1980).
\newblock \doi{10.1016/0550-3213(80)90209-6}.
\bibAnnoteFile{Shifman:1979if}

\bibitem{Dine:1981rt}
M.~Dine, W.~Fischler, and M.~Srednicki.
\newblock \enquote{{A Simple Solution to the Strong CP Problem with a Harmless
  Axion}}.
\newblock \emph{Phys. Lett. B}, \textbf{104}, 199 (1981).
\newblock \doi{10.1016/0370-2693(81)90590-6}.
\bibAnnoteFile{Dine:1981rt}

\bibitem{Zhitnitsky:1980tq}
A.~R. Zhitnitsky.
\newblock \enquote{{On Possible Suppression of the Axion Hadron Interactions.
  (In Russian)}}.
\newblock \emph{Sov. J. Nucl. Phys.}, \textbf{31}, 260 (1980).
\bibAnnoteFile{Zhitnitsky:1980tq}

\bibitem{Gorghetto:2018ocs}
M.~Gorghetto and G.~Villadoro.
\newblock \enquote{{Topological Susceptibility and QCD Axion Mass: QED and NNLO
  corrections}}.
\newblock \emph{JHEP}, \textbf{2019(03)}, 033 (2019).
\newblock \doi{10.1007/JHEP03(2019)033}.
\newblock \eprint{1812.01008}.
\bibAnnoteFile{Gorghetto:2018ocs}

\bibitem{Work22}
R.~L. Workman and Others.
\newblock \enquote{{Review of Particle Physics}}.
\newblock \emph{PTEP}, \textbf{2022}, 083C01 (2022).
\newblock \doi{10.1093/ptep/ptac097}.
\bibAnnoteFile{Work22}

\bibitem{Hook:2018jle}
A.~Hook.
\newblock \enquote{{Solving the Hierarchy Problem Discretely}}.
\newblock \emph{Phys. Rev. Lett.}, \textbf{120(26)}, 261802 (2018).
\newblock \doi{10.1103/PhysRevLett.120.261802}.
\newblock \eprint{1802.10093}.
\bibAnnoteFile{Hook:2018jle}

\bibitem{DiLuzio:2021pxd}
L.~Di~Luzio \emph{et~al.}
\newblock \enquote{{An even lighter QCD axion}}.
\newblock \emph{JHEP}, \textbf{2021(05)}, 184 (2021).
\newblock \doi{10.1007/JHEP05(2021)184}.
\newblock \eprint{2102.00012}.
\bibAnnoteFile{DiLuzio:2021pxd}

\bibitem{DiLuzio:2021gos}
L.~Di~Luzio \emph{et~al.}
\newblock \enquote{{Dark matter from an even lighter QCD axion: trapped
  misalignment}}.
\newblock \emph{JCAP}, \textbf{2021(10)}, 001 (2021).
\newblock \doi{10.1088/1475-7516/2021/10/001}.
\newblock \eprint{2102.01082}.
\bibAnnoteFile{DiLuzio:2021gos}

\bibitem{Sikivie:2020zpn}
P.~Sikivie.
\newblock \enquote{{Invisible Axion Search Methods}}.
\newblock \emph{Rev. Mod. Phys.}, \textbf{93(1)}, 015004 (2021).
\newblock \doi{10.1103/RevModPhys.93.015004}.
\newblock \eprint{2003.02206}.
\bibAnnoteFile{Sikivie:2020zpn}

\bibitem{Gr11}
P.~W. Graham and S.~Rajendran.
\newblock \enquote{Axion dark matter detection with cold molecules}.
\newblock \emph{Phys. Rev. D}, \textbf{84}, 055013 (2011).
\newblock \doi{10.1103/PhysRevD.84.055013}.
\bibAnnoteFile{Gr11}

\bibitem{Gr13}
P.~W. Graham and S.~Rajendran.
\newblock \enquote{New observables for direct detection of axion dark matter}.
\newblock \emph{Phys. Rev. D}, \textbf{88}, 035023 (2013).
\newblock \doi{10.1103/PhysRevD.88.035023}.
\bibAnnoteFile{Gr13}

\bibitem{Stadnik:2013raa}
Y.~V. Stadnik and V.~V. Flambaum.
\newblock \enquote{{Axion-induced effects in atoms, molecules, and nuclei:
  Parity nonconservation, anapole moments, electric dipole moments, and
  spin-gravity and spin-axion momentum couplings}}.
\newblock \emph{Phys. Rev. D}, \textbf{89(4)}, 043522 (2014).
\newblock \doi{10.1103/PhysRevD.89.043522}.
\newblock \eprint{1312.6667}.
\bibAnnoteFile{Stadnik:2013raa}

\bibitem{Budker:2013hfa}
D.~Budker \emph{et~al.}
\newblock \enquote{{Proposal for a Cosmic Axion Spin Precession Experiment
  (CASPEr)}}.
\newblock \emph{Phys. Rev. X}, \textbf{4(2)}, 021030 (2014).
\newblock \doi{10.1103/PhysRevX.4.021030}.
\newblock \eprint{1306.6089}.
\bibAnnoteFile{Budker:2013hfa}

\bibitem{Sikivie:1983ip}
P.~Sikivie.
\newblock \enquote{{Experimental Tests of the Invisible Axion}}.
\newblock \emph{Phys. Rev. Lett.}, \textbf{51}, 1415 (1983).
\newblock \doi{10.1103/PhysRevLett.51.1415}.
\newblock [Erratum: Phys.Rev.Lett. 52, 695 (1984)].
\bibAnnoteFile{Sikivie:1983ip}

\bibitem{Sikivie:1985yu}
P.~Sikivie.
\newblock \enquote{{Detection Rates for 'Invisible' Axion Searches}}.
\newblock \emph{Phys. Rev. D}, \textbf{32}, 2988 (1985).
\newblock \doi{10.1103/PhysRevD.36.974}.
\newblock [Erratum: Phys.Rev.D 36, 974 (1987)].
\bibAnnoteFile{Sikivie:1985yu}

\bibitem{Gr15}
P.~W. Graham \emph{et~al.}
\newblock \enquote{{Experimental Searches for the Axion and Axion-Like
  Particles}}.
\newblock \emph{Ann. Rev. Nucl. Part. Sci.}, \textbf{65}, 485 (2015).
\newblock \doi{10.1146/annurev-nucl-102014-022120}.
\newblock \eprint{1602.00039}.
\bibAnnoteFile{Gr15}

\bibitem{Bernreuther:1991mn}
W.~Bernreuther.
\newblock \enquote{{The Electric dipole moment of the muon}}.
\newblock \emph{Z. Phys. C}, \textbf{56}, S97 (1992).
\newblock \doi{10.1007/BF02426781}.
\bibAnnoteFile{Bernreuther:1991mn}

\bibitem{Ab17}
C.~Abel \emph{et~al.}
\newblock \enquote{Search for axionlike dark matter through nuclear spin
  precession in electric and magnetic fields}.
\newblock \emph{Physical Review X}, \textbf{7}, 041034 (2017).
\newblock ISSN 2160-3308.
\newblock \doi{10.1103/PhysRevX.7.041034}.
\bibAnnoteFile{Ab17}

\bibitem{Roussy:2020ily}
T.~S. Roussy \emph{et~al.}
\newblock \enquote{{Experimental Constraint on Axionlike Particles over Seven
  Orders of Magnitude in Mass}}.
\newblock \emph{Phys. Rev. Lett.}, \textbf{126}, 171301 (2021).
\newblock \doi{10.1103/PhysRevLett.126.171301}.
\newblock \eprint{2006.15787}.
\bibAnnoteFile{Roussy:2020ily}

\bibitem{Ch18}
S.~P. Chang \emph{et~al.}
\newblock \enquote{{Axion dark matter search using the storage ring EDM
  method}}.
\newblock \emph{PoS}, \textbf{PSTP2017}, 036 (2018).
\newblock \doi{10.22323/1.324.0036}.
\bibAnnoteFile{Ch18}

\bibitem{Ch19}
S.~P. Chang \emph{et~al.}
\newblock \enquote{{Axionlike dark matter search using the storage ring EDM
  method}}.
\newblock \emph{Phys. Rev. D}, \textbf{99}, 083002 (2019).
\newblock \doi{10.1103/PhysRevD.99.083002}.
\bibAnnoteFile{Ch19}

\bibitem{Ab19}
F.~Abusaif \emph{et~al.}
\newblock \emph{Storage ring to search for electric dipole moments of charged
  particles : feasibility study}.
\newblock CERN Yellow Reports. Monographs, ISSN 2519-8068, eISSN 2519-8076; 3.
  CERN, Geneva (2021).
\newblock ISBN 978-92-9083-607-0.
\newblock \doi{10.23731/CYRM-2021-003}.
\bibAnnoteFile{Ab19}

\bibitem{Kim:2021pld}
O.~Kim and Y.~K. Semertzidis.
\newblock \enquote{{New method of probing an oscillating {EDM} induced by
  axionlike dark matter using an {RF} {Wien} filter in storage rings}}.
\newblock \emph{Phys. Rev. D}, \textbf{104(9)}, 096006 (2021).
\newblock \doi{10.1103/PhysRevD.104.096006}.
\newblock \eprint{2105.06655}.
\bibAnnoteFile{Kim:2021pld}

\bibitem{Krauss:1985ub}
L.~Krauss \emph{et~al.}
\newblock \enquote{{Calculations for Cosmic Axion Detection}}.
\newblock \emph{Phys. Rev. Lett.}, \textbf{55}, 1797 (1985).
\newblock \doi{10.1103/PhysRevLett.55.1797}.
\bibAnnoteFile{Krauss:1985ub}

\bibitem{Georgi:1986df}
H.~Georgi, D.~B. Kaplan, and L.~Randall.
\newblock \enquote{{Manifesting the Invisible Axion at Low-energies}}.
\newblock \emph{Phys. Lett. B}, \textbf{169}, 73 (1986).
\newblock \doi{10.1016/0370-2693(86)90688-X}.
\bibAnnoteFile{Georgi:1986df}

\bibitem{Raffelt:1987yt}
G.~Raffelt and D.~Seckel.
\newblock \enquote{{Bounds on Exotic Particle Interactions from SN 1987a}}.
\newblock \emph{Phys. Rev. Lett.}, \textbf{60}, 1793 (1988).
\newblock \doi{10.1103/PhysRevLett.60.1793}.
\bibAnnoteFile{Raffelt:1987yt}

\bibitem{Choi:1988xt}
K.~Choi, K.~Kang, and J.~E. Kim.
\newblock \enquote{{Invisible Axion Emissions From {SN1987A}}}.
\newblock \emph{Phys. Rev. Lett.}, \textbf{62}, 849 (1989).
\newblock \doi{10.1103/PhysRevLett.62.849}.
\bibAnnoteFile{Choi:1988xt}

\bibitem{Carena:1988kr}
M.~Carena and R.~D. Peccei.
\newblock \enquote{{The Effective Lagrangian for Axion Emission From
  {SN1987A}}}.
\newblock \emph{Phys. Rev. D}, \textbf{40}, 652 (1989).
\newblock \doi{10.1103/PhysRevD.40.652}.
\bibAnnoteFile{Carena:1988kr}

\bibitem{Barbieri:1985cp}
R.~Barbieri \emph{et~al.}
\newblock \enquote{{Axion to magnon conversion: a scheme for the detection of
  galactic axions}}.
\newblock \emph{Phys. Lett. B}, \textbf{226}, 357 (1989).
\newblock \doi{10.1016/0370-2693(89)91209-4}.
\bibAnnoteFile{Barbieri:1985cp}

\bibitem{Vorobev:1989hb}
P.~V. Vorob'ev, I.~V. Kolokolov, and V.~F. Fogel.
\newblock \enquote{{Ferromagnetic detector of (pseudo)Goldstone bosons}}.
\newblock \emph{JETP Lett.}, \textbf{50}, 65 (1989).
\newblock \url{http://jetpletters.ru/ps/1125/article_17041.shtml}.
\bibAnnoteFile{Vorobev:1989hb}

\bibitem{Kakhidze:1990in}
A.~I. Kakhidze and I.~V. Kolokolov.
\newblock \enquote{{Antiferromagnetic axions detector}}.
\newblock \emph{Sov. Phys. JETP}, \textbf{72}, 598 (1991).
\bibAnnoteFile{Kakhidze:1990in}

\bibitem{Vorobev:1995pb}
P.~V. Vorob'ev, A.~I. Kakhidze, and I.~V. Kolokolov.
\newblock \enquote{{Axion wind: A Search for cosmological axion condensate}}.
\newblock \emph{Phys. Atom. Nucl.}, \textbf{58}, 959 (1995).
\bibAnnoteFile{Vorobev:1995pb}

\bibitem{Graham:2017ivz}
P.~W. Graham \emph{et~al.}
\newblock \enquote{{Spin Precession Experiments for Light Axionic Dark
  Matter}}.
\newblock \emph{Phys. Rev. D}, \textbf{97(5)}, 055006 (2018).
\newblock \doi{10.1103/PhysRevD.97.055006}.
\newblock \eprint{1709.07852}.
\bibAnnoteFile{Graham:2017ivz}

\bibitem{Stadnik:2017mid}
Y.~Stadnik.
\newblock \emph{{Manifestations of Dark Matter and Variations of the
  Fundamental Constants of Nature in Atoms and Astrophysical Phenomena}}.
\newblock Ph.D. thesis, New South Wales U. (2017).
\newblock \doi{10.1007/978-3-319-63417-3}.
\bibAnnoteFile{Stadnik:2017mid}

\bibitem{Smorra:2019qfx}
C.~Smorra \emph{et~al.}
\newblock \enquote{{Direct limits on the interaction of antiprotons with
  axion-like dark matter}}.
\newblock \emph{Nature}, \textbf{575(7782)}, 310 (2019).
\newblock \doi{10.1038/s41586-019-1727-9}.
\newblock \eprint{2006.00255}.
\bibAnnoteFile{Smorra:2019qfx}

\bibitem{Graham:2020kai}
P.~W. Graham \emph{et~al.}
\newblock \enquote{{Storage ring probes of dark matter and dark energy}}.
\newblock \emph{Phys. Rev. D}, \textbf{103(5)}, 055010 (2021).
\newblock \doi{10.1103/PhysRevD.103.055010}.
\newblock \eprint{2005.11867}.
\bibAnnoteFile{Graham:2020kai}

\bibitem{Silenko:2021qgc}
A.~J. Silenko.
\newblock \enquote{{Relativistic spin dynamics conditioned by dark matter
  axions}}.
\newblock \emph{Eur. Phys. J. C}, \textbf{82(10)}, 856 (2022).
\newblock \doi{10.1140/epjc/s10052-022-10827-7}.
\newblock \eprint{2109.05576}.
\bibAnnoteFile{Silenko:2021qgc}

\bibitem{Kolya22}
N.~N. Nikolaev.
\newblock \enquote{{Spin of protons in NICA and PTR storage rings as an axion
  antenna}}.
\newblock \emph{Pisma Zh. Eksp. Teor. Fiz.}, \textbf{115(11)}, 683 (2022).
\newblock \doi{10.1134/S0021364022600653}.
\bibAnnoteFile{Kolya22}

\bibitem{Silenko:2014kia}
A.~J. Silenko.
\newblock \enquote{{High precision description and new properties of a spin-1
  particle in a magnetic field}}.
\newblock \emph{Phys. Rev. D}, \textbf{89(12)}, 121701 (2014).
\newblock \doi{10.1103/PhysRevD.89.121701}.
\newblock \eprint{1404.4953}.
\bibAnnoteFile{Silenko:2014kia}

\bibitem{Silenko:2017iyv}
A.~J. Silenko.
\newblock \enquote{{General description of spin motion in storage rings in the
  presence of oscillating horizontal fields}}.
\newblock \emph{EPL}, \textbf{118(6)}, 61003 (2017).
\newblock \doi{10.1209/0295-5075/118/61003}.
\newblock \eprint{1706.02065}.
\bibAnnoteFile{Silenko:2017iyv}

\bibitem{Bargmann:1959gz}
V.~Bargmann, L.~Michel, and V.~L. Telegdi.
\newblock \enquote{{Precession of the polarization of particles moving in a
  homogeneous electromagnetic field}}.
\newblock \emph{Phys. Rev. Lett.}, \textbf{2}, 435 (1959).
\newblock \doi{10.1103/PhysRevLett.2.435}.
\bibAnnoteFile{Bargmann:1959gz}

\bibitem{Fukuyama:2013ioa}
T.~Fukuyama and A.~J. Silenko.
\newblock \enquote{{Derivation of Generalized Thomas-Bargmann-Michel-Telegdi
  Equation for a Particle with Electric Dipole Moment}}.
\newblock \emph{Int. J. Mod. Phys.}, \textbf{A28}, 1350147 (2013).
\newblock \doi{10.1142/S0217751X13501479}.
\newblock \eprint{1308.1580}.
\bibAnnoteFile{Fukuyama:2013ioa}

\bibitem{Stephenson:2020jzx}
E.~Stephenson.
\newblock \enquote{{A Search for Axion-like Particles with a Horizontally
  Polarized Beam In a Storage Ring}}.
\newblock \emph{PoS}, \textbf{PSTP2019}, 018 (2020).
\newblock \doi{10.22323/1.379.0018}.
\bibAnnoteFile{Stephenson:2020jzx}

\bibitem{Gu16}
G.~Guidoboni \emph{et~al.}
\newblock \enquote{How to reach a thousand-second in-plane polarization
  lifetime with $0.97\text{\ensuremath{-}}\mathrm{GeV}/c$ deuterons in a
  storage ring}.
\newblock \emph{Phys. Rev. Lett.}, \textbf{117}, 054801 (2016).
\newblock \doi{10.1103/PhysRevLett.117.054801}.
\bibAnnoteFile{Gu16}

\bibitem{Ev15}
D.~Eversmann \emph{et~al.}
\newblock \enquote{New method for a continuous determination of the spin tune
  in storage rings and implications for precision experiments}.
\newblock \emph{Phys. Rev. Lett.}, \textbf{115}, 094801 (2015).
\newblock \doi{10.1103/PhysRevLett.115.094801}.
\bibAnnoteFile{Ev15}

\bibitem{Ma97}
R.~Maier.
\newblock \enquote{{Cooler synchrotron COSY: Performance and perspectives}}.
\newblock \emph{Nucl. Instrum. Meth. A}, \textbf{390}, 1 (1997).
\newblock \doi{10.1016/S0168-9002(97)00324-0}.
\bibAnnoteFile{Ma97}

\bibitem{Ha67}
W.~Haeberli.
\newblock \enquote{Sources of polarized ions}.
\newblock \emph{Annual Review of Nuclear Science}, \textbf{17(1)}, 373 (1967).
\newblock \doi{10.1146/annurev.ns.17.120167.002105}.
\bibAnnoteFile{Ha67}

\bibitem{Ch06}
D.~Chiladze \emph{et~al.}
\newblock \enquote{Determination of deuteron beam polarizations at {COSY}}.
\newblock \emph{Phys. Rev. ST Accel. Beams}, \textbf{9}, 050101 (2006).
\newblock \doi{10.1103/PhysRevSTAB.9.050101}.
\bibAnnoteFile{Ch06}

\bibitem{Ta18}
M.~Tanifuji.
\newblock \emph{Polarization Phenomena in Physics: Applications to Nuclear
  Reactions}.
\newblock World Scientific, Singapore (2018).
\newblock ISBN 987-981-3230-88-0.
\newblock \doi{10.1142/10731}.
\bibAnnoteFile{Ta18}

\bibitem{MC71}
H.~H. Barschall and W.~Haeberli (eds.).
\newblock \emph{The Madison Convention, Polarization Phenomena in Nuclear
  Reactions}.
\newblock University of Wisconsin Press, Madison, WI (1971).
\newblock \url{https://www.osti.gov/biblio/4726823}.
\bibAnnoteFile{MC71}

\bibitem{Mu20}
F.~Müller \emph{et~al.}
\newblock \enquote{{Measurement of deuteron carbon vector analyzing powers in
  the kinetic energy range 170–380 MeV}}.
\newblock \emph{Eur. Phys. J. A}, \textbf{56}, 1 (2020).
\newblock ISSN 1434-601X.
\newblock \doi{10.1140/EPJA/S10050-020-00215-8}.
\bibAnnoteFile{Mu20}

\bibitem{Ba14}
Z.~Bagdasarian \emph{et~al.}
\newblock \enquote{Measuring the polarization of a rapidly precessing deuteron
  beam}.
\newblock \emph{Phys. Rev. Accel. Beams}, \textbf{17}, 052803 (2014).
\newblock \doi{10.1103/PhysRevSTAB.17.052803}.
\bibAnnoteFile{Ba14}

\bibitem{Be12}
P.~Benati \emph{et~al.}
\newblock \enquote{Synchrotron oscillation effects on an rf-solenoid spin
  resonance}.
\newblock \emph{Phys. Rev. Accel. Beams}, \textbf{15}, 124202 (2012).
\newblock \doi{10.1103/PhysRevSTAB.15.124202}.
\bibAnnoteFile{Be12}

\bibitem{Be13}
P.~Benati \emph{et~al.}
\newblock \enquote{{Erratum: Synchrotron oscillation effects on an rf-solenoid
  spin resonance [Phys. Rev. ST Accel. Beams 15, 124202 (2012)]}}.
\newblock \emph{Phys. Rev. Accel. Beams}, \textbf{16}, 049901 (2013).
\newblock \doi{10.1103/PhysRevSTAB.16.049901}.
\bibAnnoteFile{Be13}

\bibitem{Gu18}
G.~Guidoboni \emph{et~al.}
\newblock \enquote{Connection between zero chromaticity and long in-plane
  polarization lifetime in a magnetic storage ring}.
\newblock \emph{Phys. Rev. Accel. Beams}, \textbf{21}, 024201 (2018).
\newblock \doi{10.1103/PhysRevAccelBeams.21.024201}.
\bibAnnoteFile{Gu18}

\bibitem{Sl16}
J.~Slim \emph{et~al.}
\newblock \enquote{{Electromagnetic Simulation and Design of a Novel Waveguide
  RF Wien Filter for Electric Dipole Moment Measurements of Protons and
  Deuterons}}.
\newblock \emph{Nuclear Instruments and Methods in Physics Research Section A:
  Accelerators, Spectrometers, Detectors and Associated Equipment},
  \textbf{828}, 116 (2016).
\newblock \doi{10.1016/j.nima.2016.05.012}.
\bibAnnoteFile{Sl16}

\bibitem{Slim_2020}
J.~Slim \emph{et~al.}
\newblock \enquote{The driving circuit of the waveguide {RF} {Wien} filter for
  the deuteron {EDM} precursor experiment at {COSY}}.
\newblock \emph{Journal of Instrumentation}, \textbf{15(03)}, P03021 (2020).
\newblock \doi{10.1088/1748-0221/15/03/p03021}.
\bibAnnoteFile{Slim_2020}

\bibitem{Rat2020}
F.~Rathmann, N.~N. Nikolaev, and J.~Slim.
\newblock \enquote{Spin dynamics investigations for the electric dipole moment
  experiment}.
\newblock \emph{Phys. Rev. Accel. Beams}, \textbf{23}, 024601 (2020).
\newblock \doi{10.1103/PhysRevAccelBeams.23.024601}.
\bibAnnoteFile{Rat2020}

\bibitem{Slim_2021}
J.~Slim \emph{et~al.}
\newblock \enquote{{First detection of collective oscillations of a stored
  deuteron beam with an amplitude close to the quantum limit}}.
\newblock \emph{Phys. Rev. Accel. Beams}, \textbf{24(12)}, 124601 (2021).
\newblock \doi{10.1103/PhysRevAccelBeams.24.124601}.
\newblock \eprint{2101.07582}.
\bibAnnoteFile{Slim_2021}

\bibitem{rawData}
{JEDI Collaboration}.
\newblock \enquote{{Replication Data for: First Search for Axion-Like Particles
  in a Storage Ring Using a Polarized Deuteron Beam}} (2022).
\newblock \doi{10.26165/JUELICH-DATA/HHNVQ3}.
\bibAnnoteFile{rawData}

\bibitem{Fe98}
G.~J. Feldman and R.~D. Cousins.
\newblock \enquote{Unified approach to the classical statistical analysis of
  small signals}.
\newblock \emph{Phys. Rev. D}, \textbf{57}, 3873 (1998).
\newblock \doi{10.1103/PhysRevD.57.3873}.
\bibAnnoteFile{Fe98}

\bibitem{SPCthesis}
S.~P. Chang.
\newblock \emph{{Studies on Axion-EDM experiment using the storage ring
  method}}.
\newblock {Ph.D. dissertation},
  \url{https://library.kaist.ac.kr/search/ctlgSearch/posesn/view.do?bibctrlno=996474&se=t0&ty=B&_csrf=cdfcdff0-c27d-4b60-9bef-cbb0967f7d6d},
  {The Korea Advanced Institute of Science and Technology (KAIST), Daedeok
  Innopolis, Daejeon, South Korea} (2022).
\bibAnnoteFile{SPCthesis}

\bibitem{Pl14}
S.~Plaszczynski \emph{et~al.}
\newblock \enquote{{A novel estimator of the polarization amplitude from
  normally distributed Stokes parameters}}.
\newblock \emph{Monthly Notices of the Royal Astronomical Society},
  \textbf{439(4)}, 4048 (2014).
\newblock ISSN 0035-8711.
\newblock \doi{10.1093/mnras/stu270}.
\bibAnnoteFile{Pl14}

\bibitem{Ev16}
D.~Eversmann, J.~Pretz, and M.~Rosenthal.
\newblock \enquote{Amplitude estimation of a sine function based on confidence
  intervals and {Bayes}{\textquotesingle} theorem}.
\newblock \emph{Journal of Instrumentation}, \textbf{2016(11)}, P05003 (2016).
\newblock \doi{10.1088/1748-0221/11/05/p05003}.
\bibAnnoteFile{Ev16}

\bibitem{Ay21}
D.~Aybas \emph{et~al.}
\newblock \enquote{Search for axionlike dark matter using solid-state nuclear
  magnetic resonance}.
\newblock \emph{Phys. Rev. Lett.}, \textbf{126}, 141802 (2021).
\newblock \doi{10.1103/PhysRevLett.126.141802}.
\bibAnnoteFile{Ay21}

\bibitem{Schulthess:2022pbp}
I.~Schulthess \emph{et~al.}
\newblock \enquote{{New Limit on Axionlike Dark Matter Using Cold Neutrons}}.
\newblock \emph{Phys. Rev. Lett.}, \textbf{129(19)}, 191801 (2022).
\newblock \doi{10.1103/PhysRevLett.129.191801}.
\newblock \eprint{2204.01454}.
\bibAnnoteFile{Schulthess:2022pbp}

\bibitem{AxLim}
C.~O'Hare.
\newblock \enquote{cajohare/axionlimits: Axionlimits}.
\newblock \url{https://cajohare.github.io/AxionLimits/} (2020).
\newblock \doi{10.5281/zenodo.3932430}.
\bibAnnoteFile{AxLim}

\bibitem{Bar:2019ifz}
N.~Bar, K.~Blum, and G.~D'Amico.
\newblock \enquote{{Is there a supernova bound on axions?}}
\newblock \emph{Phys. Rev. D}, \textbf{101(12)}, 123025 (2020).
\newblock \doi{10.1103/PhysRevD.101.123025}.
\newblock \eprint{1907.05020}.
\bibAnnoteFile{Bar:2019ifz}

\bibitem{Caloni_2022}
L.~Caloni \emph{et~al.}
\newblock \enquote{Novel cosmological bounds on thermally-produced axion-like
  particles}.
\newblock \emph{Journal of Cosmology and Astroparticle Physics},
  \textbf{2022(09)}, 021 (2022).
\newblock \doi{10.1088/1475-7516/2022/09/021}.
\bibAnnoteFile{Caloni_2022}

\bibitem{Pospelov:1999ha}
M.~Pospelov and A.~Ritz.
\newblock \enquote{{Theta induced electric dipole moment of the neutron via QCD
  sum rules}}.
\newblock \emph{Phys. Rev. Lett.}, \textbf{83}, 2526 (1999).
\newblock \doi{10.1103/PhysRevLett.83.2526}.
\newblock \eprint{hep-ph/9904483}.
\bibAnnoteFile{Pospelov:1999ha}

\bibitem{Crewther:1979pi}
R.~J. Crewther \emph{et~al.}
\newblock \enquote{{Chiral Estimate of the Electric Dipole Moment of the
  Neutron in Quantum Chromodynamics}}.
\newblock \emph{Phys. Lett. B}, \textbf{88}, 123 (1979).
\newblock \doi{10.1016/0370-2693(79)90128-X}.
\newblock [Erratum: Phys.Lett.B 91, 487 (1980)].
\bibAnnoteFile{Crewther:1979pi}

\bibitem{Baluni:1978rf}
V.~Baluni.
\newblock \enquote{{CP-nonconserving effects in quantum chromodynamics}}.
\newblock \emph{Phys. Rev. D}, \textbf{19}, 2227 (1979).
\newblock \doi{10.1103/PhysRevD.19.2227}.
\bibAnnoteFile{Baluni:1978rf}

\bibitem{Ottnad:2009jw}
K.~Ottnad \emph{et~al.}
\newblock \enquote{{New insights into the neutron electric dipole moment}}.
\newblock \emph{Phys. Lett. B}, \textbf{687}, 42 (2010).
\newblock \doi{10.1016/j.physletb.2010.03.005}.
\newblock \eprint{0911.3981}.
\bibAnnoteFile{Ottnad:2009jw}

\bibitem{Yamanaka:2015qfa}
N.~Yamanaka and E.~Hiyama.
\newblock \enquote{{Enhancement of the CP-odd effect in the nuclear electric
  dipole moment of $^6$Li}}.
\newblock \emph{Phys. Rev. C}, \textbf{91(5)}, 054005 (2015).
\newblock \doi{10.1103/PhysRevC.91.054005}.
\newblock \eprint{1503.04446}.
\bibAnnoteFile{Yamanaka:2015qfa}

\bibitem{Bsaisou:2014zwa}
J.~Bsaisou \emph{et~al.}
\newblock \enquote{{Nuclear Electric Dipole Moments in Chiral Effective Field
  Theory}}.
\newblock \emph{JHEP}, \textbf{2015(03)}, 104 (2015).
\newblock \doi{10.1007/JHEP03(2015)104}.
\newblock {[Erratum: \href{https://doi.org/10.1007/JHEP05(2015)083}{JHEP 05,
  083 (2015)}.]}, \eprint{1411.5804}.
\bibAnnoteFile{Bsaisou:2014zwa}

\bibitem{Khriplovich:1999qr}
I.~B. Khriplovich and R.~A. Korkin.
\newblock \enquote{{P and T odd electromagnetic moments of deuteron in chiral
  limit}}.
\newblock \emph{Nucl. Phys. A}, \textbf{665}, 365 (2000).
\newblock \doi{10.1016/S0375-9474(99)00403-0}.
\newblock \eprint{nucl-th/9904081}.
\bibAnnoteFile{Khriplovich:1999qr}

\bibitem{Lebedev:2004va}
O.~Lebedev \emph{et~al.}
\newblock \enquote{{Probing CP violation with the deuteron electric dipole
  moment}}.
\newblock \emph{Phys. Rev. D}, \textbf{70}, 016003 (2004).
\newblock \doi{10.1103/PhysRevD.70.016003}.
\newblock \eprint{hep-ph/0402023}.
\bibAnnoteFile{Lebedev:2004va}

\bibitem{Liu:2004tq}
C.~P. Liu and R.~G.~E. Timmermans.
\newblock \enquote{{P- and T-odd two-nucleon interaction and the deuteron
  electric dipole moment}}.
\newblock \emph{Phys. Rev. C}, \textbf{70}, 055501 (2004).
\newblock \doi{10.1103/PhysRevC.70.055501}.
\newblock \eprint{nucl-th/0408060}.
\bibAnnoteFile{Liu:2004tq}

\bibitem{Afnan:2010xd}
I.~R. Afnan and B.~F. Gibson.
\newblock \enquote{{Model Dependence of the 2H Electric Dipole Moment}}.
\newblock \emph{Phys. Rev. C}, \textbf{82}, 064002 (2010).
\newblock \doi{10.1103/PhysRevC.82.064002}.
\newblock \eprint{1011.4968}.
\bibAnnoteFile{Afnan:2010xd}

\bibitem{deVries:2011re}
J.~de~Vries \emph{et~al.}
\newblock \enquote{{Parity- and Time-Reversal-Violating Form Factors of the
  Deuteron}}.
\newblock \emph{Phys. Rev. Lett.}, \textbf{107}, 091804 (2011).
\newblock \doi{10.1103/PhysRevLett.107.091804}.
\newblock \eprint{1102.4068}.
\bibAnnoteFile{deVries:2011re}

\bibitem{Bsaisou:2012rg}
J.~Bsaisou \emph{et~al.}
\newblock \enquote{{The electric dipole moment of the deuteron from the QCD
  $\theta$-term}}.
\newblock \emph{Eur. Phys. J. A}, \textbf{49}, 31 (2013).
\newblock \doi{10.1140/epja/i2013-13031-x}.
\newblock \eprint{1209.6306}.
\bibAnnoteFile{Bsaisou:2012rg}

\bibitem{Bsaisou:2014oka}
J.~Bsaisou \emph{et~al.}
\newblock \enquote{{P- and T-Violating Lagrangians in Chiral Effective Field
  Theory and Nuclear Electric Dipole Moments}}.
\newblock \emph{Annals Phys.}, \textbf{359}, 317 (2015).
\newblock \doi{10.1016/j.aop.2015.04.031}.
\newblock \eprint{1412.5471}.
\bibAnnoteFile{Bsaisou:2014oka}

\bibitem{Wirzba:2016saz}
A.~Wirzba, J.~Bsaisou, and A.~Nogga.
\newblock \enquote{{Permanent Electric Dipole Moments of Single-, Two-, and
  Three-Nucleon Systems}}.
\newblock \emph{Int. J. Mod. Phys. E}, \textbf{26(01n02)}, 1740031 (2017).
\newblock \doi{10.1142/S0218301317400316}.
\newblock \eprint{1610.00794}.
\bibAnnoteFile{Wirzba:2016saz}

\bibitem{Hook:2017psm}
A.~Hook and J.~Huang.
\newblock \enquote{{Probing axions with neutron star inspirals and other
  stellar processes}}.
\newblock \emph{JHEP}, \textbf{2018(06)}, 036 (2018).
\newblock \doi{10.1007/JHEP06(2018)036}.
\newblock \eprint{1708.08464}.
\bibAnnoteFile{Hook:2017psm}

\bibitem{Blum:2014vsa}
K.~Blum \emph{et~al.}
\newblock \enquote{{Constraining Axion Dark Matter with Big Bang
  Nucleosynthesis}}.
\newblock \emph{Phys. Lett. B}, \textbf{737}, 30 (2014).
\newblock \doi{10.1016/j.physletb.2014.07.059}.
\newblock \eprint{1401.6460}.
\bibAnnoteFile{Blum:2014vsa}

\bibitem{Raffelt:2006cw}
G.~G. Raffelt.
\newblock \enquote{{Astrophysical axion bounds}}.
\newblock \emph{Lect. Notes Phys.}, \textbf{741}, 51 (2008).
\newblock \doi{10.1007/978-3-540-73518-2_3}.
\newblock \eprint{hep-ph/0611350}.
\bibAnnoteFile{Raffelt:2006cw}

\bibitem{Carenza:2019pxu}
P.~Carenza \emph{et~al.}
\newblock \enquote{{Improved axion emissivity from a supernova via
  nucleon-nucleon bremsstrahlung}}.
\newblock \emph{JCAP}, \textbf{2019(10)}, 016 (2019).
\newblock \doi{10.1088/1475-7516/2019/10/016}.
\newblock [Erratum: JCAP 05, E01 (2020)], \eprint{1906.11844}.
\bibAnnoteFile{Carenza:2019pxu}

\bibitem{Bloch:2019lcy}
I.~M. Bloch \emph{et~al.}
\newblock \enquote{{Axion-like Relics: New Constraints from Old Comagnetometer
  Data}}.
\newblock \emph{JHEP}, \textbf{2020(01)}, 167 (2020).
\newblock \doi{10.1007/JHEP01(2020)167}.
\newblock \eprint{1907.03767}.
\bibAnnoteFile{Bloch:2019lcy}

\bibitem{Bloch:2021vnn}
I.~M. Bloch \emph{et~al.}
\newblock \enquote{{New constraints on axion-like dark matter using a Floquet
  quantum detector}}.
\newblock \emph{Sci. Adv.}, \textbf{8(5)}, abl8919 (2022).
\newblock \doi{10.1126/sciadv.abl8919}.
\newblock \eprint{2105.04603}.
\bibAnnoteFile{Bloch:2021vnn}

\bibitem{JacksonKimball:2017elr}
D.~F. Jackson~Kimball \emph{et~al.}
\newblock \enquote{{Overview of the Cosmic Axion Spin Precession Experiment
  (CASPEr)}}.
\newblock \emph{Springer Proc. Phys.}, \textbf{245}, 105 (2020).
\newblock \doi{10.1007/978-3-030-43761-9_13}.
\newblock \eprint{1711.08999}.
\bibAnnoteFile{JacksonKimball:2017elr}

\bibitem{Garcon:2019inh}
A.~Garcon \emph{et~al.}
\newblock \enquote{{Constraints on bosonic dark matter from ultralow-field
  nuclear magnetic resonance}}.
\newblock \emph{Sci. Adv.}, \textbf{5(10)}, eaax4539 (2019).
\newblock \doi{10.1126/sciadv.aax4539}.
\newblock \eprint{1902.04644}.
\bibAnnoteFile{Garcon:2019inh}

\bibitem{Adelberger:2006dh}
E.~G. Adelberger \emph{et~al.}
\newblock \enquote{{Particle Physics Implications of a Recent Test of the
  Gravitational Inverse Sqaure Law}}.
\newblock \emph{Phys. Rev. Lett.}, \textbf{98}, 131104 (2007).
\newblock \doi{10.1103/PhysRevLett.98.131104}.
\newblock \eprint{hep-ph/0611223}.
\bibAnnoteFile{Adelberger:2006dh}

\bibitem{Vasilakis2009}
G.~{Vasilakis} \emph{et~al.}
\newblock \enquote{{Limits on New Long Range Nuclear Spin-Dependent Forces Set
  with a K-He3 Comagnetometer}}.
\newblock \emph{Phys. Rev. Lett.}, \textbf{103(26)}, 261801 (2009).
\newblock \doi{10.1103/PhysRevLett.103.261801}.
\newblock \eprint{0809.4700}.
\bibAnnoteFile{Vasilakis2009}

\bibitem{Bhusal:2020bvx}
A.~Bhusal, N.~Houston, and T.~Li.
\newblock \enquote{{Searching for Solar Axions Using Data from the Sudbury
  Neutrino Observatory}}.
\newblock \emph{Phys. Rev. Lett.}, \textbf{126(9)}, 091601 (2021).
\newblock \doi{10.1103/PhysRevLett.126.091601}.
\newblock \eprint{2004.02733}.
\bibAnnoteFile{Bhusal:2020bvx}

\bibitem{Buschmann:2021juv}
M.~Buschmann \emph{et~al.}
\newblock \enquote{{Upper Limit on the QCD Axion Mass from Isolated Neutron
  Star Cooling}}.
\newblock \emph{Phys. Rev. Lett.}, \textbf{128(9)}, 091102 (2022).
\newblock \doi{10.1103/PhysRevLett.128.091102}.
\newblock \eprint{2111.09892}.
\bibAnnoteFile{Buschmann:2021juv}

\bibitem{Pretz:2019ham}
J.~Pretz \emph{et~al.}
\newblock \enquote{{Statistical sensitivity estimates for oscillating electric
  dipole moment measurements in storage rings}}.
\newblock \emph{Eur. Phys. J. C}, \textbf{80(2)}, 107 (2020).
\newblock \doi{10.1140/epjc/s10052-020-7664-9}.
\newblock \eprint{1908.09678}.
\bibAnnoteFile{Pretz:2019ham}

\end{thebibliography}

\end{document}